\documentclass[twocolumn]{aastex6}
\usepackage{natbib}
\usepackage{color}
\usepackage{hyperref} 
\usepackage{subfig}

\def    \apjl  		{\rm {ApJL}}
\def    \apj  		{\rm {ApJ}}
\def    \mnras  	{\rm {MNRAS}}

\def    \apjl  		{\rm {ApJL}}

\def	\cm		{\,{\rm {cm}}}
\def	\K		{\,{\rm K}}

\def	\mum	{\,{\mu \rm{m}}}

\def \bea {\begin{eqnarray}}
\def \ena {\end{eqnarray}}

\def	\ted	{{\tau_{\rm ed}}}
\def	\tH	{{\tau_{\rm H}}}

\def	\cm	{\,{\rm cm}}

\def	\D	{{\rm D}}

\def	\erg	{\,{\rm erg}}

\def	\gas	{\,{\rm gas}}

\def	\H	{{\rm H}}

\def	\Hz	{\,{\rm Hz}}

\def	\pc	{\,{\rm pc}}

\def	\s	{\,{\rm s}}

\def	\rot {{\rm rot}}

\def	\km 	{\,{\rm km}}
\def	\yr 	{\,{\rm yr}}

\def    \gas     	{{\rm gas}}

\begin{document}
\shorttitle{Dust rotational dynamics in non-stationary shocks}
\shortauthors{Tram and Hoang}
\title{Dust rotational dynamics in non-stationary shock: Rotational disruption of nanoparticles by stochastic mechanical torques and spinning dust emission}
\author{Le Ngoc Tram}
\affil{Korea Astronomy and Space Science Institute, Daejeon 34055, Republic of Korea; \href{mailto:lengoctramlyk31@gmail.com}{ngoctram.le@nasa.gov}}
\affil{SOFIA-USRA, Nasa Ames Research Center, Ms 232-11, Moffett Field, CA 94035, USA\\}
\affil{University of Science and Technology of Hanoi, VAST, 18 Hoang Quoc Viet, Hanoi, Vietnam}
\author{Thiem Hoang}
\affil{Korea Astronomy and Space Science Institute, Daejeon 34055, Republic of Korea\\}
\affil{Korea University of Science and Technology, 217 Gajeong-ro, Yuseong-gu, Daejeon, 34113, Republic of Korea}

\begin{abstract}
In a previous work, Hoang and Tram discovered a new mechanism for destruction of nanoparticles due to suprathermal rotation of grains in stationary C-shocks, which is termed rotational disruption. In this paper, we extend our previous study for non-stationary shocks driven by outflows and young supernovae remnants that have dynamical ages shorter than the time required to establish a stationary C-shock, which is composed of a C-shock and a J-shock tail (referred as CJ-shock). For the C-shock component, we find that smallest nanoparticles (size $\lesssim 1$ nm) of weak materials (i.e., tensile strength $S_{\rm max}\lesssim 10^{9}\erg\cm^{-3}$) can be rotationally disrupted due to suprathermal rotation induced by supersonic neutral drift. For the J-shock component, although nanoparticles are rotating thermally, the smallest ones can still be disrupted because the gas is heated to higher temperatures by J-shocks. We then model microwave emission from rapidly spinning nanoparticles where the grain size distribution has the lower cutoff determined by rotational disruption for the different shock models. We also calculate the spectral flux of microwave emission from a shocked region at distance of 100 pc from the observer for the different gas density, shock age, and shock velocities. We suggest that microwave emission from spinning dust can be used to trace nanoparticles and shock velocities in dense molecular outflows. Finally, we discuss a new way that can release molecules from the nanoparticle surface into the gas in the shocked regions, which we name rotational desorption. 

\end{abstract}

\section{Introduction\label{sec:intro}}
Very small dust grains, whose size is below 10 nm (hereafter referred to as nanoparticles), including Polycyclic Aromatic Hydrocarbons (PAHs), play an important role in the evolution of the interstellar medium (ISM). \cite{2001ApJS..134..263W} and \cite{2013ApJ...766....8A} demonstrated that nanoparticles can control the heating process of gas, while other studies (e.g., \citealt{1956MNRAS.116..503M}, \citealt{2016MNRAS.460.2050Z}) found that nanoparticles can influence the dynamics of molecular clouds and star formation due to their dominant charge carrier in dense and low ionization molecular clouds. 

PAHs and nanoparticles are expected to be abundant in shocked regions due to shattering of large grains via grain-grain collisions (e.g., \citealt{1994ApJ...431..321T}; \citealt{1994ApJ...433..797J}). Nevertheless, most of observations show the lack of strong PAH emission features in supernova remnants (\citealt{2009ApJ...693..713S}) and outflows of massive young stellar objects (YSOs) (\citealt{2006ApJ...645.1264S}) where shocks are present. It suggests that PAHs/nanoparticles are perhaps efficiently destroyed in the shocked regions. 

For stationary magnetized shock models (\citealt{2015A&A...578A..63F}), \cite{2019ApJ...877...36H} found that the supersonic drift of charged nanoparticles relative to neutrals can rapidly spin nanoparticles up to suprathermal rotation. As a result, smallest nanoparticles ($a\lesssim 1$ nm) can be disrupted into tiny fragments when the centrifugal stress induced by grain rotation exceeds the maximum tensile strength of the grain material. This mechanism is found to be the most efficient in destroying nanoparticles in C-shocks compared to previously known mechanisms such as thermal sputtering and grain shattering (see \citealt{2019ApJ...877...36H} for details).

In the case of shocks driven by outflows from (YSOs) and young supernova remnants (SNRs), shocks cannot reach the steady stage because the required timescale is longer than the dynamical age of outflows and SNRs. For instance, the dynamical age of the BHR 71 bipolar outflow is $\sim 4000\yr$ (\citealt{2015A&A...575A..98G}), of the blue lobe of the L1157 outflow is $\sim 3000\yr$ (\citealt{1998A&A...333..287G}), of SNR N132D is $\sim 2500\yr$ (\citealt{2006ApJ...653..267T}) or SNR IC443 is $\sim 4000\yr$ (\citealt{2008A&A...485..777T}), while it requires $\sim 10^{4}\yr$ for the C-shock to reache the steady state. Therefore, it is necessary to account for non-stationary shocks (e.g., \citealt{2003MNRAS.341...70F}; \citealt{2004A&A...419..999G, 2006A&A...459..821G}; \citealt{2008A&A...490..695G, 2015A&A...575A..98G}). 

 \cite{1998MNRAS.295..672C} and \cite{2004A&A...427..147L, 2004A&A...427..157L} discovered that non-stationary shocks are composed of a magnetic precursor and a J-type tail, so-called CJ-shock (see Section \ref{sec:model}). The key difference of the J-shock from the C-shock stage is that the gas can be heated to higher temperatures, resulting in an enhanced rotation rate of nanoparticles compared to the C-shock stage in the absence of supersonic drift. As a result, we expect that rotational disruption is also efficient in J-shock tails even without supersonic neutral drift. The goal of this paper is to quantify the efficiency of rotational disruption and model microwave emission from spinning dust for CJ-shocks. 
 
 
 The paper is structured as follows. In Section \ref{sec:model}, We describe the profiles of gas temperature, velocities of neutral, ion, and charged nanoparticles in CJ-type shock model, which are specially computed in dense magnetized clouds conditions. Section \ref{sec:rot} is a brief description of the rotational dynamics of dust grains in CJ-shocks. The mechanism of rotational disruption, and therefore the minimum size of survival nanoparticles through the shock are presented in Section \ref{sec:disrupt}. In Section \ref{sec:spindust}, we calculate spinning dust emissivity and emission flux from nanoparticles in the CJ-shock regions. In Section \ref{sec:discuss}, we discuss the importance of extremely fast rotating nanoparticles on grain surface chemistry, and potential application for probing nanoparticles in shocks, as well as shock tracing. Section \ref{sec:sum} summarizes our main findings.

\section{Structures of non-stationary shocks}\label{sec:model}
\subsection{C-shock, J-shock and CJ-shock}
The existence of magnetic fields and the ionization fraction affect the shock structure. When the magnetic field is weak or the ionization fraction is large, the shocks behave like hydrodynamic one since all of its neutral and charged particle components have the same velocity, including an extra contribution of the magnetic pressure. Because shocks are faster than the signal speed (e.g., sound speed) in the pre-shock medium, the signal speed thus can not "feel" the shock wave before it arrives. The shock properties (e.g., temperature, velocity, density) then abruptly vary as a viscous discontinuity jump (the so-called J-type shock) across the shock front. 

When the magnetic field is significant, it interacts directly with the charged component and slows it down, which makes the neutral and charged components have different velocities ($v_{n}> v_{i}$). If the ionization fraction is small, the speed of magnetosonic waves of the charged component $v_{m}$ is greater than the entrance shock speed. This forms a magnetic precursor upstream of the discontinuity, in which the charged and neutral fluids dynamically decouple. Thus, the neutral fluid is heated up and accelerated due to the consequent friction between the two fluids. The precursor size increases with increasing the magnetic field intensity, and hence the neutrals are compressed sooner before the arrival of the shock front. Eventually, the discontinuity disappears, and the shock properties now change continuously (the so-called C-type shock). In this case, the kinetic energy dissipates much more gradually because of the friction between the neutral and charged components, and the C-shock volume is therefore much larger.

For young C-type shocks (at early age), the shock is actually composed of a magnetic precursor and a J-type tail \cite{1998MNRAS.295..672C} (the so-called time-dependent CJ-type shock). \cite{2004A&A...427..157L} indicated that in the large compression case, which is appropriate in dense media, the J-type front in the young C-type shock is inserted when the flow time in the charged fluid is equal to the shock age. The J-type shock ends when the total neutral flow time across the J-type part reaches the age of the shock. 

\subsection{CJ-Shock structures and physical parameters}
As in our previous work, we model non-stationary shock structure with different shocked parameters (see Table \ref{tab:ISM}) using the 1D MHD shock code, namely Paris-Durham \citep{2015A&A...578A..63F}. Note that the C-type shock reaches the steady state at a typical time about $t_{\s}=10^{6}\yr/(n_\H/10^{2}\cm^{-3})$ for the standard ISM with the scaled radiation field $G_0=1$ \citep{2004A&A...427..147L}. So, in order to take into account the effect of the finite shock age, we consider two different values of age: $10^{2}\yr$ and $10^{3}\yr$ for $n_\H=10^{4}\cm^{-3}$, a ten times shorter for a density of $n_\H=10^{5}\cm^{-3}$, and hundred times shorter for a density of $n_\H=10^{6}\cm^{-3}$. The magnetic field strength is evaluated for $b=2$ (see \citealt{2019ApJ...877...36H} for an explanation of this choice). 

\begin{figure*}
  \centering
    \subfloat{
        \includegraphics[width=0.45\textwidth]{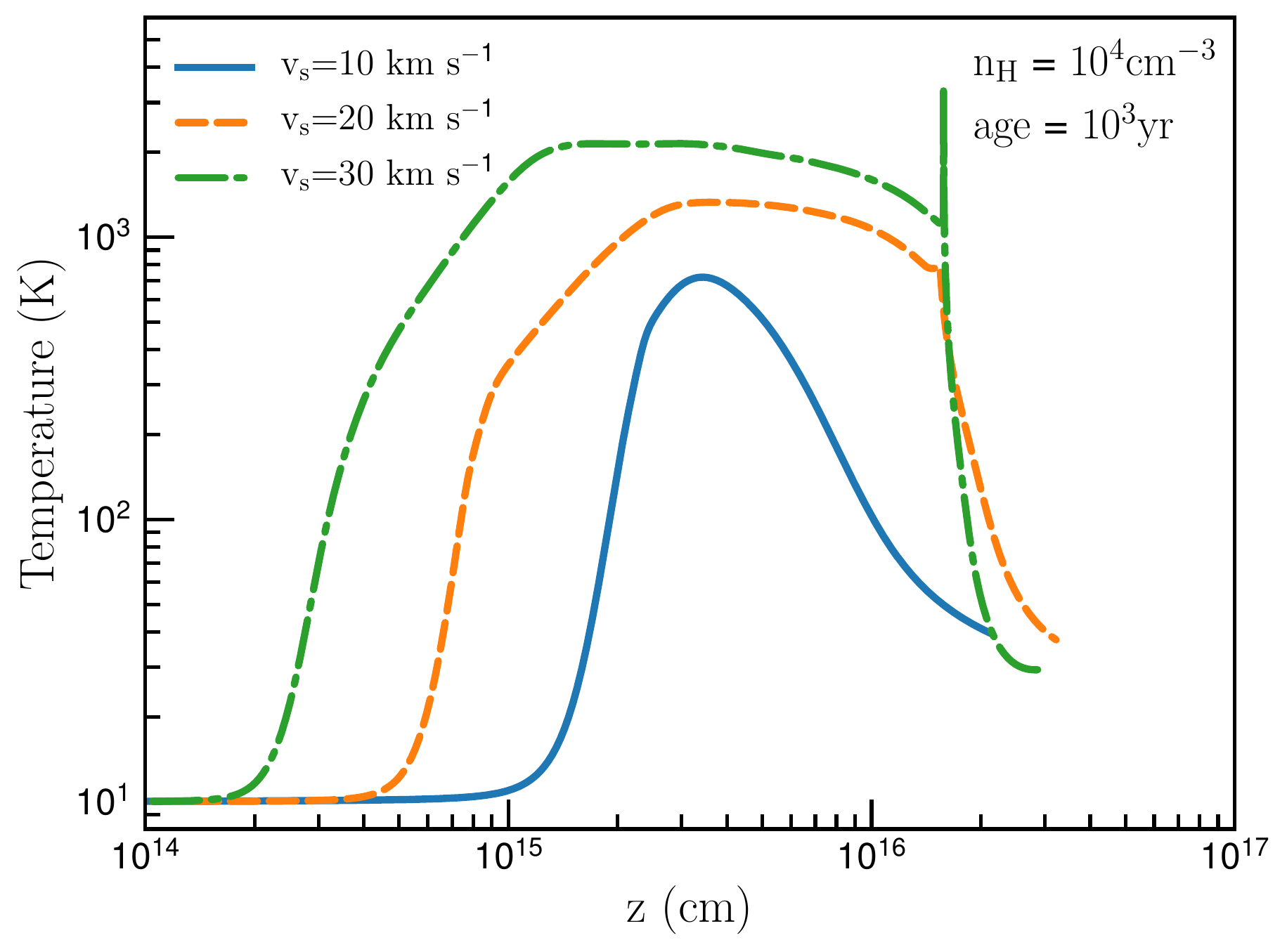}
        \includegraphics[width=0.44\textwidth]{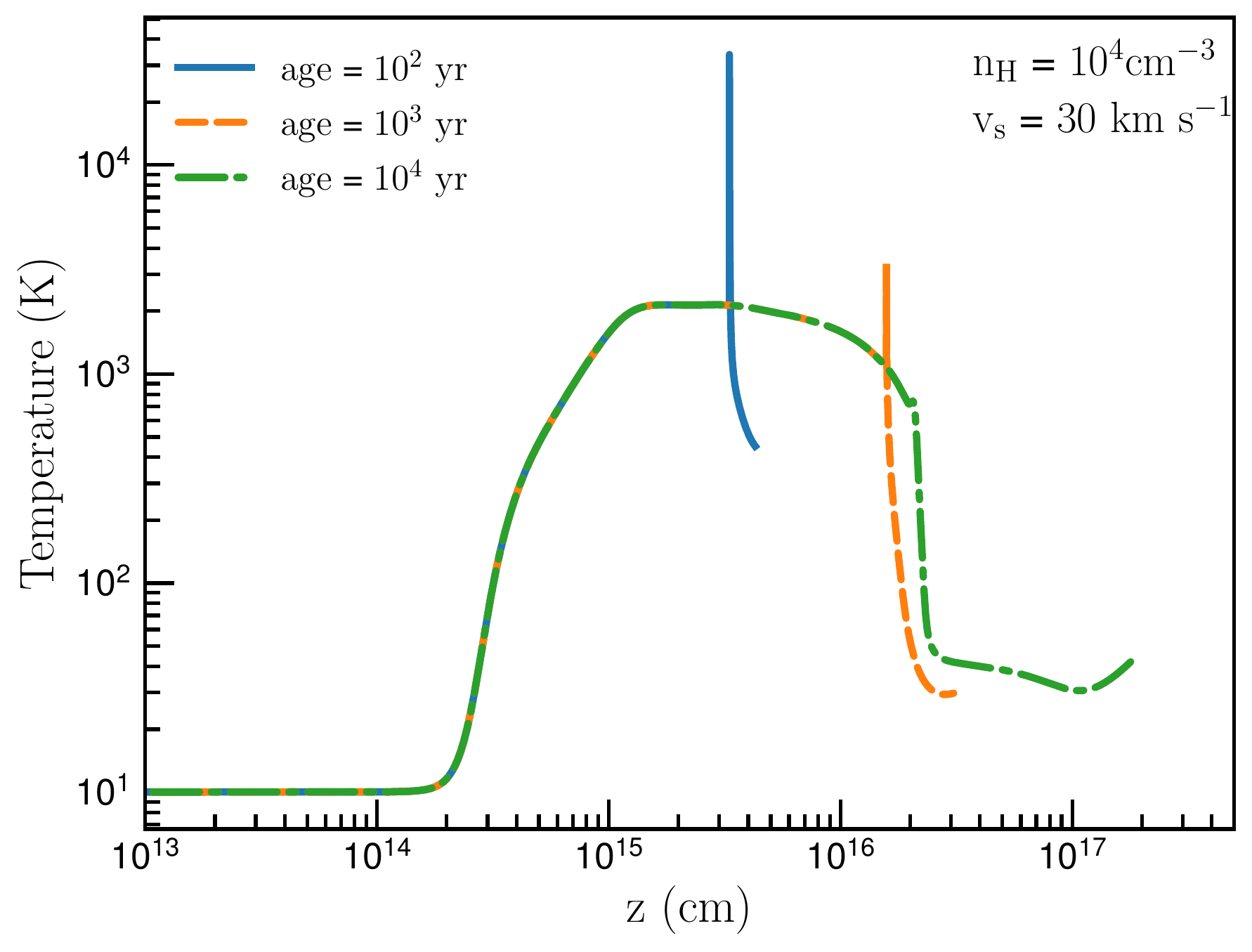}
    }\\
    \subfloat{
        \includegraphics[width=0.45\textwidth]{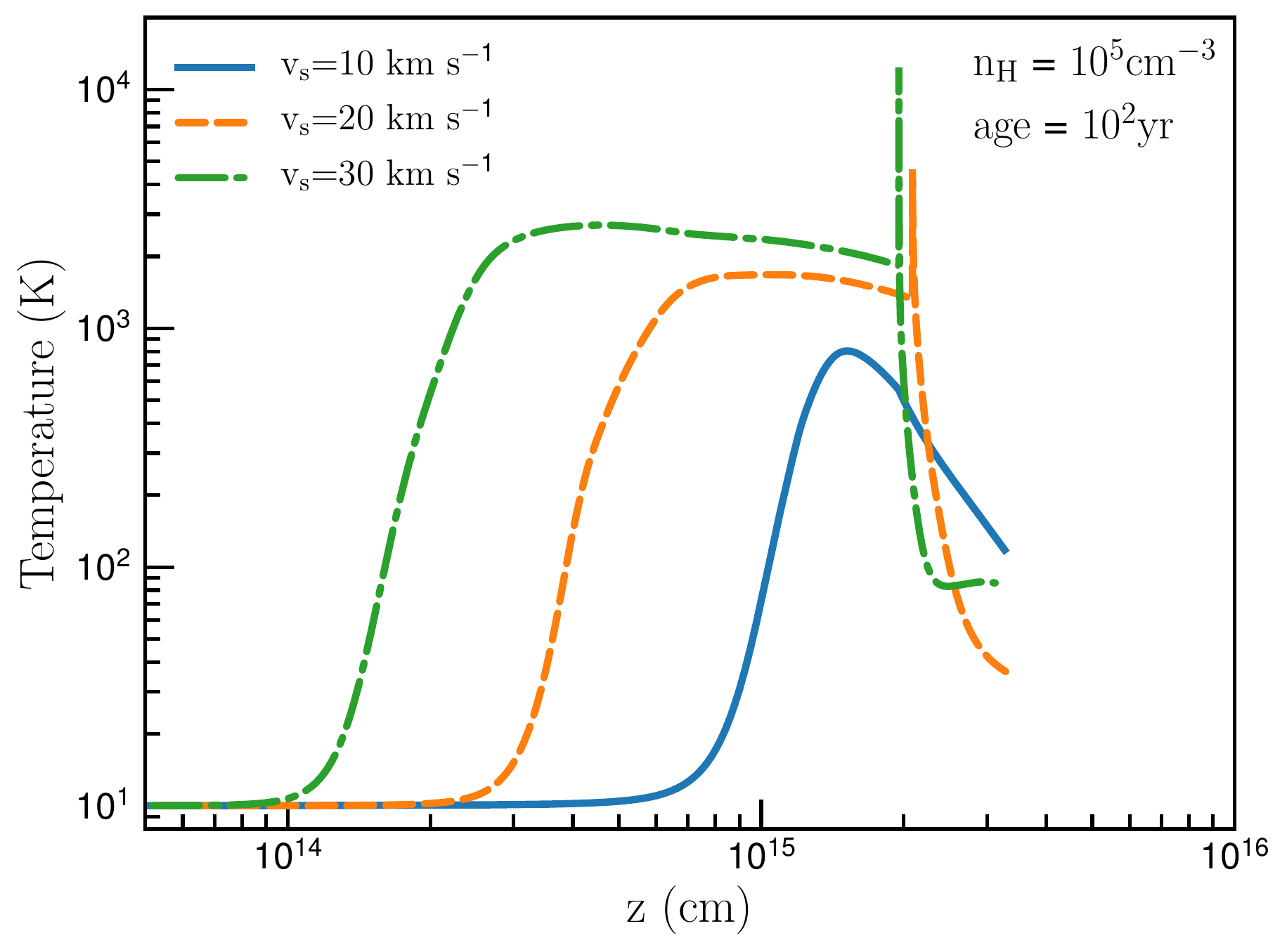}
        \includegraphics[width=0.45\textwidth]{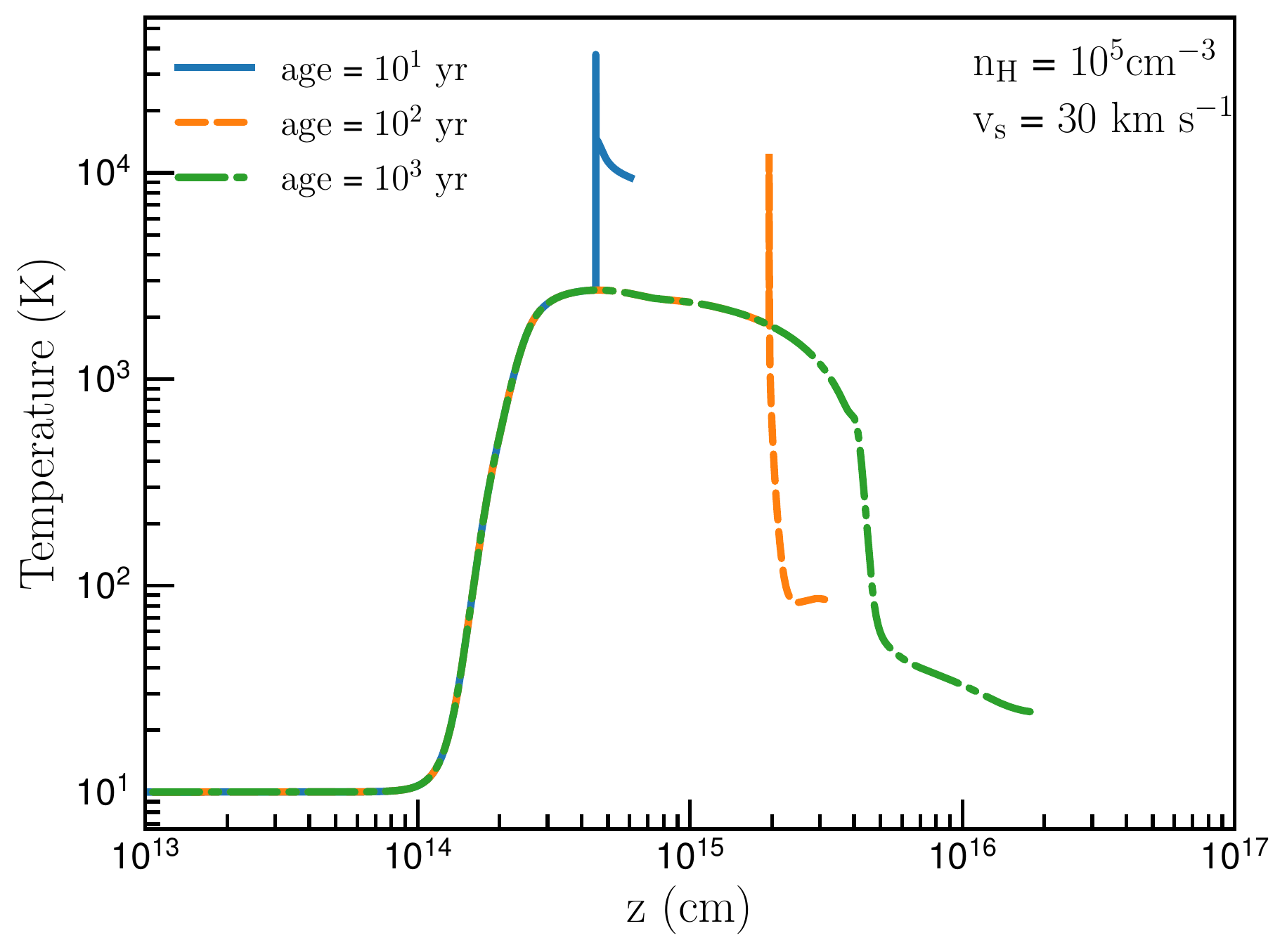}
    }\\
    \subfloat{
        \includegraphics[width=0.45\textwidth]{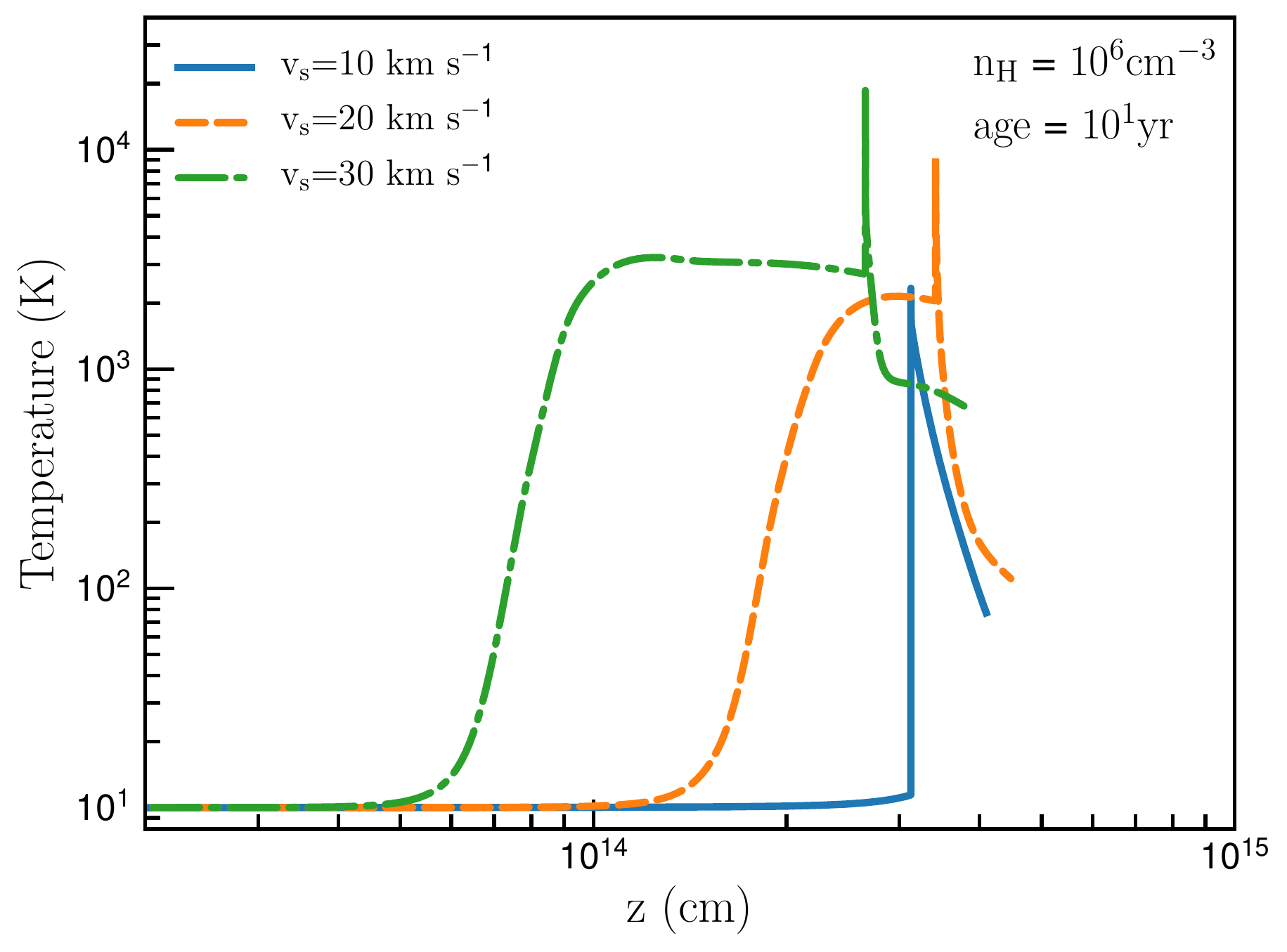}
        \includegraphics[width=0.44\textwidth]{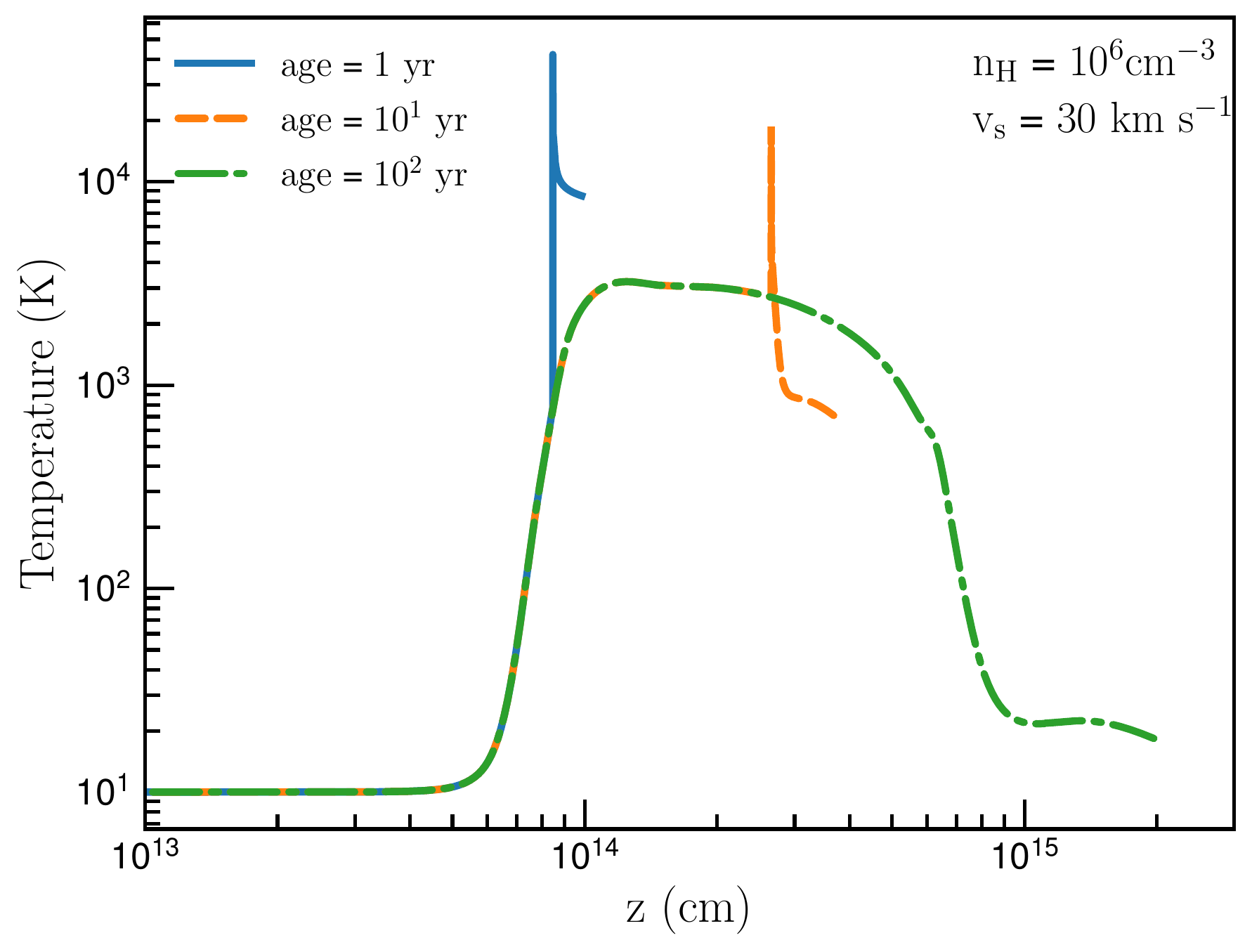}
    }
    \caption{Profiles of gas temperature for $\rm n_{\H}=10^{4}$ cm$^{-3}$, $\rm n_{\H}=10^{5}$ cm$^{-3}$, and $\rm n_{\H}=10^{6}$ cm$^{-3}$. Left panels: computed with three different shock velocities (shock age fixed). Right panels: computed with three different shock ages (shock velocity fixed). The maximum gas temperatures increases with $\rm v_{s}$ and decreases with $\rm age$.}
    \label{fig:profile_temp}
\end{figure*}

Profiles of the gas temperature in non-stationary shocks are shown in Figures \ref{fig:profile_temp} for the different gas densities and shock ages. For a given dynamical age, shocks in a denser cloud sweep and compress the gas stronger and earlier than in a less dense cloud, such that the peak temperature of the shocked gas in the former is higher and increases much sooner than in the latter case. The shock age also influences the gas temperature. Indeed, as the shock gets older, the magnetic precursor grows larger, and the entrance shock speed into the J-shock front decreases due to ion-neutral collisions. As a result, the maximum temperature of the J-shock component decreases with the shock age. When the age equals to $t_{\s}$, the J-shock tail disappears, and the shock returns to a stationary C-shock (see the green dashed-dotted lines on the right panels). 

\begin{table}
\begin{center}
\caption{Shock Model Parameters}\label{tab:ISM}
\begin{tabular}{l l l l } \hline \hline \\
{\it Parameters} & {Model A} & {Model B} & {Model C}\cr
\hline\\
$v_{\rm s}$(km$\,$s$^{-1}$) &$5-30$ &$5-30$ &$5-30$\cr
$n_{\rm H}(\cm^{-3})$ & $10^{4}$ & $10^{5}$ & $10^{6}$\cr
$age(\yr)$ & $10^{2},\,10^{3}$ & $10^{1},\,10^{2}$ & $10^{0},\,10^{1}$\cr
$T_{\rm gas}$(K)& 10 & 10 & 10 \cr
$T_{\rm d}$(K)& 10 & 10 & 10 \cr
$\chi$ &$0.01$ & 0.01 & 0.01\cr
$x_{\rm H}\equiv n(\H^{+})/n_{\H}$ &$0$ &$0$ &$0$\cr
$x_{\rm M}\equiv n(\rm M^{+})/n_{\H}$ &$10^{-6}$ &$10^{-6}$ &$10^{-6}$\cr
$x_{\rm PAH}\equiv n(\rm PAH)/n_{\H}$ &$10^{-6}$ &$10^{-6}$ &$10^{-6}$\cr
$y=2n(\H_{2})/n_{\H}$ & {$0.999$}\cr
$B(\mu G)=bn_{\H}^{1/2}$ & 200 & 632 & 2000\cr
\hline
\multicolumn{4}{l}{Here b=2 is assumed.}
\cr\cr
\hline\hline
\end{tabular}
\end{center}
\end{table}

\subsection{Drifting velocities in C-shock component}
\begin{figure}
\includegraphics[width=0.45\textwidth]{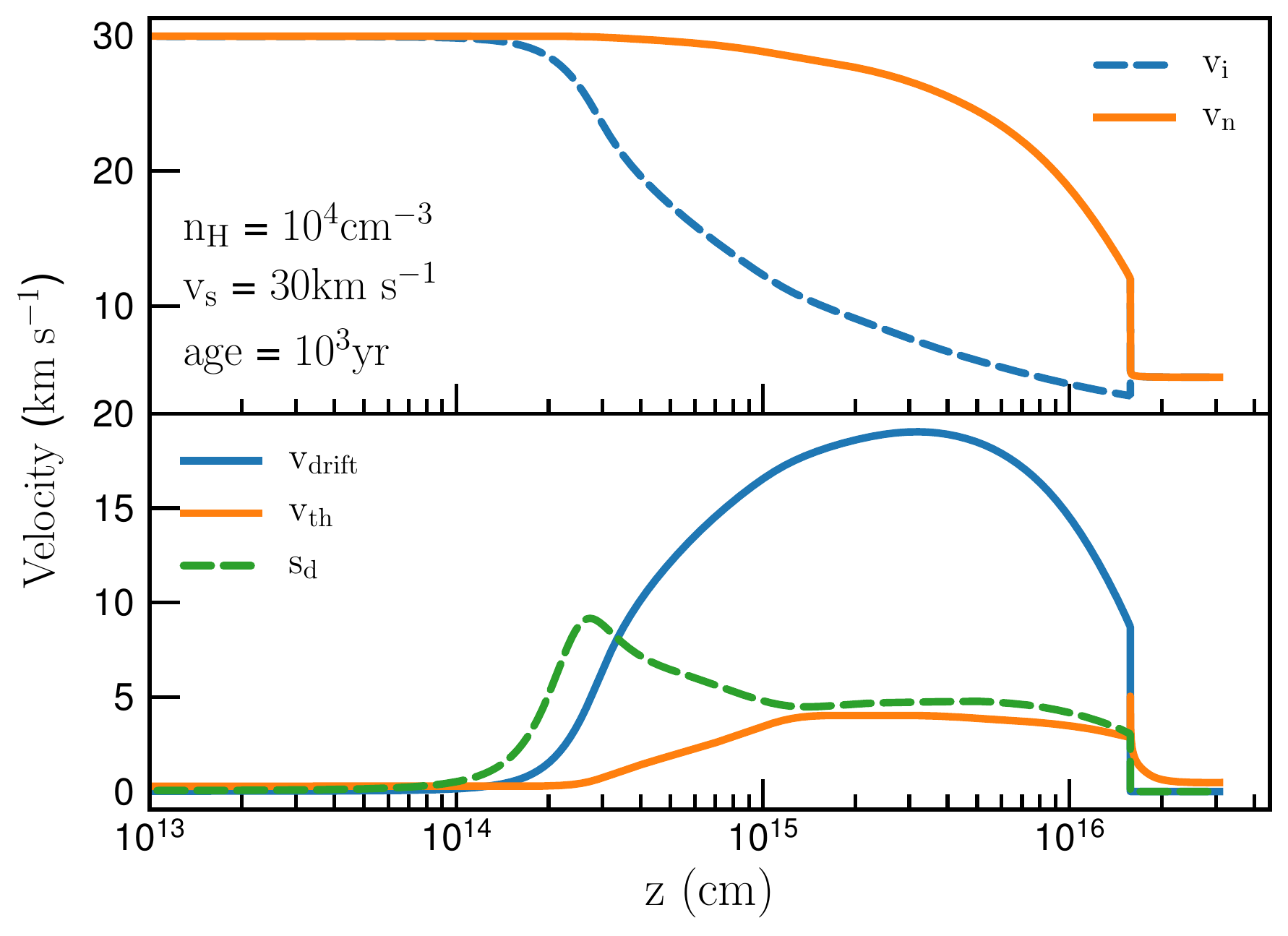}
\includegraphics[width=0.45\textwidth]{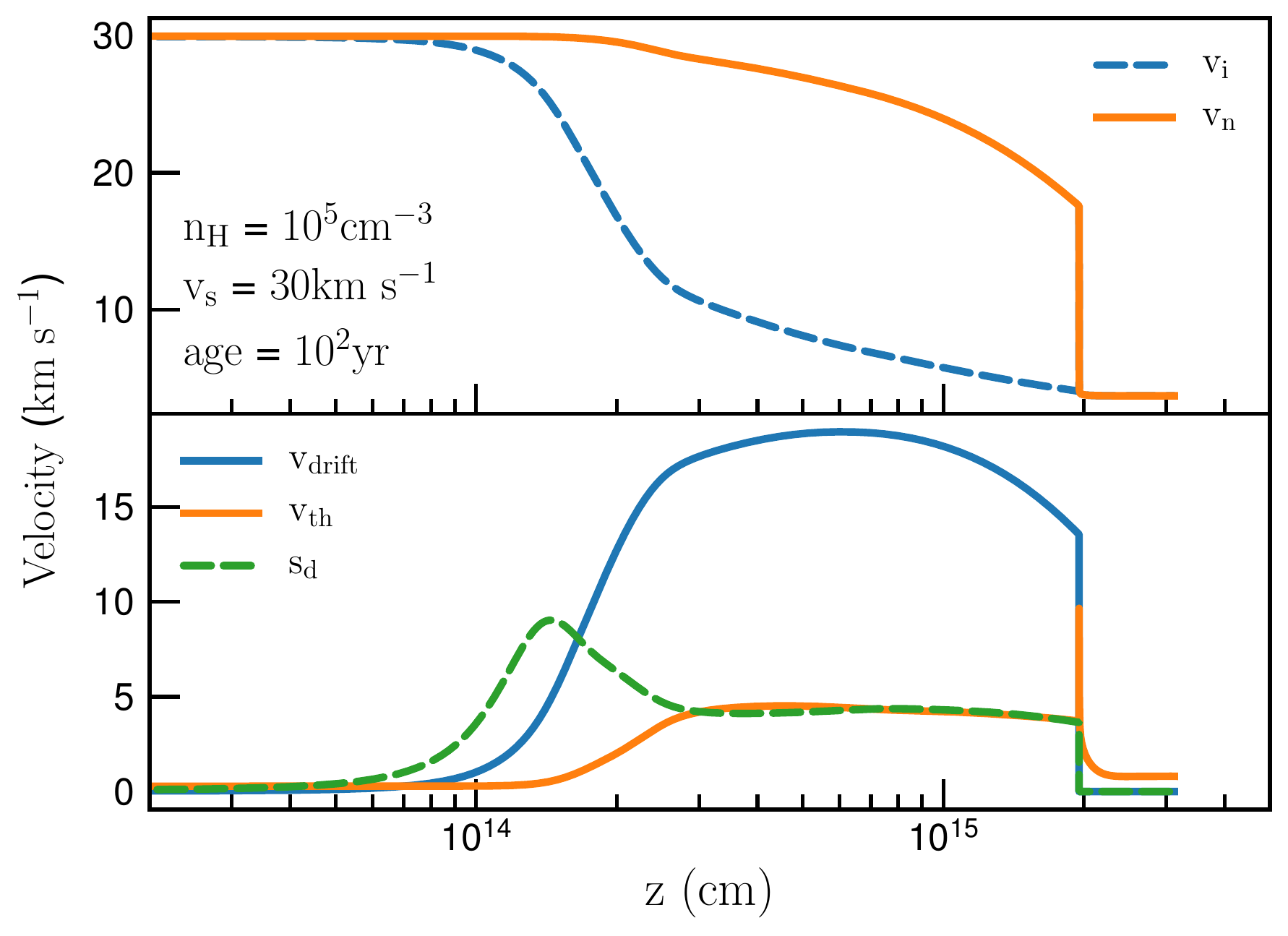}
\includegraphics[width=0.45\textwidth]{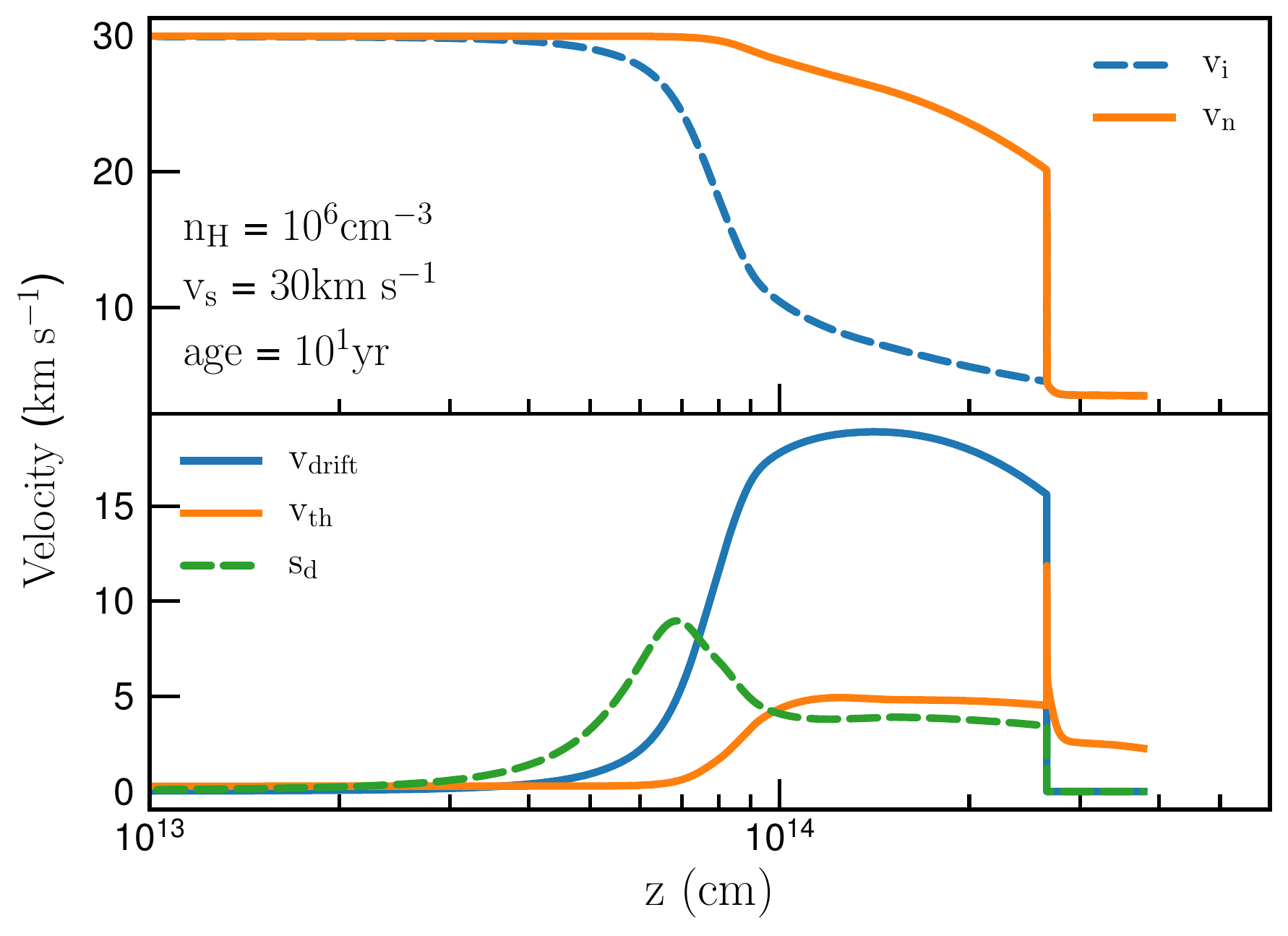}
\caption{Profiles of neutral velocity ($v_{n})$, ion and charged grains velocity ($v_{i}$), and their relative velocity ($v_{\rm drift})$ in the different CJ-shock models. Dashed line is the corresponding dimensionless drifting parameter $s_{d}=v_{\rm drift}/v_{\rm th}$, where $v_{\rm th}$ is the thermal gas velocity.}
\label{fig:CJshock-velo-old}
\end{figure}

In the C-shock component, ions and charged grains are coupled to the magnetic field and move slower than neutrals, resulting in the drift of neutral gas relative to charged grains. Figure \ref{fig:CJshock-velo-old} shows the profiles of neutral and ions velocities, including the drift velocity $v_{\rm drift} = |v_{n}-v_{i}|$, and the dimensionless drifting parameter $s_{d}=v_{\rm drift}/v_{\rm th}$ with $v_{\rm th}$ the gas thermal velocity as functions of distance ($z$) for the different gas density, assuming a shock velocity $v_{\s}=30\km\s^{-1}$. In this component, $v_{\rm drift}$ exists and reaches the maximum value at the middle of its part. The corresponding value of $s_{d}$ increases rapidly with $z$ and then declines when the gas is heated to high temperatures. In the J-shock component, on the other hand, the drifting velocity is vanished because grains and neutrals move at the same velocity. For the same shock velocity and age, the shock length decreases with increasing the gas density as a result of faster radiative cooling.

\begin{figure}
\includegraphics[width=0.45\textwidth]{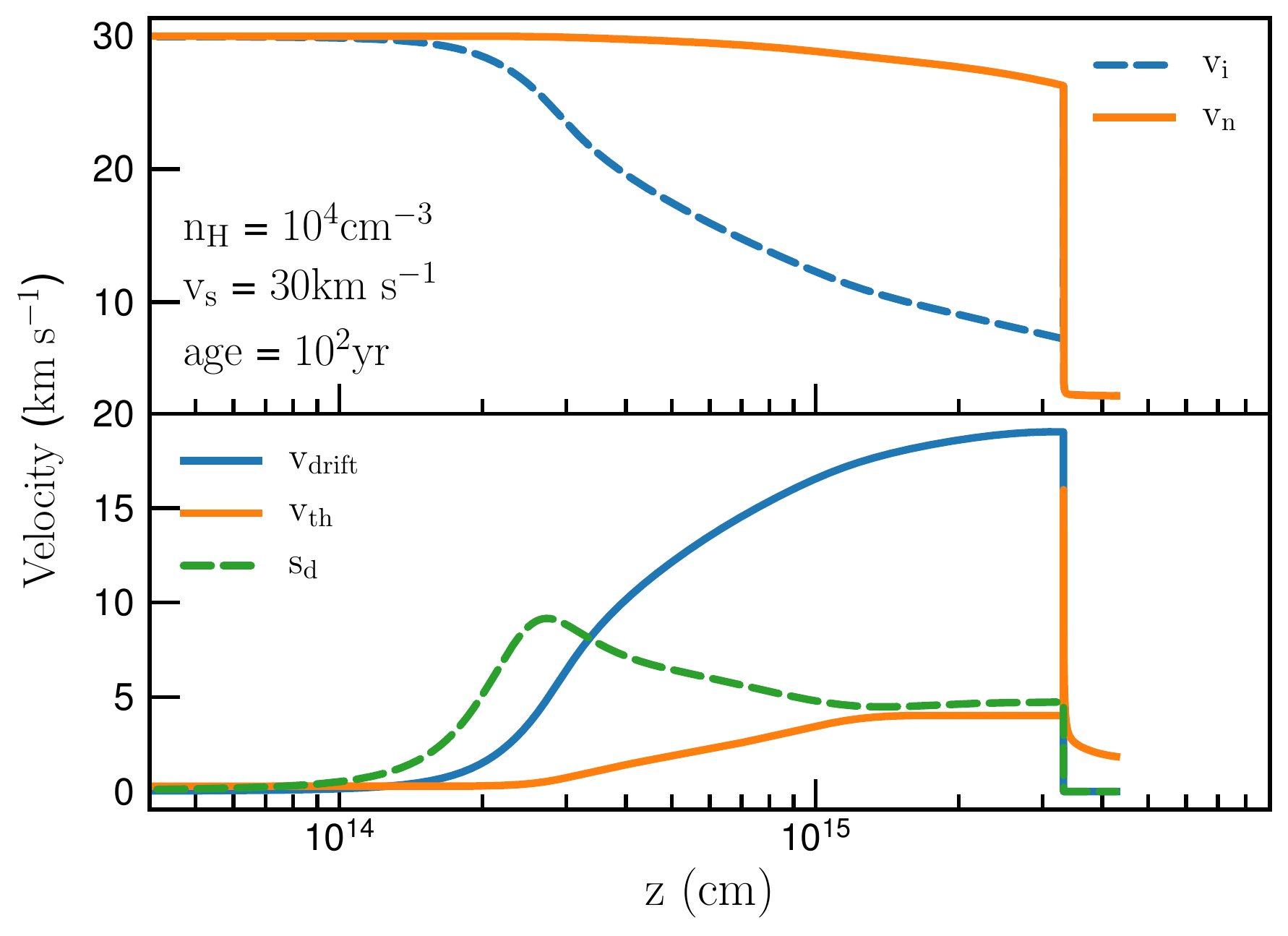}
\includegraphics[width=0.45\textwidth]{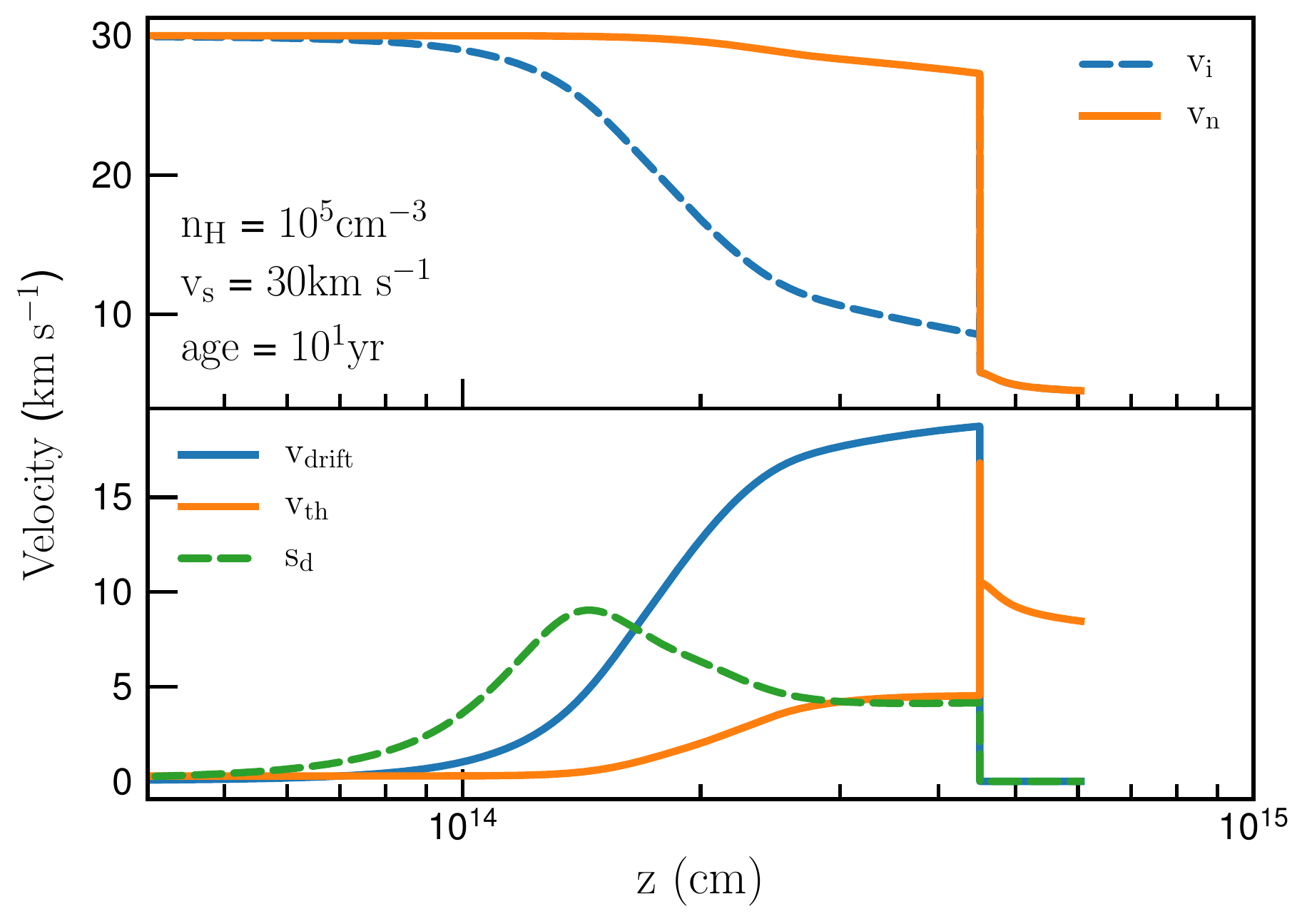}
\includegraphics[width=0.45\textwidth]{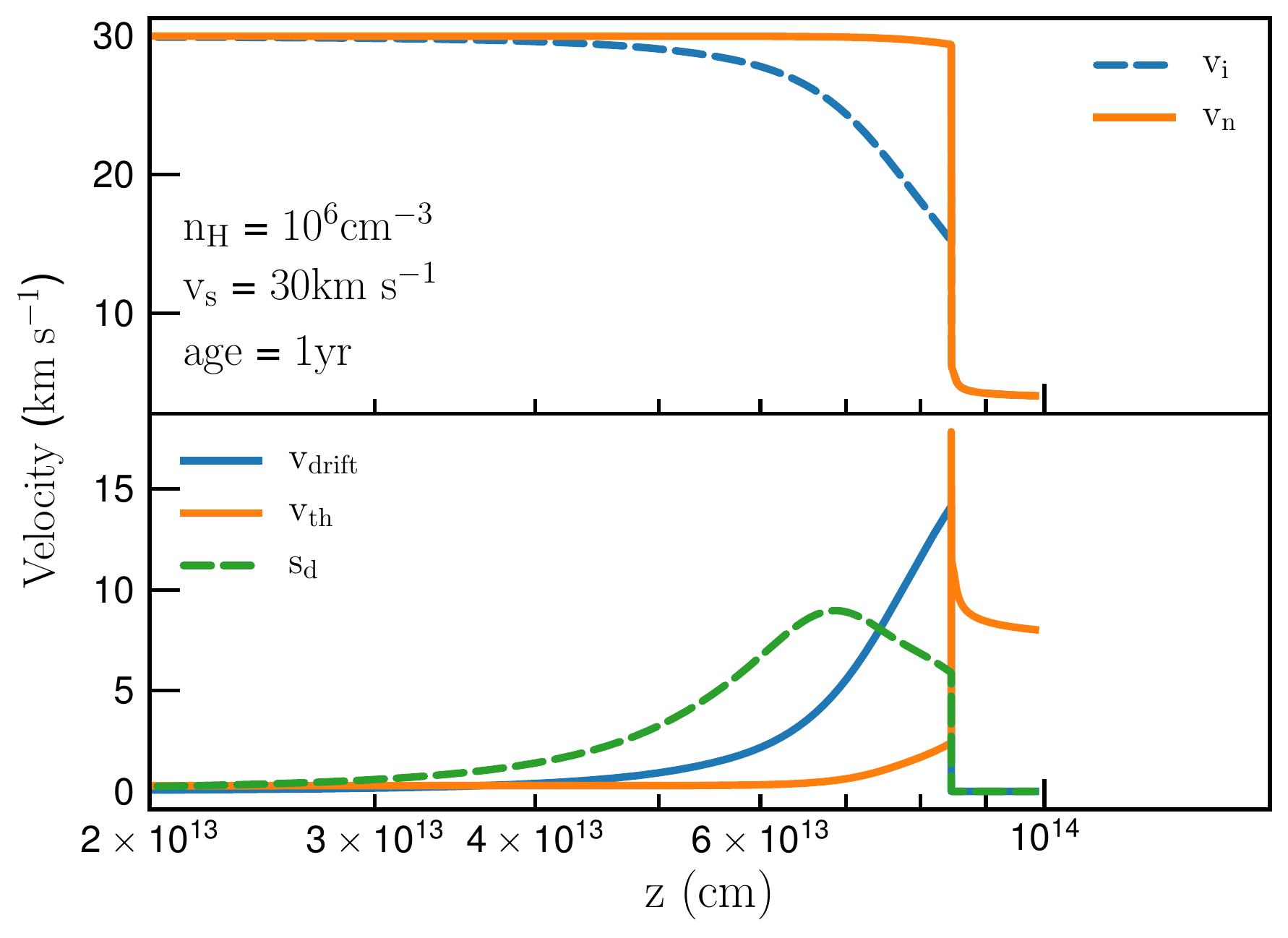}
\caption{Same as Figure \ref{fig:CJshock-velo-old} but for a younger shock age. The J-shock component dominates.}
\label{fig:CJshock-velo-young}
\end{figure}

Figure \ref{fig:CJshock-velo-young} shows the results for a younger shock age. The apparent difference is that the dimensionless drifting parameter $s_{d}$ is vanished much sooner because the J-shock part occurs earlier and dominates over the C-shock component. The shock length is also narrower due to the fast radiative cooling of J-shocks.

\section{Rotational dynamics of dust grains in non-stationary shocks}\label{sec:rot}
\subsection{Rotational temperature and rate}
In shock regions, rotational dynamics of nanoparticles is controlled by collisions with gas atoms and molecules, supersonic neutral drift relative to charged grains, bombardment of ions, long-distant interaction with passing ions, and photon absorption and re-emission (see e.g., \citealt{2019ApJ...877...36H}). The rotational temperature of spinning nanoparticles is given by (\citealt{1998ApJ...508..157D}):
\bea
\frac{T_{\rm rot}}{T_{\rm gas}}=\frac{G}{F}\frac{2}{1 + [1+ (G/F^2)(20\tH/3\ted)]^{1/2}},\label{eq:Trot}
\ena
where $\tH$ and $\ted$ are the characteristic damping times due to gas collisions and electric dipole emission, which are respectively defined as:
\bea
&\tau_{\H}& \simeq 0.067 a_{-7} \left(\frac{T_{\rm gas}}{1000\ \rm K} \right)^{-1/2} \left(\frac{n_{\H}}{10^5 \rm cm^{-3}}\right)^{-1} \rm yr, \\
&\tau_{ed}& \simeq 225 \left[ \frac{a^7_{-7}}{3.8 \times (\beta/0.4D)}\right] \left(\frac{T_{\rm gas}}{1000\K}\right)^{-1} \rm yr,
\ena
where $a_{-7}=a/(10^{-7}\cm)$ and $\mu$ is the dipole moment, $\beta$ the dipole moment per structure of grain, and $G$ and $F$ are the dimensionless coefficients which describe the total rotational excitation and damping, respectively. 

The rotational damping and excitation of grains can arise from various processes, including collisions with neutral atoms, ions, distant interaction with ions (plasma drag), and infrared emission. Thus, the coefficients are denoted by $F_{j},G_{j}$ for $j=n,ion, p, IR$  (see \citealt{2019ApJ...877...36H}). 

For the neutral-grain interaction, the damping and excitation coefficients $G_{n}$ and $F_{n}$ depend on thermal motion as well as drift motion of neutrals with respect to the grain, which depends on the grain charge state that fluctuates over time in the shock. As in \citealt{2019ApJ...877...36H}, we assume that dust grains in the shock can be statistically described by two populations of neutral and charged grains. The former population only interacts thermally with the neutrals, while the later population experience both thermal and drift motion (see Section \ref{sec:chargefluc} for an extended discussion). Therefore, the average coefficients are given by
\bea \label{eq:F}
&F_{n}&= F_{n,sd=0} f_{Z}(Z=0) + \sum _{Z \neq 0} F_{n,sd \neq 0} f_{Z}(Z), \\ \label{eq:G}
&G_{n}&= G_{n,sd=0} f_{Z}(Z=0) + \sum _{Z \neq 0} G_{n,sd \neq 0} f_{Z}(Z).
\ena
The first terms of the coefficients are calculated for several processes as in \cite{1998ApJ...508..157D}, while the second term is the additional effect due to the neutral drift in shocks (see \citealt{2019ApJ...877...36H} for more details). For our calculations, we assume that the evaporation temperature of stuck atoms from the grain surface is equal to the gas temperature (i.e., $T_{ev}=T_{\rm gas}$). This choice is valid for dense regions of low radiation intensity where all active sites on the grain surface are quickly occupied by impinging atoms and molecules such that subsequent incoming species just bounce back upon collisions (see e.g., \citealt{2009MNRAS.395.1055A}). 

\begin{figure}
\includegraphics[width=0.45\textwidth]{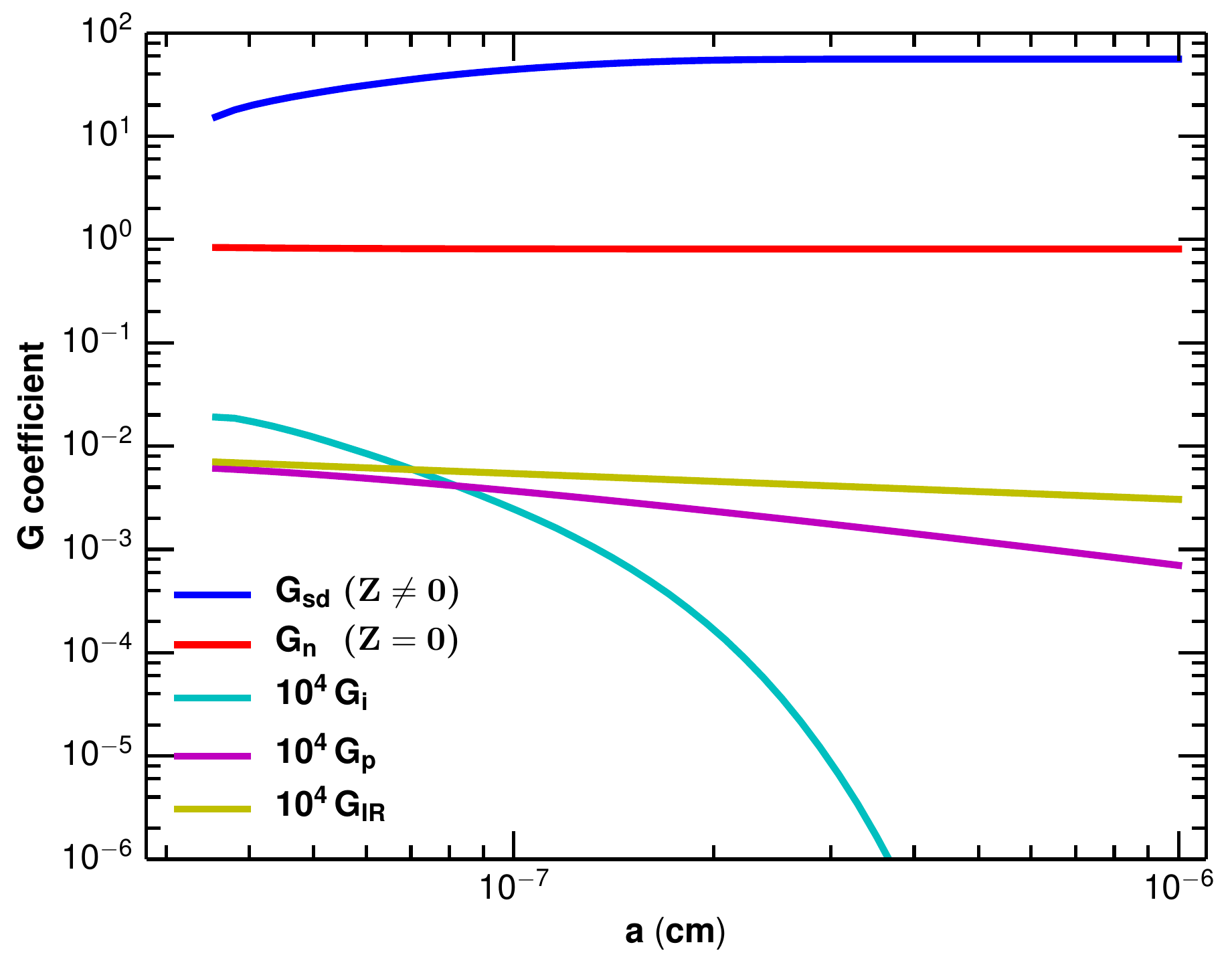}
\includegraphics[width=0.45\textwidth]{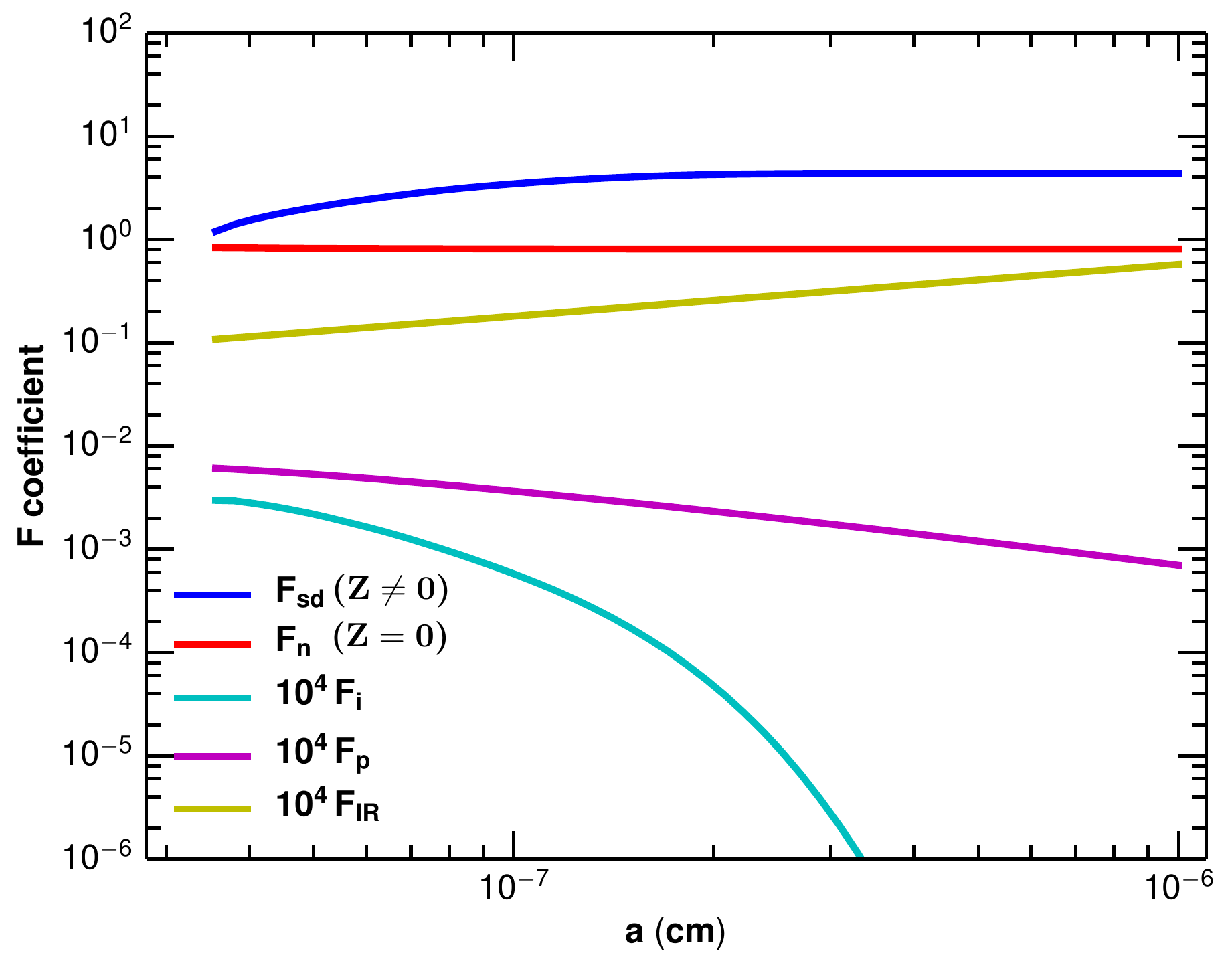}
\caption{Damping and excitation coefficients from various interaction processes in the CJ-shock of $n_{\H}=10^{4}\cm^{-3}$, $age=10^{3}\yr$ and $v_{s}=30\km\s^{-1}$ computed at location $z=10^{15}\cm$ (C-shock part). Collisional excitation ($G_{sd}$) and damping ($F_{sd}$) by supersonic neutral drift is dominant.}
\label{fig:FG_coeff_Cshock}
\end{figure}

\begin{figure}
\includegraphics[width=0.45\textwidth]{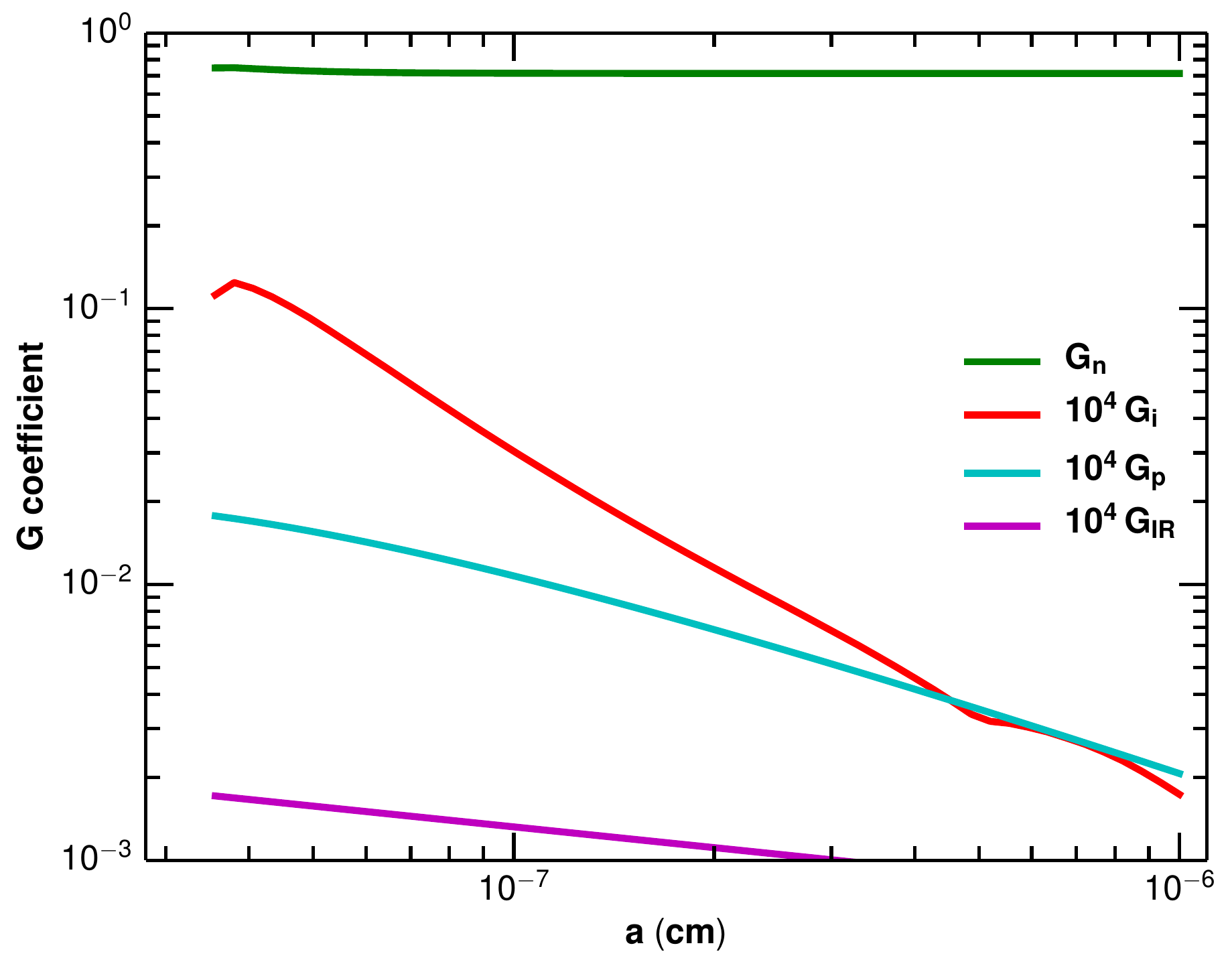}
\includegraphics[width=0.45\textwidth]{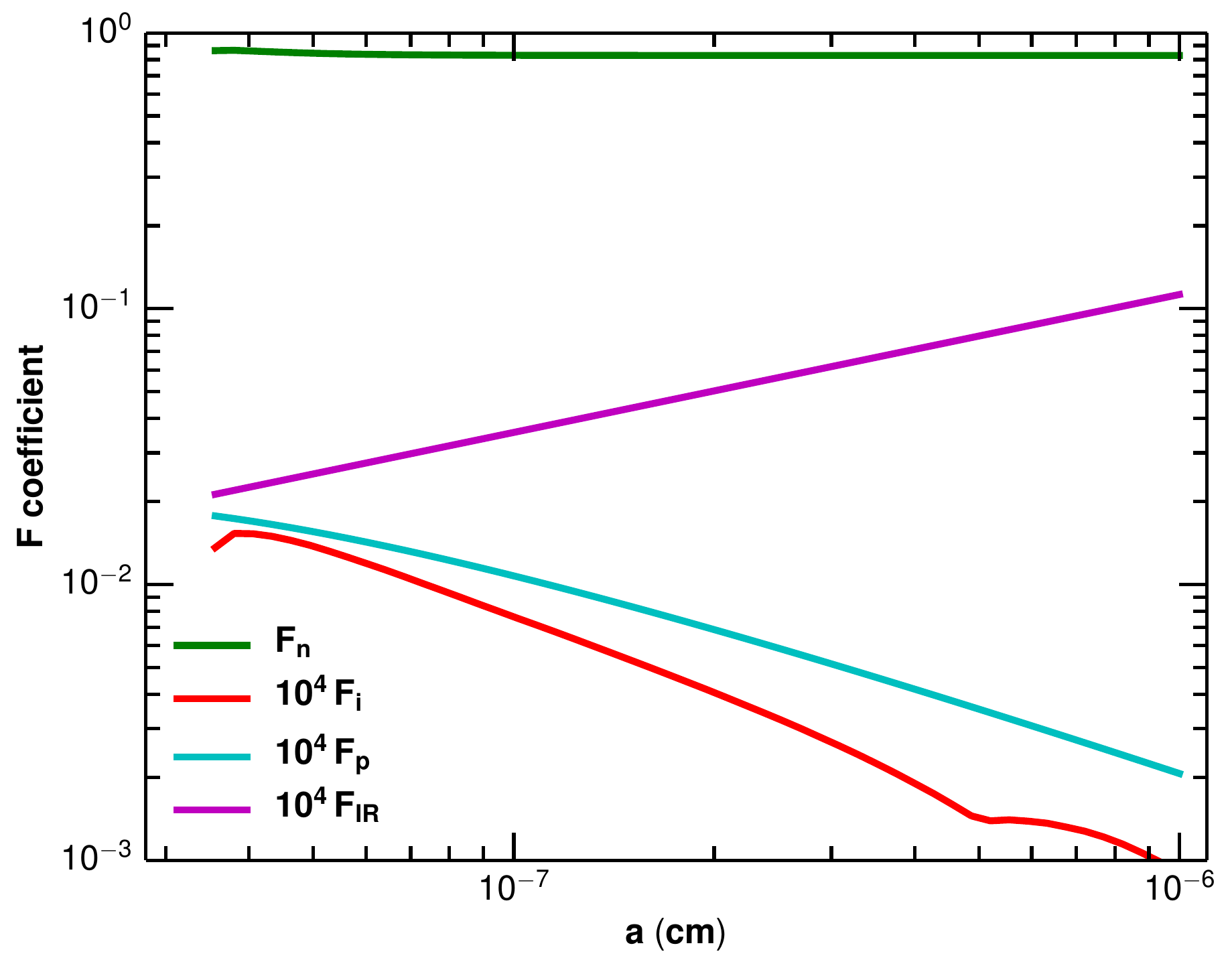}
\caption{Same as Figure \ref{fig:FG_coeff_Cshock} but computed at location $z=1.58\times 10^{16}\cm$ (J-shock tail). Excitation and damping by thermal collisions ($G_{n},F_{n}$) are dominant.}
\label{fig:FG_coeff_Jshock}
\end{figure}

Figure \ref{fig:FG_coeff_Cshock} shows an example of the rotational excitation and damping coefficients in the C-shock component of the CJ-shock, in which the supersonic drifting is $s_{d}\simeq5$. As expected, the excitation and rotational damping by neutral-grain drift, denoted by $G_{sd}$ and $F_{sd}$ are dominant. Obviously, $F_{sd}$ and $G_{sd}$ increase with increasing of the grain size due to the increase of the fraction of the negative charge states of grains (see Fig. 4 in \citealt{2019ApJ...877...36H}). The contribution from plasma drag, ion collisions, and IR emission is negligible. In this case, the dominance of the excitation by neutral-grain drift leads to suprathermal rotation of nanoparticles. On the other hand, Figure \ref{fig:FG_coeff_Jshock} shows the excitation and damping coefficients in the J-shock component of the CJ-shock, in which the neutral-grain drift is vanished. Therefore, rotational excitation and damping of grains are completely influenced by collisions with thermal gas.

\begin{figure}
\includegraphics[width=0.45\textwidth]{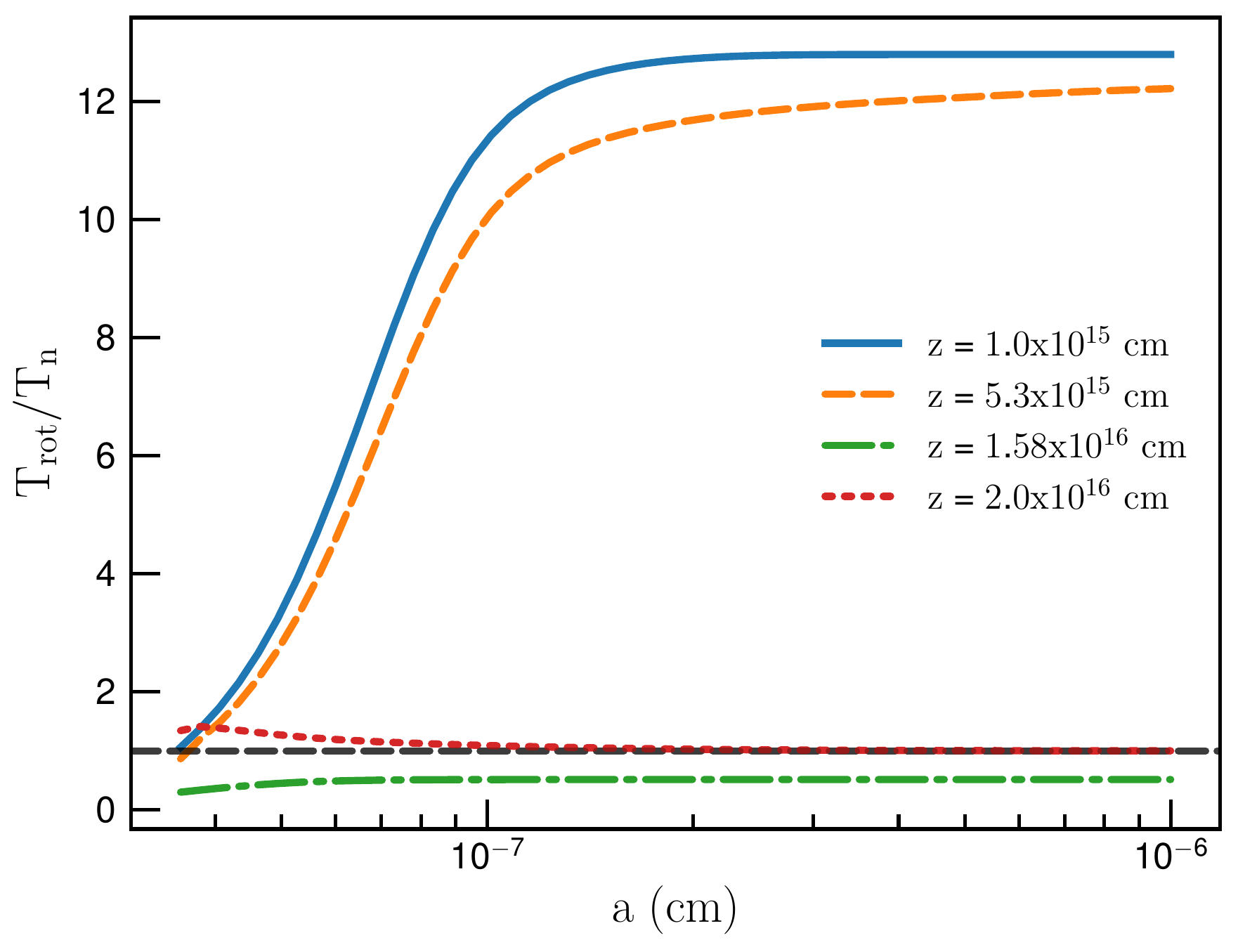}
\includegraphics[width=0.45\textwidth]{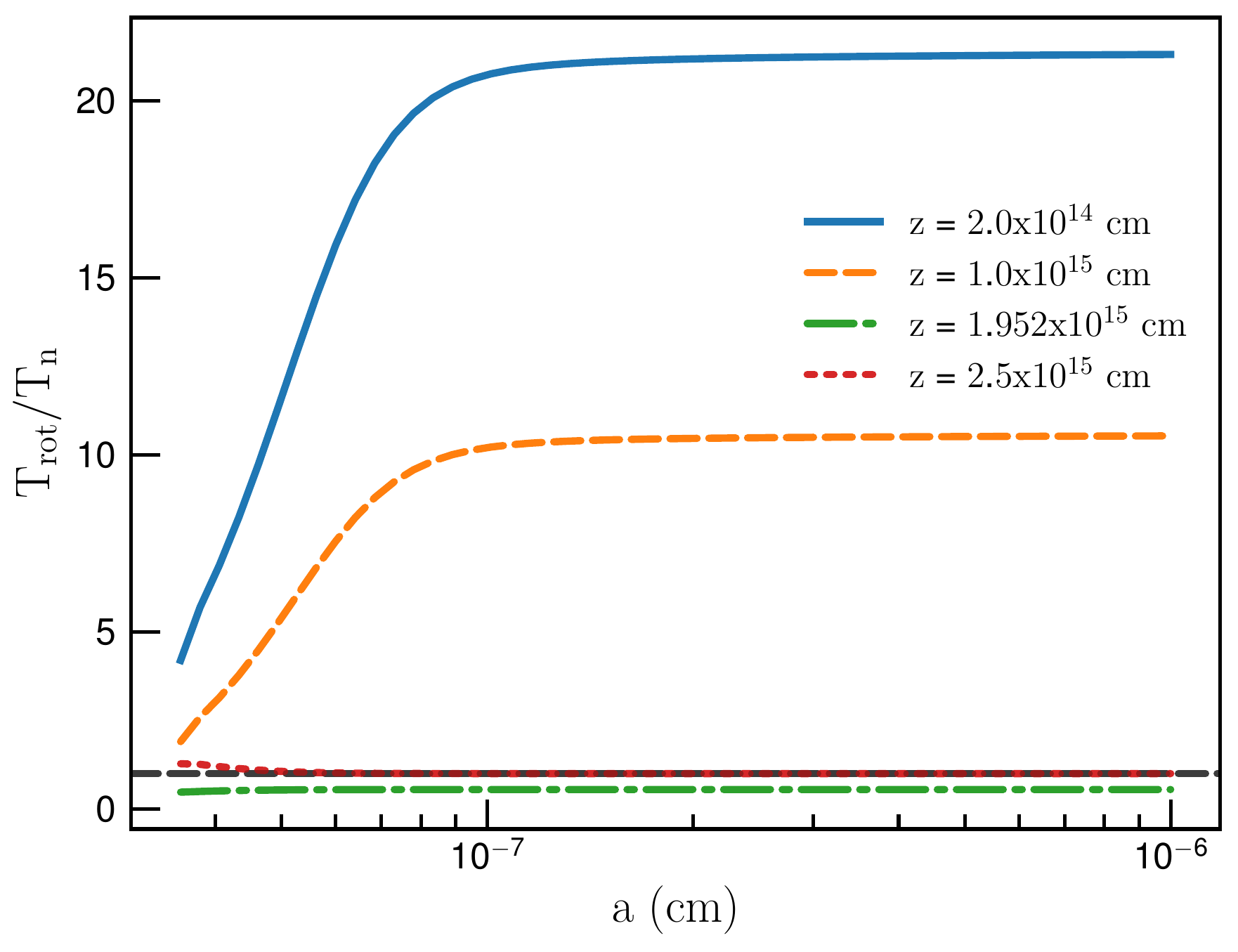}
\includegraphics[width=0.45\textwidth]{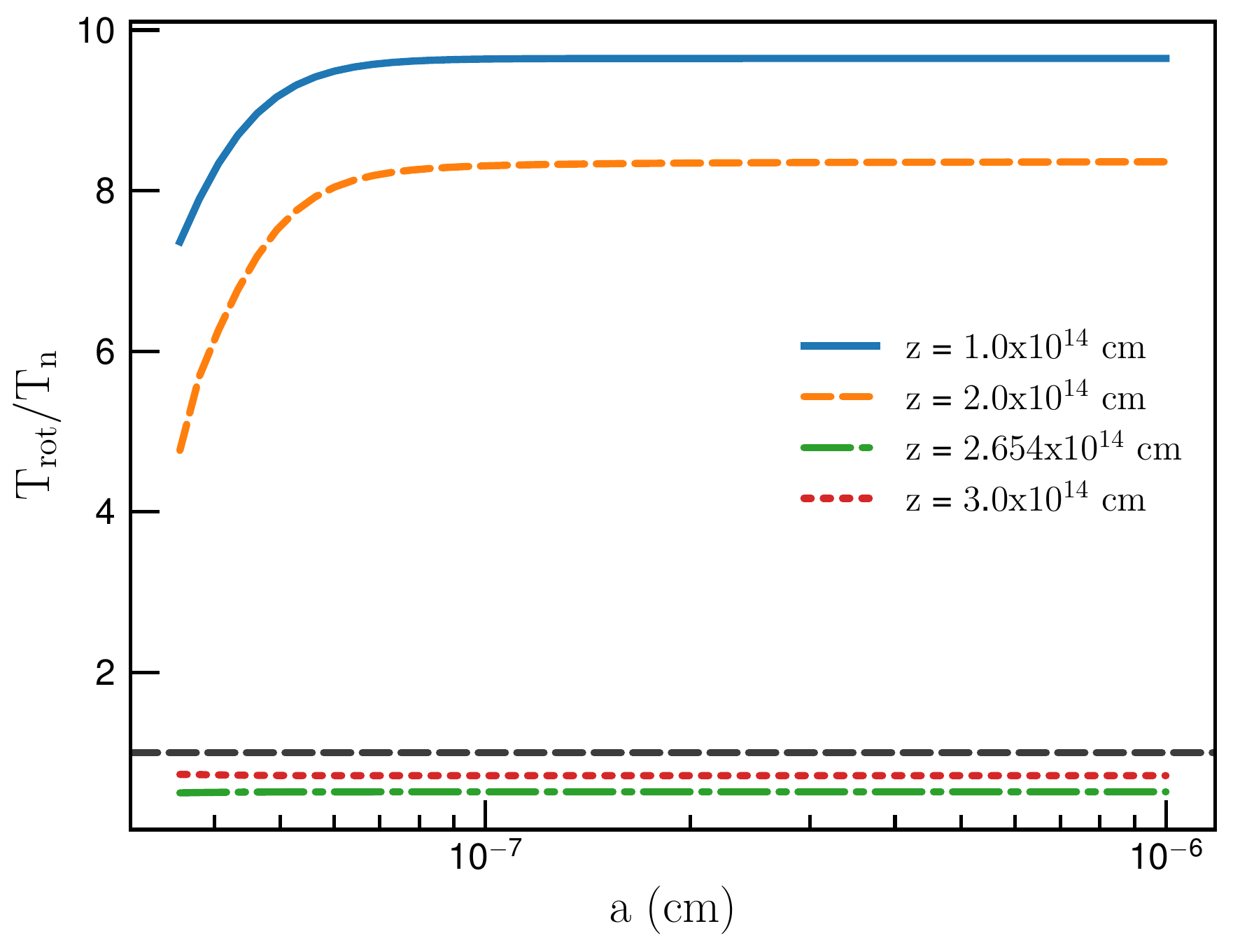}
\caption{The variation of $T_{\rm rot}/T_{\rm n}$ vs. the grain size for $n_{\H}=10^{4}\cm^{-3}$, v$_{s}$=30 km$\,$s$^{-1}$ and $age=10^{3}\yr$ (top panel), $n_{\H}=10^{5}\cm^{-3}$, v$_{s}$=30 km$\,$s$^{-1}$ and $age=10^{2}\yr$ (middle panel), and for $n_{\H}=10^{6}\cm^{-3}$, v$_{s}$=30 km$\,$s$^{-1}$ and $age=10^{1}\yr$ (bottom panel). Suprathermal rotation is observed at the locations of C-part (solid blue and dashed orange lines), while subthermal rotation is observed at the location of J-part (dotted red and dashed dotted green lines).}
\label{fig:Trot_a}
\end{figure}

Figure \ref{fig:Trot_a} shows the rotational temperature of spinning nanoparticles normalized to the gas temperature as a function of the grain size at four different locations in the shock. In the C-shock component, nanoparticles rotate suprathermal velocities due to supersonic drift velocity. The ratio of $T_{\rm rot}/T_{n}$ increases with the grain size and saturates dues to the increasing fraction of grains on the negative charge states \citep{2019ApJ...877...36H}. In the J-shock component, on contrary, nanoparticles rotate subthermally due to the dominance of thermal gas collisions.  

The rotational rate of nanoparticles corresponding to the rotational temperature $\rm T_{rot}$ is:
\bea
\frac{\omega_{\rm rot}}{2\pi}=\frac{1}{2\pi}\left(\frac{3kT_{\rm rot}}{I}\right)^{1/2}\Hz,\label{eq:omega_Trot}
\ena
where $I=8\pi\rho a^{5}/15 $ is the inertial moment of grain of size $a$ with ${\rho}$ the bulk density of grain, and $k$ is the Boltzmann constant.

\subsection{Dynamic and disruption timescales} \label{sec:timescale}
To understand the efficiency of the grain rotational excitation by stochastic gas bombardment in the shock, we need to quantify this timescale in term of comparison with the timescale of grain flow.  

The timescale to spin the grain at rest up to an angular momentum $J$ is:
\bea
\tau_{\rm spin-up}=\frac{J^2}{(\Delta J)^2/\Delta t},
\ena
where $(\Delta J)^2/\Delta t$ is the increase of grain rotational energy per unit time. Here one can assume $J = I\omega_{T}$, with $\omega_{T}=(2kT_{\rm gas}/I)^{1/2}$ the thermal angular velocity.

As mentioned in Section \ref{sec:model}, the structure of the CJ-shock approximates a composition of the C-shock and J-shock components. In the C-shock component, in which charged grains move slower than neutrals, the spin-up time by neutral gas drift is equal to
\bea
\tau_{\rm spin-up}^{C}&=&\frac{16\rho k T_{\gas} a}{15n_{\H} m_{\H} ^{2}v_{\rm drift}^{3}} \nonumber\\
&=&0.005a_{-7}\left(\frac{T_{\gas}}{10^{3}\K}\right)
\left(\frac{n_{\H}}{10^{5}\cm^{-3}}\right)^{-1}\nonumber\\
&&\times\left(\frac{10\km\s^{-1}}{v_{\rm drift}}\right)^{3}\yr.\label{eq:tspinup_C}
\ena
In the J-shock component, on the other hand, grains move with the same velocity as gas, the spin-up time is characterized by thermal collision, which yields:
\bea
\tau_{\rm spin-up}^{J}&=&\frac{16\rho k T_{\gas} a}{15n_{\H} m_{\H} ^{2}v_{\rm th}^{3}} \nonumber\\
&\simeq &0.016a_{-7}\left(\frac{T_{\gas}}{10^{4}\K}\right)^{-1/2}
\left(\frac{n_{\H}}{10^{5}\cm^{-3}}\right)^{-1}\yr.\label{eq:tspinup_J}
\ena

The dynamical flow time of grains in the C-shock and J-shock components can be respectively estimated as
\bea
\tau_{\rm flow}^{C}=\frac{L^C}{v_{\rm drift}}\simeq 30\left(\frac{L^C}{10^{15}\cm}\right)\left(\frac{10\km \s^{-1}}{v_{\rm drift}}\right)\rm yr,\label{eq:tflow_C}
\ena
and
\bea
\tau_{\rm flow}^{J}=\frac{L^J}{v_{\rm gas}}\simeq 60\left(\frac{L^J}{10^{15}\cm}\right)\left(\frac{5\km \s^{-1}}{v_{\rm gas}}\right)\rm yr.\label{eq:tflow_J}
\ena
where $L^C$ and $L^J$ are the widths of C-shock and J-shock components (see Section \ref{sec:model}).

By comparing Equation (\ref{eq:tspinup_C}) with (\ref{eq:tflow_C}), one can see that the spin-up timescale by stochastic gas collisions is shorter than the time passing the dense C-shock structure for drift velocity $v_{\rm drift}> 1\km\s^{-1}$. Therefore, supersonic gas flow can rapidly spin up nanoparticles to suprathermal rotation in the C-shock component.

From Equation (\ref{eq:tspinup_J}) and (\ref{eq:tflow_J}) it follows that the spin-up timescale by thermal collisions is also shorter than the time passing the J-shock structure for a typical gas velocity of $5\km\s^{-1}$. Therefore, thermal gas collisions can rapidly spin up nanoparticles to thermal rotation in the J-shock component.  



\section{Rotational disruption mechanism}\label{sec:disrupt} \label{sec:rotational_disruption}
\subsection{Rotational disruption}
In this section, we briefly describe the rotational disruption mechanism of nanoparticles in shocks (look \citealt{2019ApJ...877...36H} for details).

In shocks, nanoparticles can be excited to very fast rotations with the rate given by Equation (\ref{eq:omega_Trot}). When the grain rotation rate becomes sufficiently large such that the centrifugal stress ($S=\rho a\omega^{2}/4$) exceeds the maximum tensile strength of the grain material ($S_{\max}$), nanoparticles are disrupted instantaneously. The critical angular velocity for the disruption is obtained by setting $S\equiv S_{\max}$ (\citealt{Hoang:2018es}; \citealt{Hoang:2019da}):
\bea
\frac{\omega_{\rm cri}}{2\pi}=\frac{1}{\pi a}\left(\frac{S_{\rm max}}{\rho} \right)^{1/2}\Hz.~~~~\label{eq:omega_cri}
\ena

The exact value of $S_{\max}$ depends on the composition and internal structure of dust grains, which is unfortunately poorly known. Nevertheless, \cite{2018arXiv181208391H} suggested that the tensile strength and internal structure of dust can be constrained with observations of grain disruption in strong radiation fields such as near supernovae. Previously, \cite{1974ApJ...190....1B} and \cite{1979ApJ...231...77D} suggested that $S_{\max}\sim 10^{9}-10^{10}\erg\cm^{-3}$ for polycrystalline bulk solid. Ideal materials, such as diamond can have $S_{\max}\ge 10^{11}\erg\cm^{-3}$, which is considered strongest material. We will consider several values of $S_{\max}=10^{9}-10^{11}\erg\cm^{-3}$, which are expected for nanoparticles. From now on, nanoparticles whose $S_{\max}\gtrsim 10^{10}\erg\cm^{-3}$ are rendered to {\it strong materials}, and whose $S_{\max}<10^{10}\erg\cm^{-3}$ are referred to {\it weak materials}.

Comparing Equations (\ref{eq:omega_Trot}) and (\ref{eq:omega_cri}), we can derive the critical rotational temperature required for grain disruption:
\bea
T_{\rm rot}\ge 1.6\times10^{5}a_{-7}^{3}S_{\rm max,10}\K,
\ena
where $S_{\rm max,10}=S_{\rm max}/10^{10}\erg\cm^{-3}$. It means that to destroy strong nanoparticles (e.g., $S_{\max}=10^{10}\erg\cm^{-3}$) of $a\lesssim 0.5$ nm, we need the rotational temperature of $T_{\rm rot}\sim 2\times 10^{4}\K$. But for weak materials, e.g., $S_{\max}=10^{9}\erg\cm^{-3}$, the disruption can occur at lower temperatures of $T_{\rm rot}\sim 2\times 10^{3}\K$. 

\subsection{Grain disruption size} \label{sec:size_dist}
To obtain the disruption size of nanoparticles by centrifugal stress in the shock, we compute $\langle \omega^{2}\rangle^{1/2}\equiv \omega_{\rm rot}$ for a range of grain sizes and compare it with $\omega_{\rm cri}$.

\begin{figure}
\includegraphics[width=0.45\textwidth]{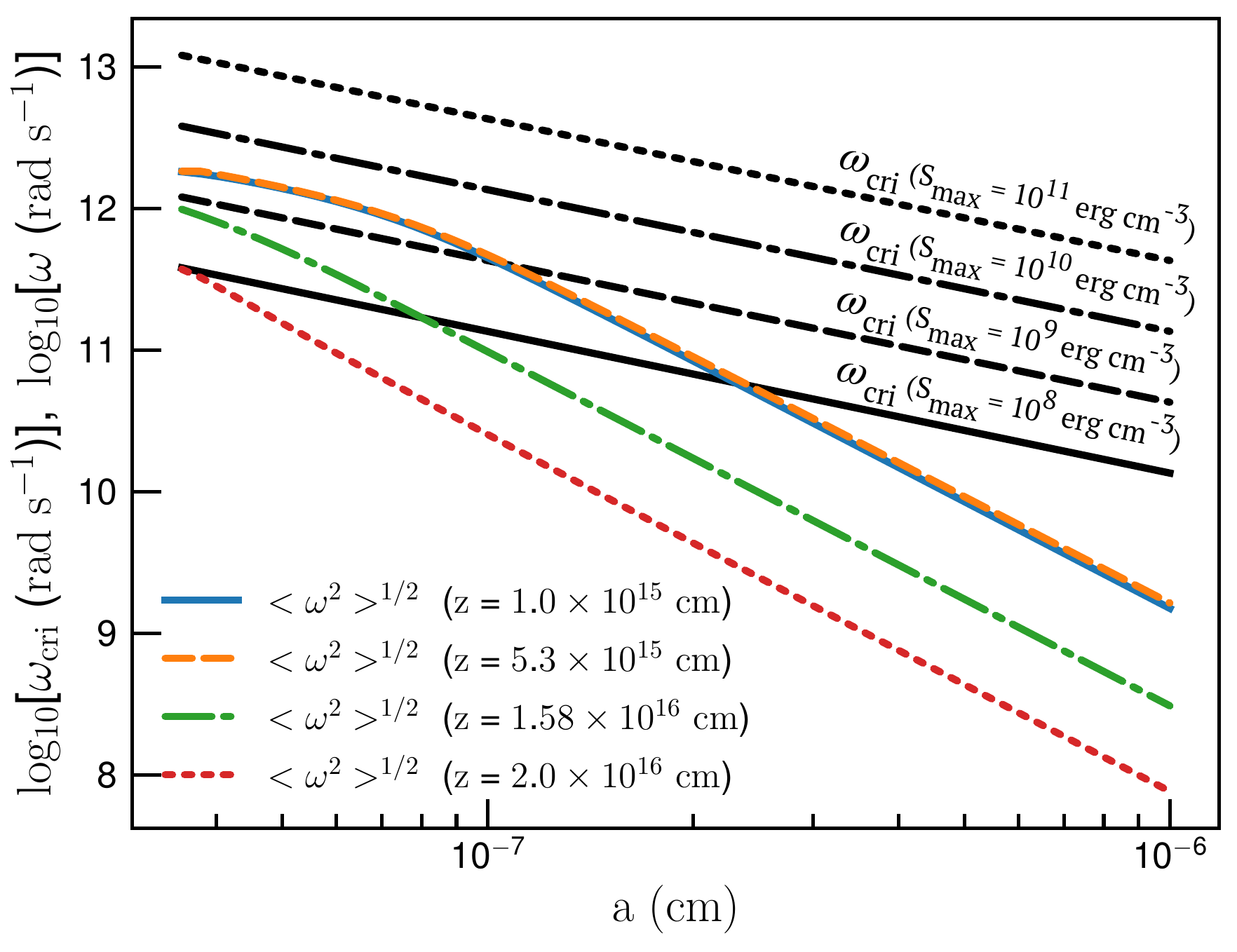}
\includegraphics[width=0.45\textwidth]{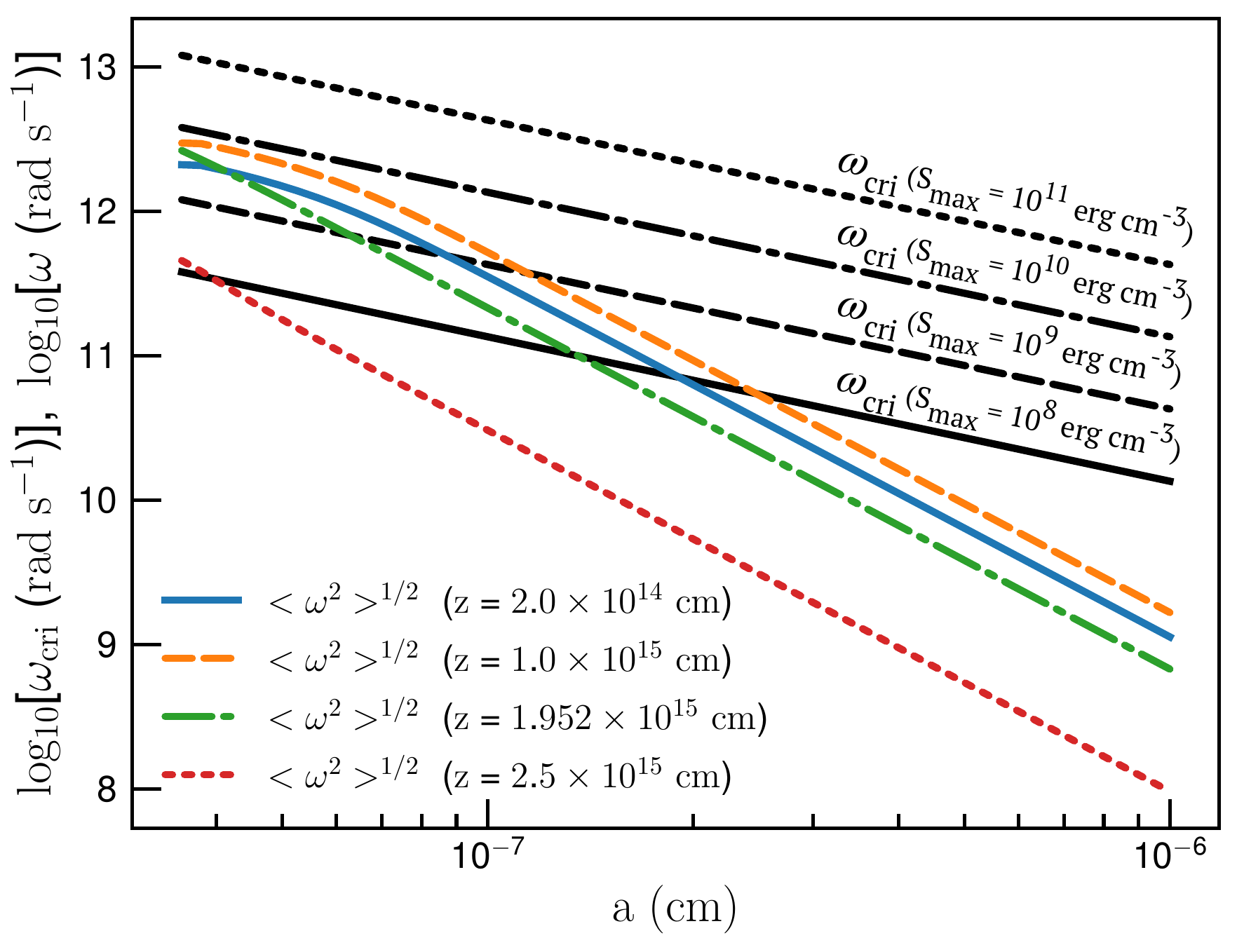}
\includegraphics[width=0.45\textwidth]{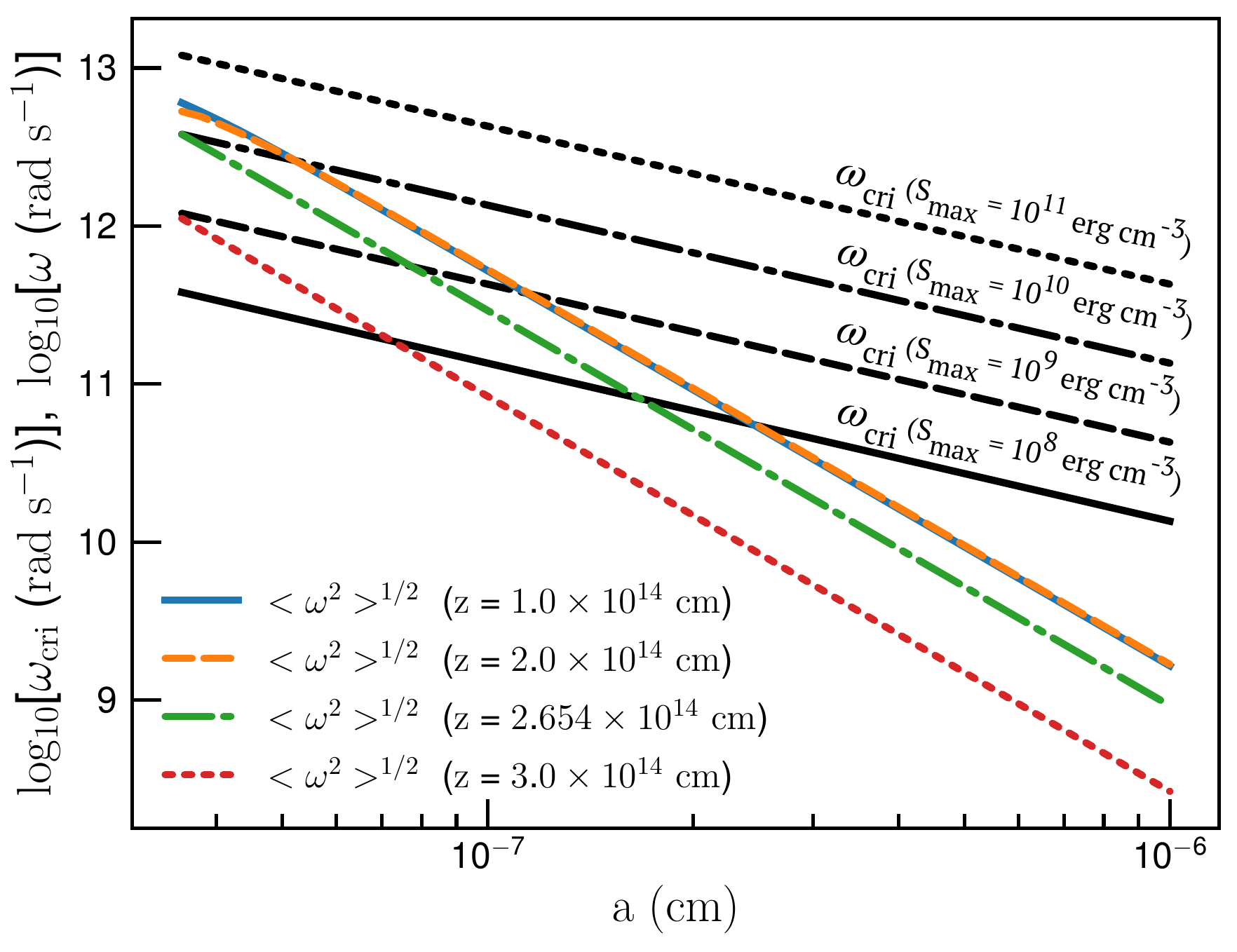}
\caption{Angular velocity of nanoparticles computed at several positions in the shock as in Figure \ref{fig:Trot_a} in comparison with the disruption critical velocity, assuming the CJ-shocks with $n_{\H}=10^{4}\cm^{-3}$, v$_{s}$=30 km$\,$s$^{-1}$ and $age=10^{3}\yr$ (top panel), with $n_{\H}=10^{5}\cm^{-3}$, v$_{s}$=30 km$\,$s$^{-1}$ and $age=10^{2}\yr$ (middle panel), and with $n_{\H}=10^{6}\cm^{-3}$, v$_{s}$=30 km$\,$s$^{-1}$ and $age=10^{1}\yr$ (bottom panel).}
\label{fig:omega_cri}
\end{figure}

\begin{figure*}
    \centering
    \subfloat{
    \includegraphics[width=0.45\textwidth]{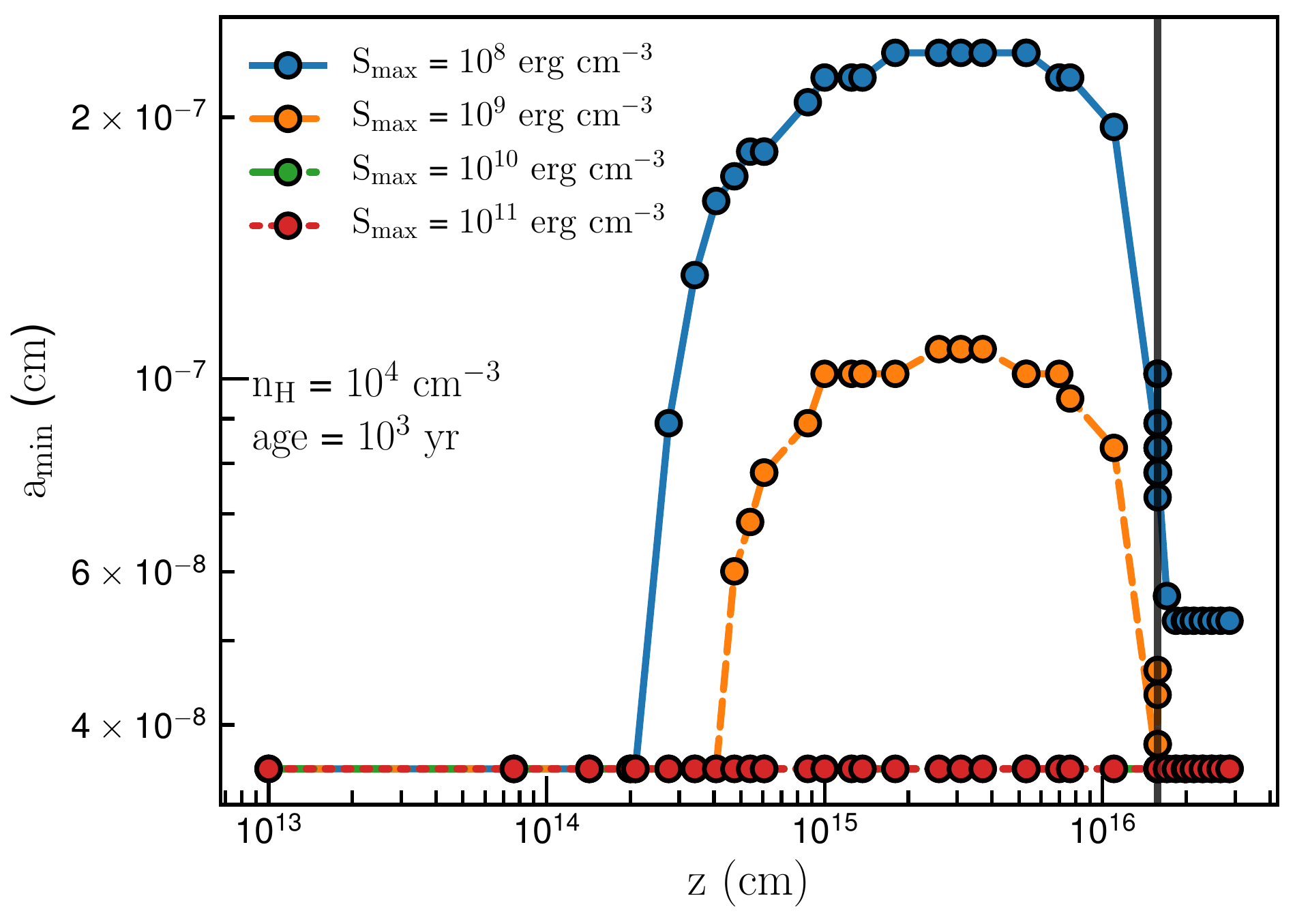}
    \includegraphics[width=0.45\textwidth]{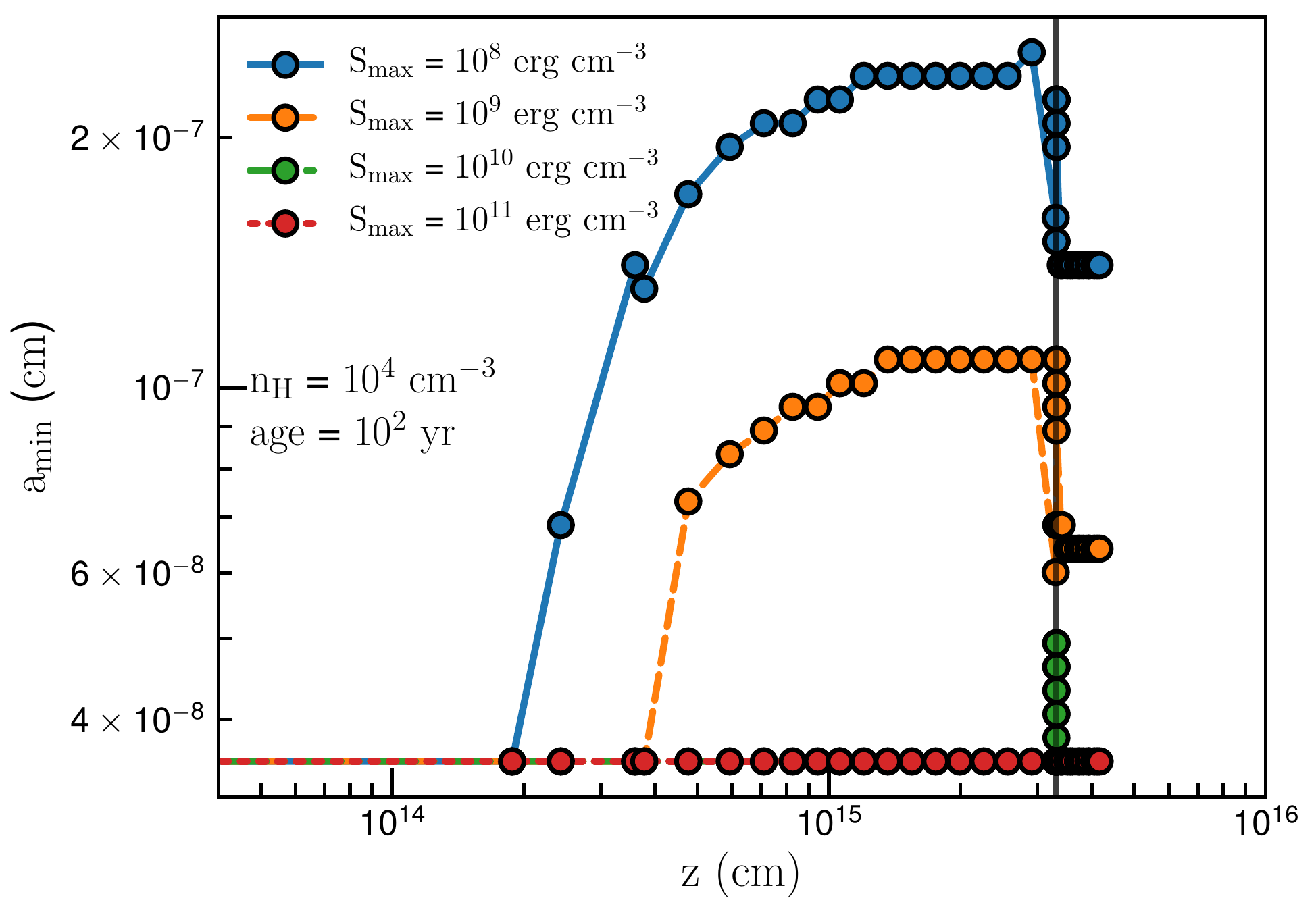}
    }\\
    \subfloat{
    \includegraphics[width=0.45\textwidth]{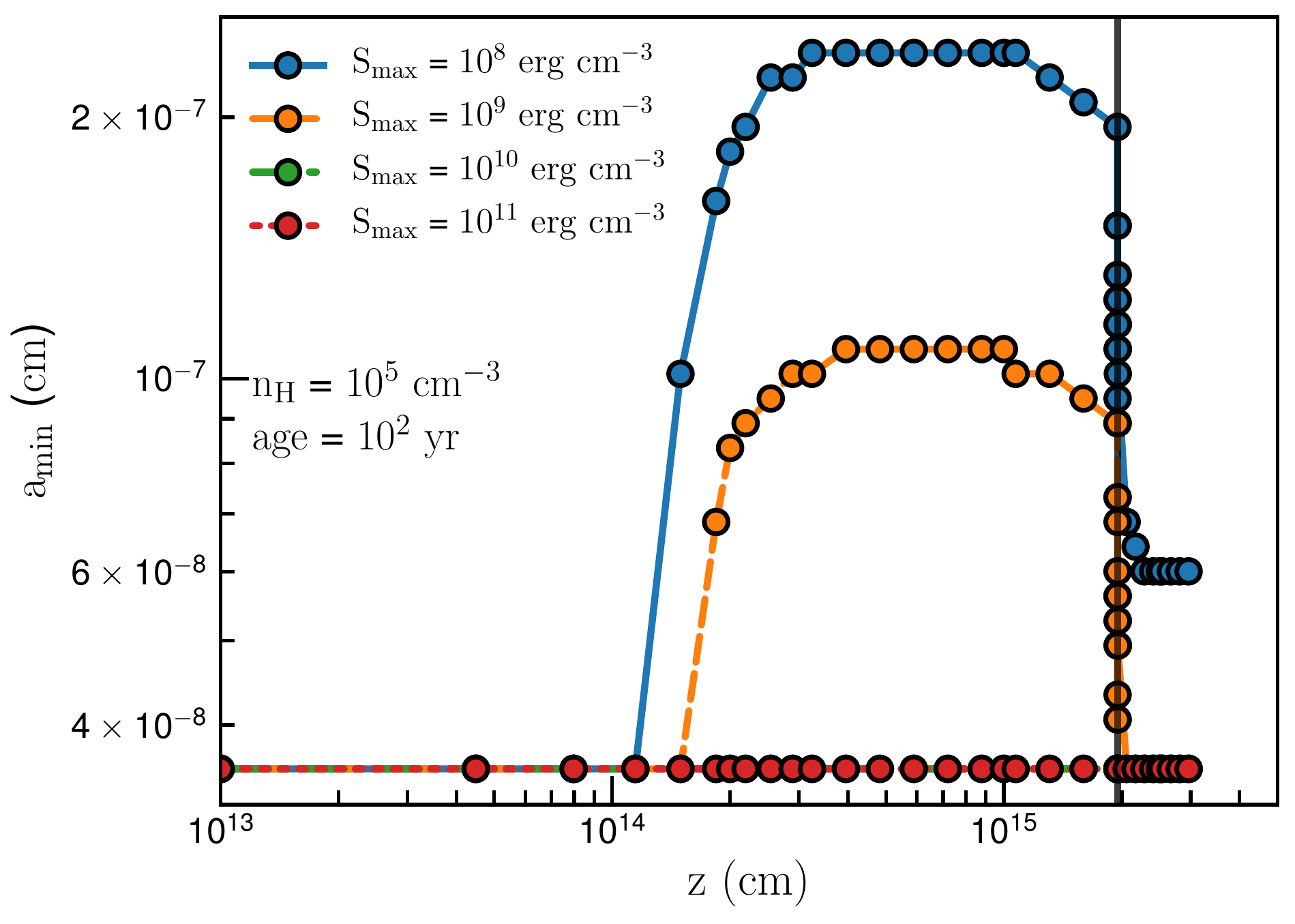}
    \includegraphics[width=0.45\textwidth]{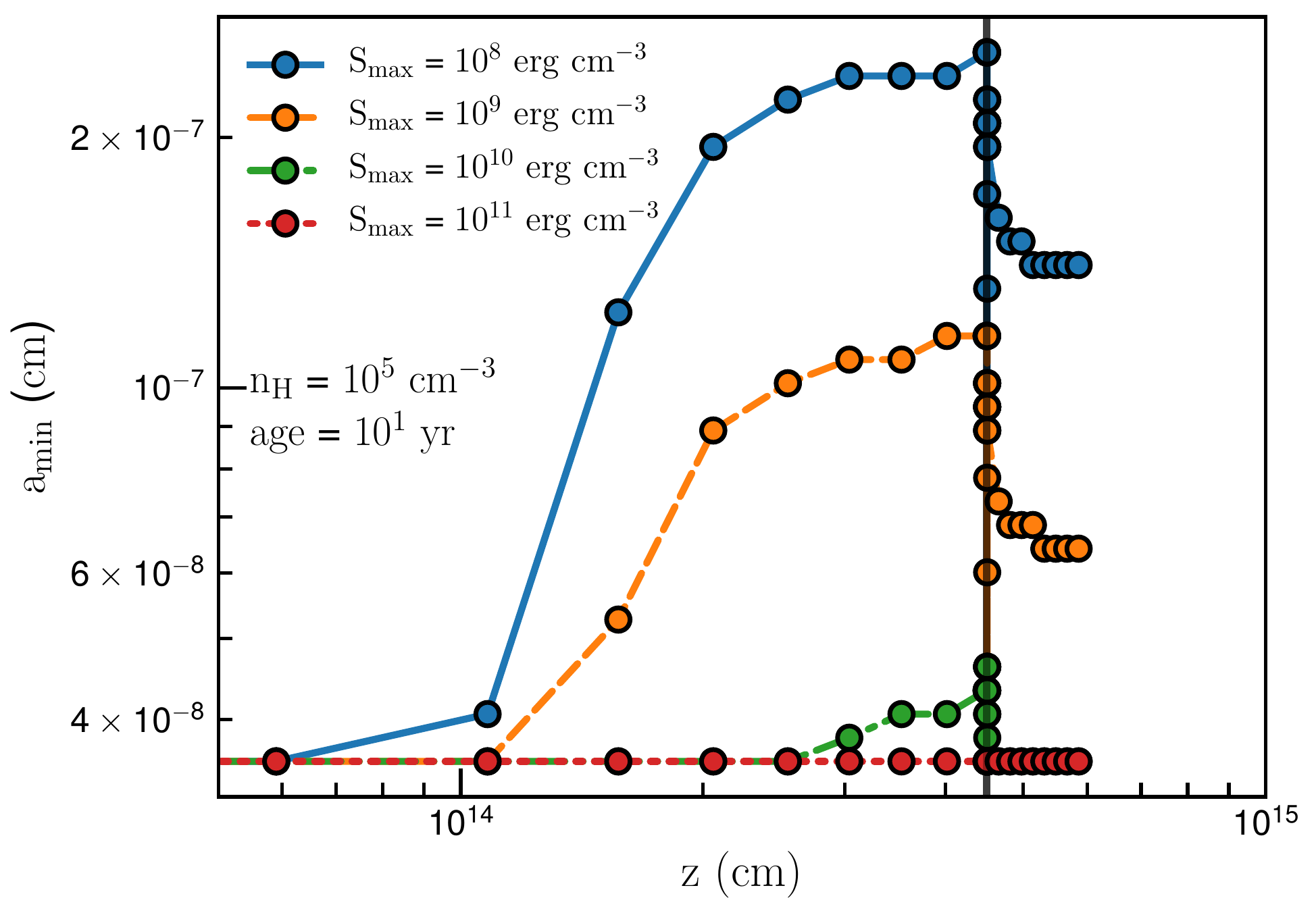}
    }\\
    \subfloat{    
    \includegraphics[width=0.45\textwidth]{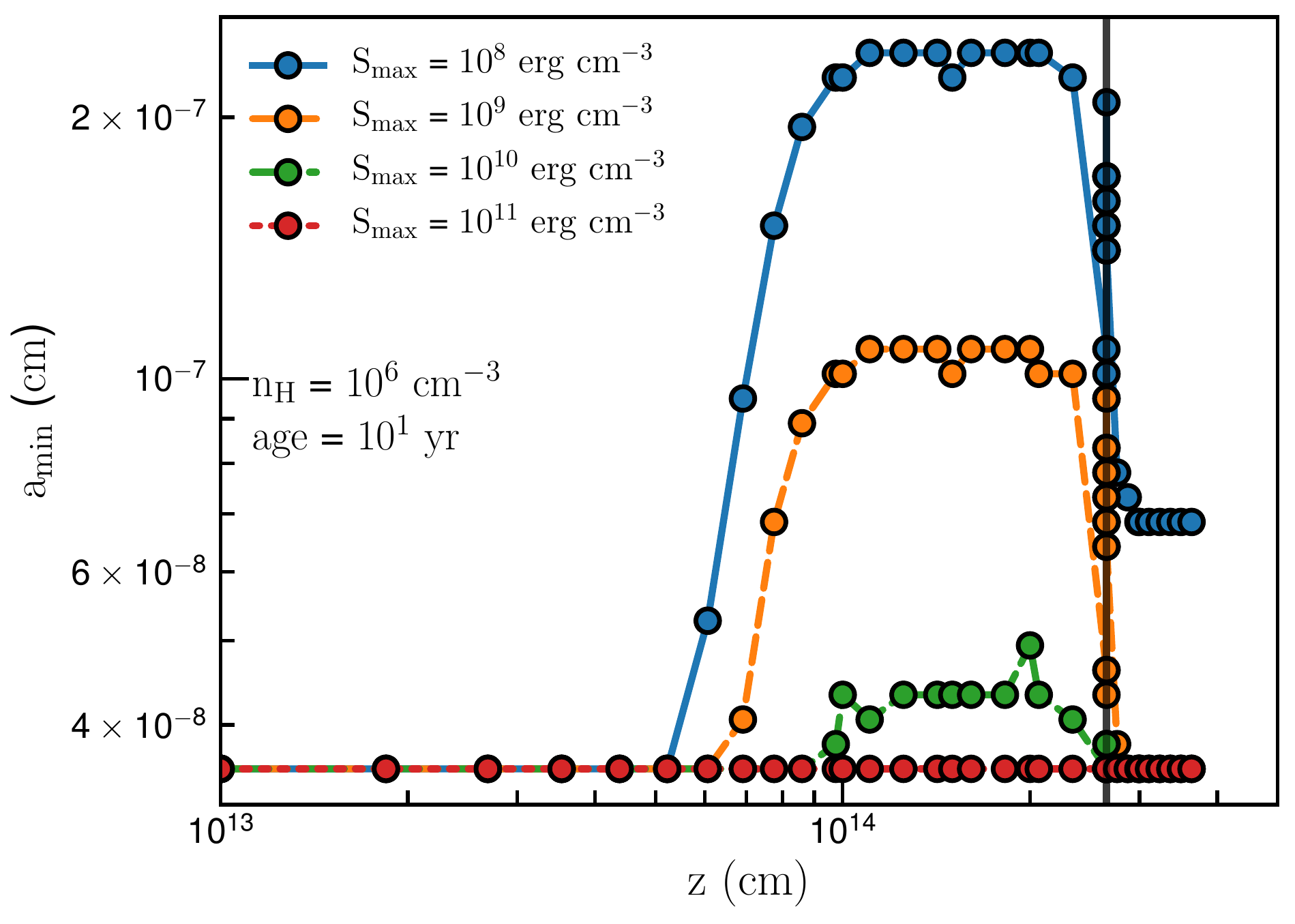}
    \includegraphics[width=0.45\textwidth]{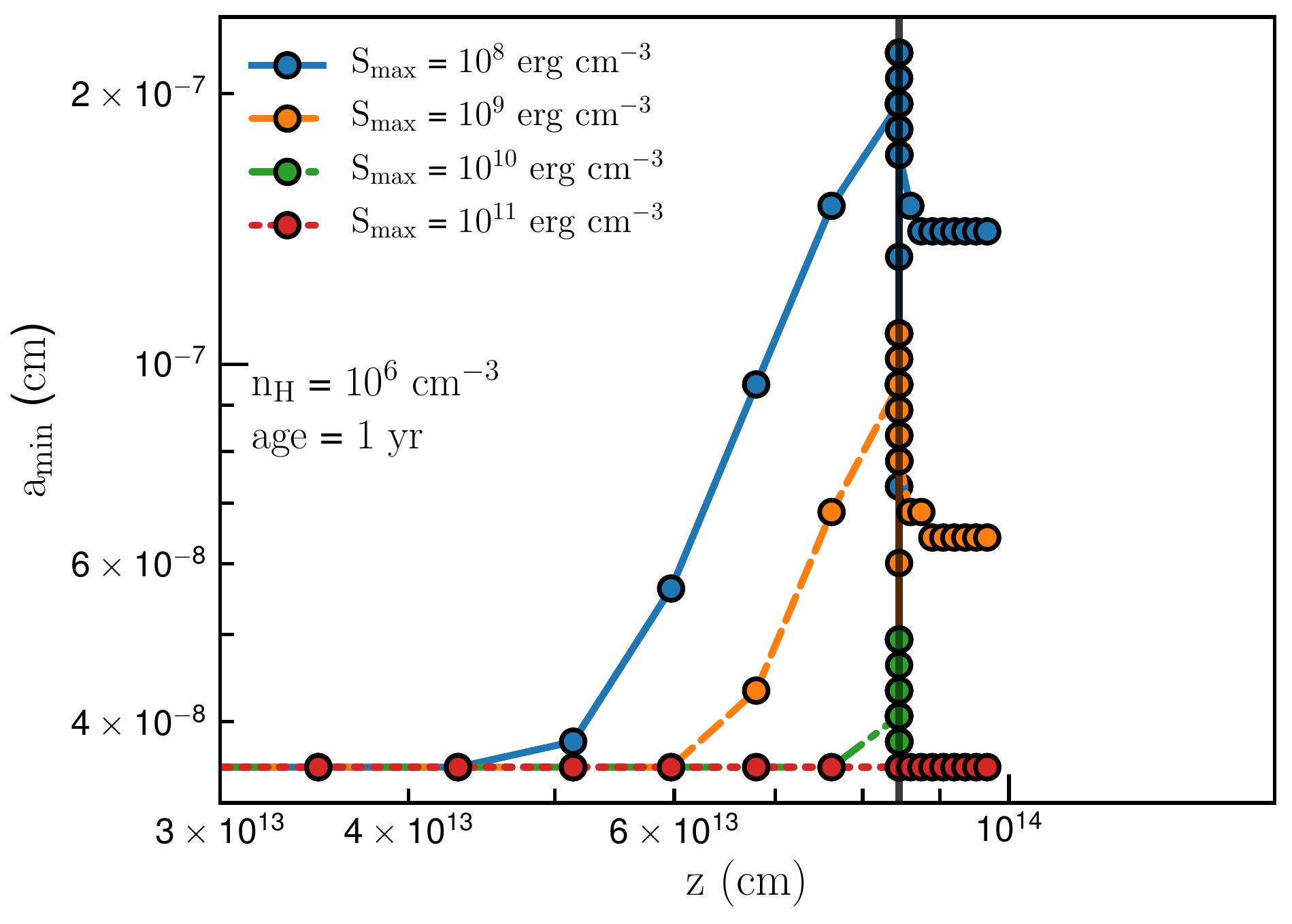}
    }
    \caption{Grain disruption size of PAHs vs. the location in the shock of $v_s=30\km\s^{-1}$, assuming the different tensile strengths and $n_{\H}=10^{4}\cm^{-3}$ (top panel), $n_{\H}=10^{5}\cm^{-3}$ (middle panel) and $n_{\H}=10^{6}\cm^{-3}$ (bottom panel). For each gas density, the results for a younger shock are shown in the right hand side. The black vertical solid line is the J-shock front.}
    \label{fig:a_cri}
\end{figure*}

Figure \ref{fig:omega_cri} illustrates the rotation rate $\langle\omega^{2}\rangle^{1/2}$ as a function of the grain size. The intersection between the rotation rate and the critical disruption rate $\omega_{\rm cri}$ determines the critical disruption size of nanoparticles, denoted by $a_{\rm disr}$ or $a_{\rm min}$. Larger nanoparticles ($a>a_{\rm disr}$) can survive the shock passage, but smaller nanoparticles of $a\lesssim 2$ nm are disrupted due to its faster rotation. The disruption is efficient in the C-shock component, which is representative in blue and orange lines, because of the suprathermal rotation of nanoparticles. For example, nanoparticles of $a< 0.5$ nm with $S_{\max}=10^{10}\erg\cm^{-3}$ can be disrupted (Figure \ref{fig:omega_cri}, bottom panel). In the J-shock component, despite their thermal/subthermal rotation (i.e., $T_{\rot} \leq T_{\gas}$, see Figure \ref{fig:Trot_a}), nanoparticles can be still disrupted by centrifugal force due to high gas temperatures of $\simeq 10^{4}\K$ (see Section \ref{sec:model}), which is shown by green and red lines. However, the disruption effect in J-shocks is much less efficient than in C-shocks. In general, the efficiency of rotational disruption in the CJ-shock actually depends on the shock density, velocity, and age.

Figure \ref{fig:a_cri} illustrates the grain disruption size $a_{\rm min}$ as a function of $z$ for the different values of $S_{\rm max}$. The vertical black line represents for the location of the J-shock front. We see the same effect as reported in \cite{2019ApJ...877...36H} that strong nanoparticles can survive the shock passage (red dotted line), while weak nanoparticles can be destroyed (other colored lines). The disruption size increases for weaker materials (see blue, orange and green dotted lines). The apparent difference from \cite{2019ApJ...877...36H} is the existence of the disruption by thermal collisions in the J-shock component. The disruption mechanism is stronger in younger shocks (Figure \ref{fig:a_cri}, right panel) because the J-shock component dominates and induces higher gas temperature in this case (see Section \ref{sec:model}).  

\section{Spinning dust emission from nanoparticles in non-stationary shocks}\label{sec:spindust}
Rapidly spinning nanoparticles that own permanent electric dipole moment is known to emit electric dipole radiation at microwave frequencies (\citealt{1998ApJ...508..157D}; \citealt{Hoang:2010jy}). In this section, we model the electric dipole emission from nanoparticles that survive the rotational disruption, whose smallest size a$_{\rm min}$ is determined by the disruption mechanism (see Section \ref{sec:rotational_disruption}). Here the net abundance of nanoparticles is assumed to be constant throughout the shock. Such an assumption is not implausible because grain shattering (see e.g, \citealt{2011A&A...527A.123G}) can reproduce nanoparticles to compensate for rotational disruption.

\subsection{Emission spectrum}
As introduced in \cite{2019ApJ...877...36H}, the rotational emissivity of nanoparticles weighted by H nucleon is:
\bea \label{eq:jnu_w}
\frac{j_{\nu}(\mu, T_{\rm rot})}{n_{\H}}=\int_{a_{\min}}^{a_{\max}}j_{\nu}^{a}(\mu,T_{\rm rot})\frac{1}{n_{\H}} \frac{dn}{da} da,\label{eq:jem}
\ena 
where $dn/da$ is grain size distribution of nanoparticles which is given by (\citealt{Li:2001p4761}):
\bea
    \frac{1}{n_H}\frac{dn_j}{da}=\frac{B_j}{a} \exp{ \left( -0.5 \left[ \frac{\log(a/a_{0,j})}{\sigma_{j}}\right]^{2}\right)}
\ena
with j stands for PAHs and silicate nanoparticles. The corresponding constant $B$, and the parameters $a_{0}$, $\sigma$ are adopted as in \cite{2019ApJ...877...36H}. 

Above, $j_{\nu}^{a}(\mu, T_{\rm rot})$ is the emissivity from an individual spinning nanoparticle of size $a$ at $\rm T_{rot}$, which is statistically determined by
\bea
j_{\nu}^{a}(\mu, T_{rot}) = \frac{1}{2}P(\omega,\mu)f_{MW}(\omega, T_{rot}),
\ena
where $P(\omega,\mu)$ is the emission power emitted by a rotating dipole moment $\mu$ at angular velocity $\omega$, and $f_{MW}$ is the Maxwellian distribution of angular velocity of a rotating grain at $T_{rot}$. Respectively, they are given as:
\bea
P(\omega,\mu)& = &\frac{4}{9}\frac{\omega^4 \mu^2}{c^3} \\
f_{MW}(\omega,T_{rot})& =& \frac{4\pi}{(2\pi)^{3/2}} \frac{I^{3/2} \omega^2}{(k T_{rot})^{3/2}} \exp{\left( -\frac{I\omega^2}{2kT_{rot}} \right)}
\ena
with $c$ speed of light, dipole moment $\rm \mu \simeq 9.3 (\beta/0.4\D) a^{3/2}_{-7}\D$ for PAH particles and $\rm \mu \simeq 8.2(\beta/0.4\D) a^{3/2}_{-7}\D$ for nanosilicates. In this work, $\beta=0.4\D$ is adopted as a typical value. 

\begin{figure}
\includegraphics[width=0.45\textwidth]{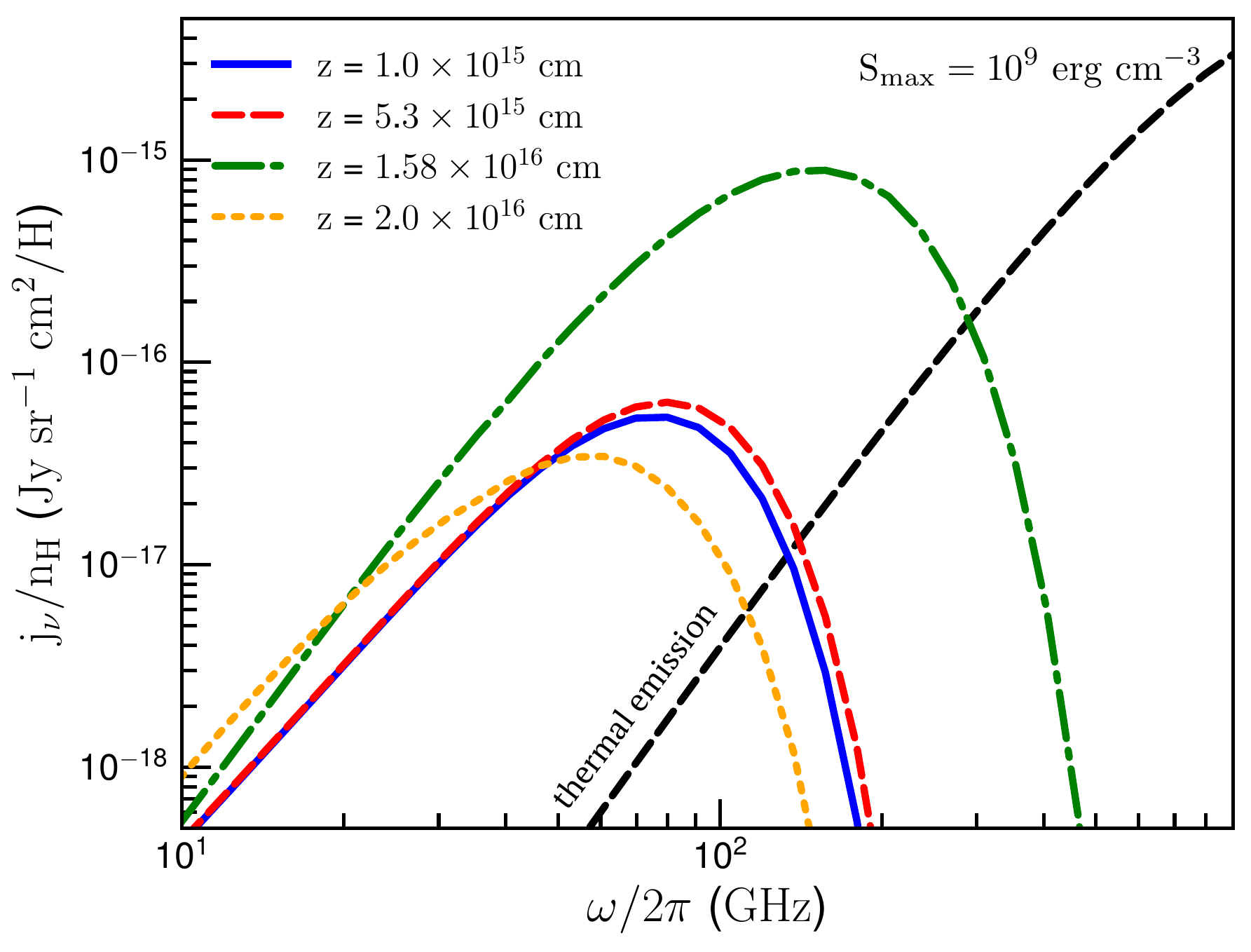}
\includegraphics[width=0.45\textwidth]{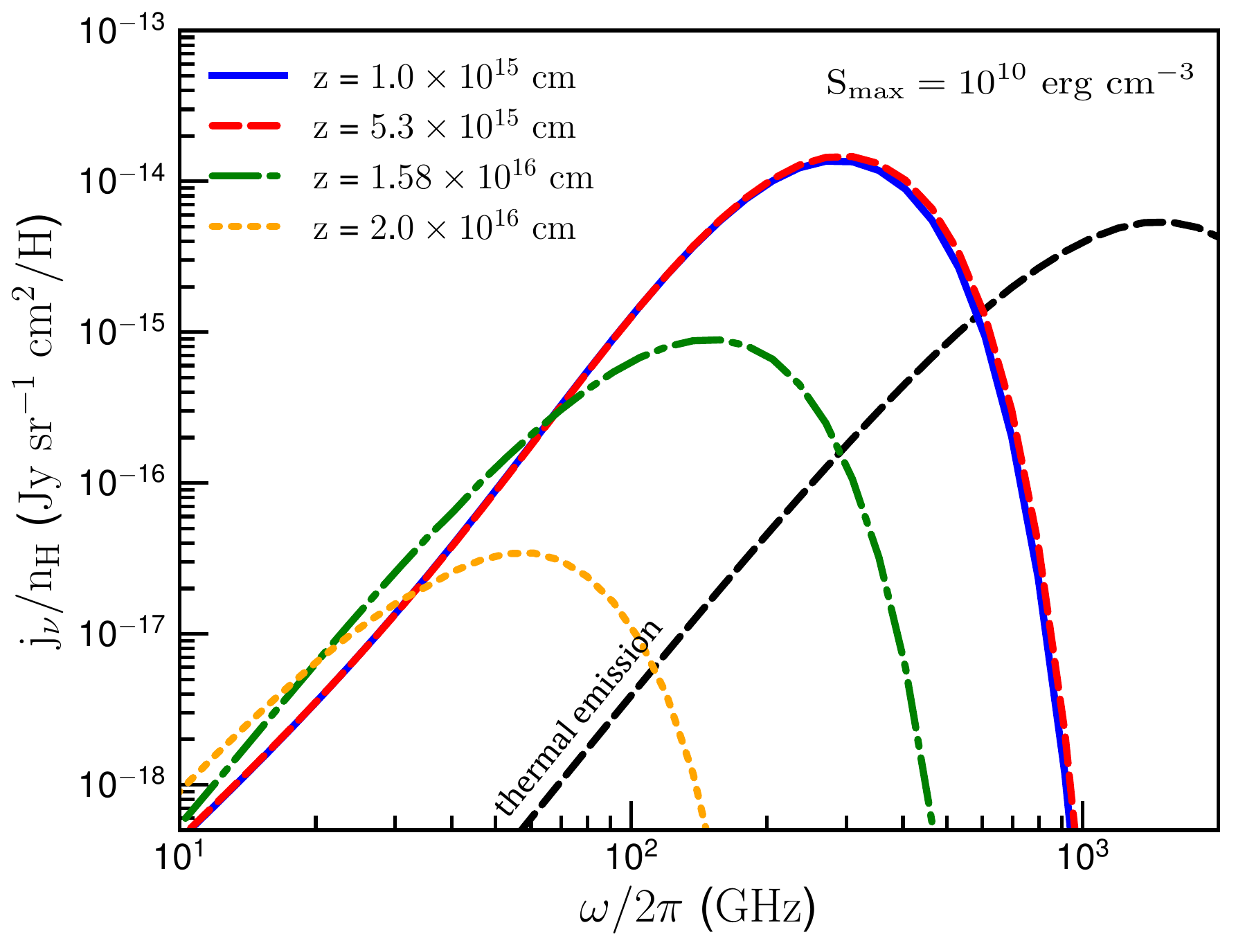}
\caption{Rotational emission spectrum of spinning nanoparticles for n$_{\H}$=10$^{4}\,$cm$^{-3}$, v$_{s}$=30 km$\,$s$^{-1}$, and $age=10^{3}\yr$ computed at several positions in the shock with $S_{\rm max}=10^{9}\erg\cm^{-3}$ (upper panel) and $S_{\rm max}=10^{10}\erg\cm^{-3}$ (lower panel). Dust is considered of $90\%$ of PAHs and $10\%$ of silicates. Thermal dust emissivity is also shown for comparison (black dashed line).}
\label{fig:spindust_n4}
\end{figure}

\begin{figure}
\includegraphics[width=0.45\textwidth]{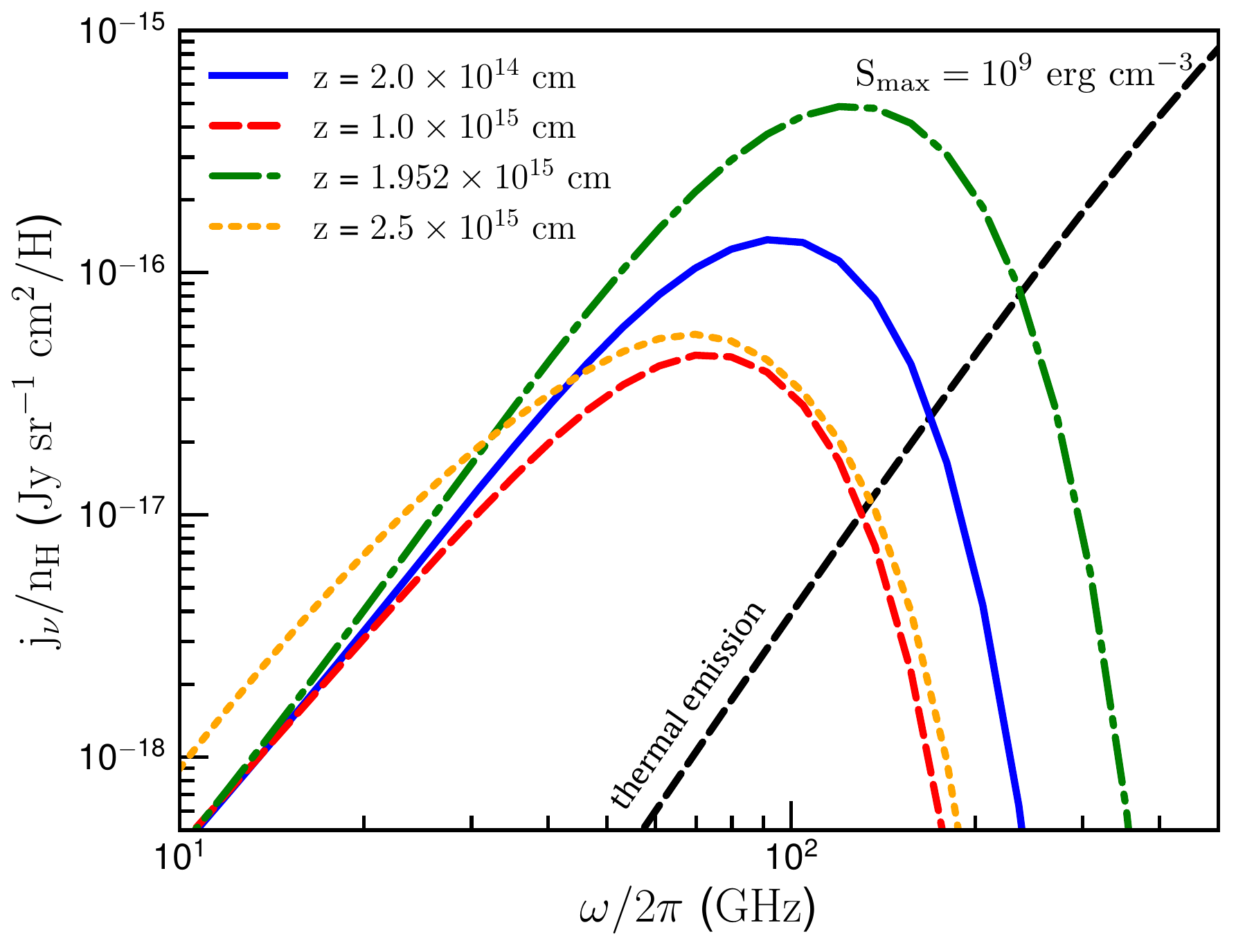}
\includegraphics[width=0.45\textwidth]{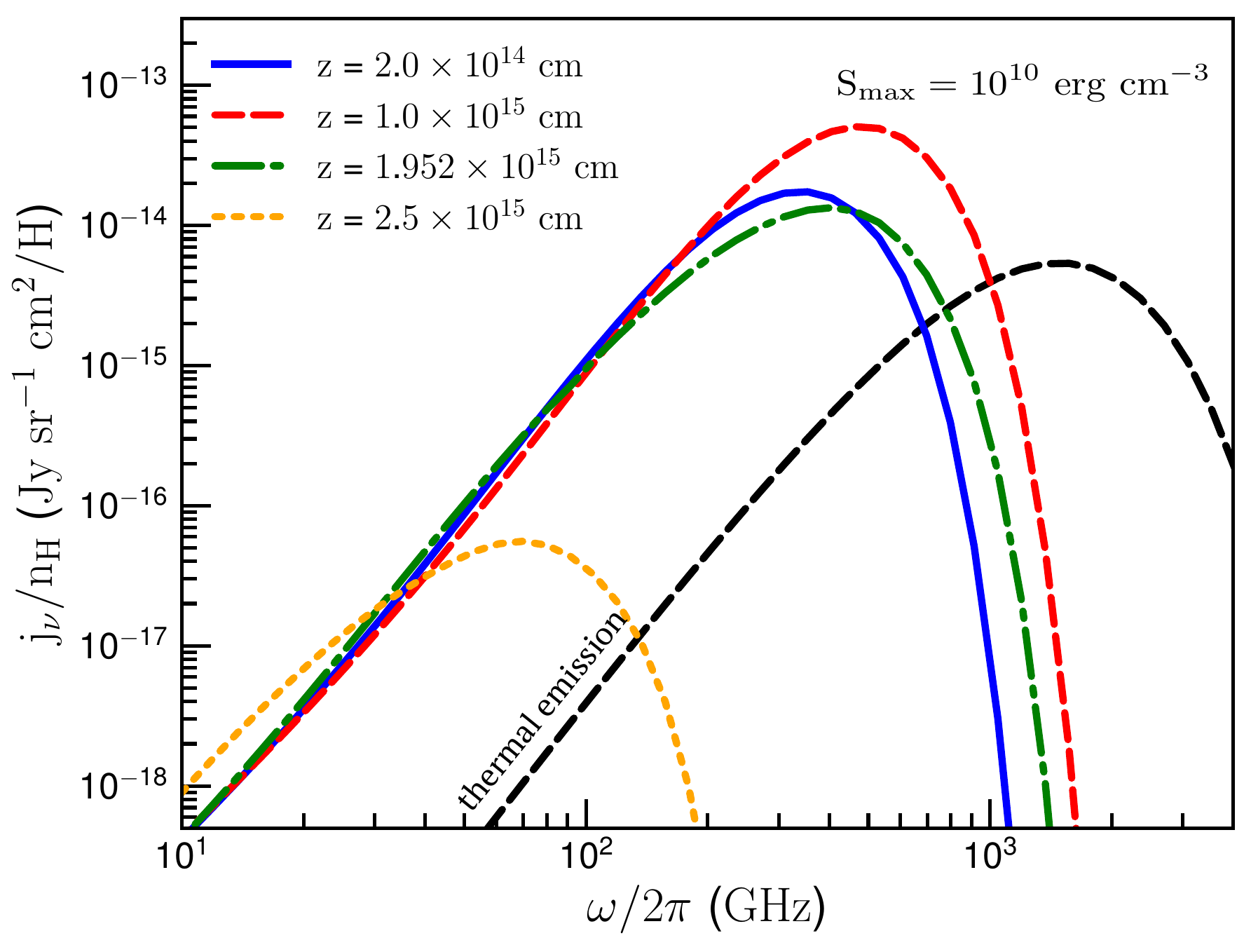}
\caption{Same as Figure \ref{fig:spindust_n4} but for n$_{\H}$=10$^{5}\,$cm$^{-3}$, v$_{s}$=30 km$\,$s$^{-1}$, and $age=10^{2}\yr$ computed at several positions in the shock with $S_{\rm max}=10^{9}\erg\cm^{-3}$ (upper panel) and $S_{\rm max}=10^{10}\erg\cm^{-3}$ (lower panel).}
\label{fig:spindust_n5}
\end{figure}

\begin{figure}
\includegraphics[width=0.45\textwidth]{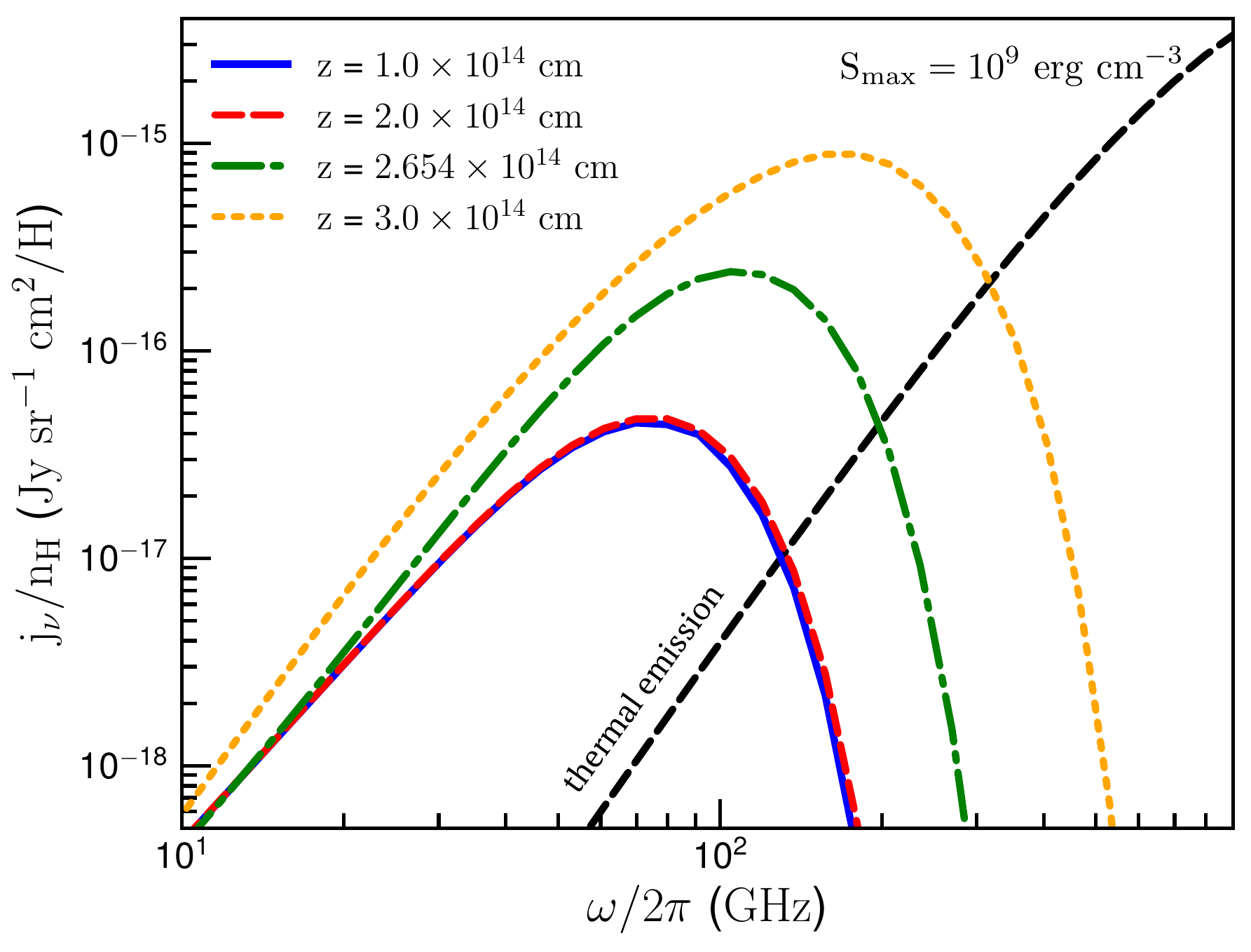}
\includegraphics[width=0.45\textwidth]{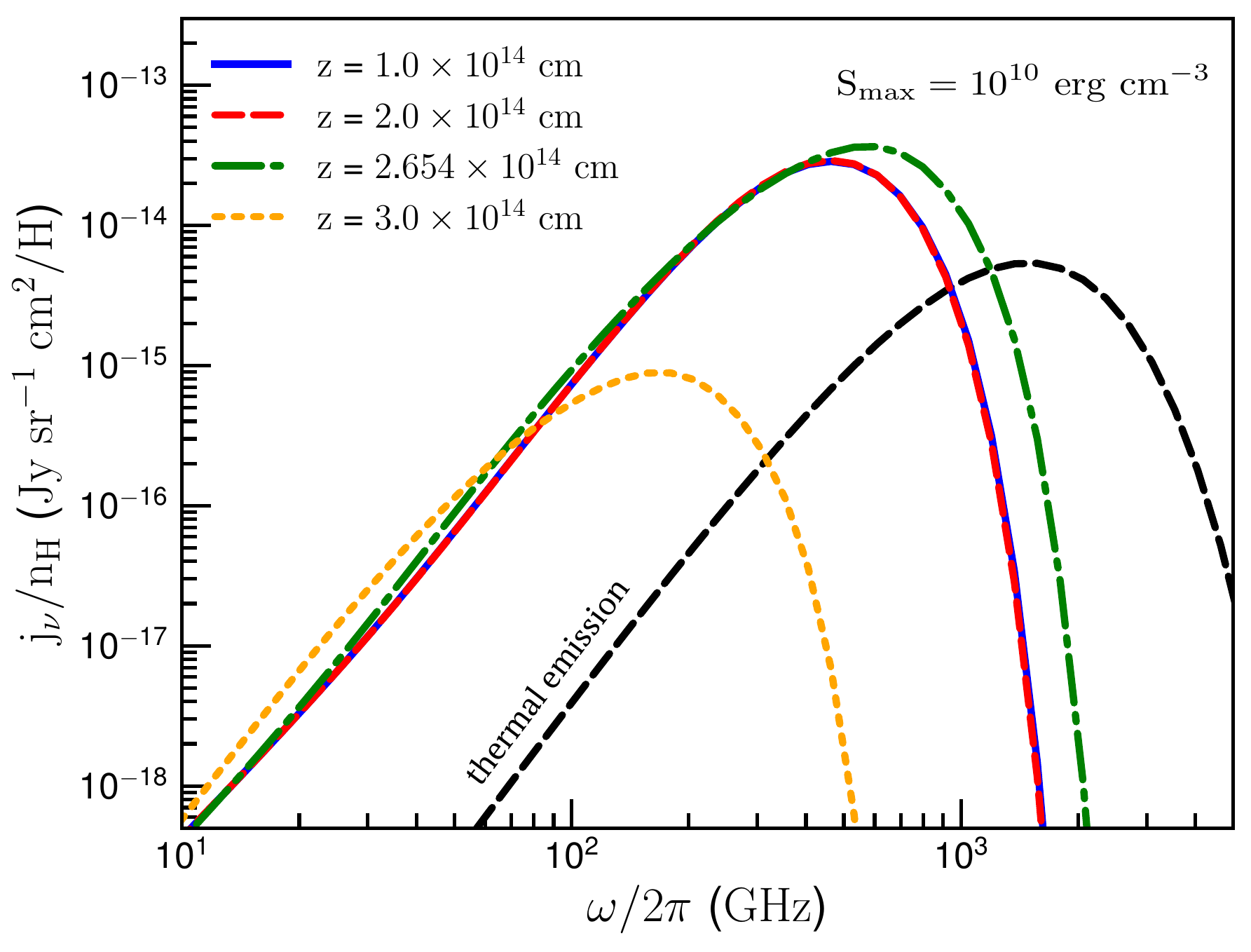}
\caption{Same as Figure \ref{fig:spindust_n4} but for n$_{\H}$=10$^{6}\,$cm$^{-3}$, v$_{s}$=30 km$\,$s$^{-1}$, and $age=10\yr$ computed at several positions in the shock with $S_{\rm max}=10^{9}\erg\cm^{-3}$ (upper panel) and $S_{\rm max}=10^{10}\erg\cm^{-3}$ (lower panel).}
\label{fig:spindust_n6}
\end{figure}

Figure \ref{fig:spindust_n4} shows an example of the emission spectrum of spinning dust as a function of frequency computed at several locations $z$ in the shock as considered in Figure \ref{fig:omega_cri}. The dashed black dashed line shows the thermal dust emissivity from dust grains (see \citealt{Hoang:2018el}). For weak material (e.g., $S_{\rm max}=10^{9}\erg\cm^{-3}$), when the smallest grains are remarkably suppressed by rotational disruption, the rotational emissivity is much less than the thermal emission, but it is dominant over thermal dust at frequencies below $\nu<100 $GHz at most considered shock locations (upper panel). For stronger material (e.g., $S_{\rm max}=10^{10}\erg\cm^{-3}$), whose the smallest grains are not or unremarkable enhanced by rotational disruption (see Fig. \ref{fig:a_cri}, left panel), the rotational emissivity is much stronger and comparable with the thermal emission (lower panel). Figure \ref{fig:spindust_n5} and Figure \ref{fig:spindust_n6} illustrate the same phenomena but for denser shocked medium with $n_{\H}=10^{5}\cm^{-3}$ and $n_{\H}=10^{6}\cm^{-3}$, respectively. 

\begin{figure*}
    \centering
    \subfloat{
     \includegraphics[width=0.45\textwidth]{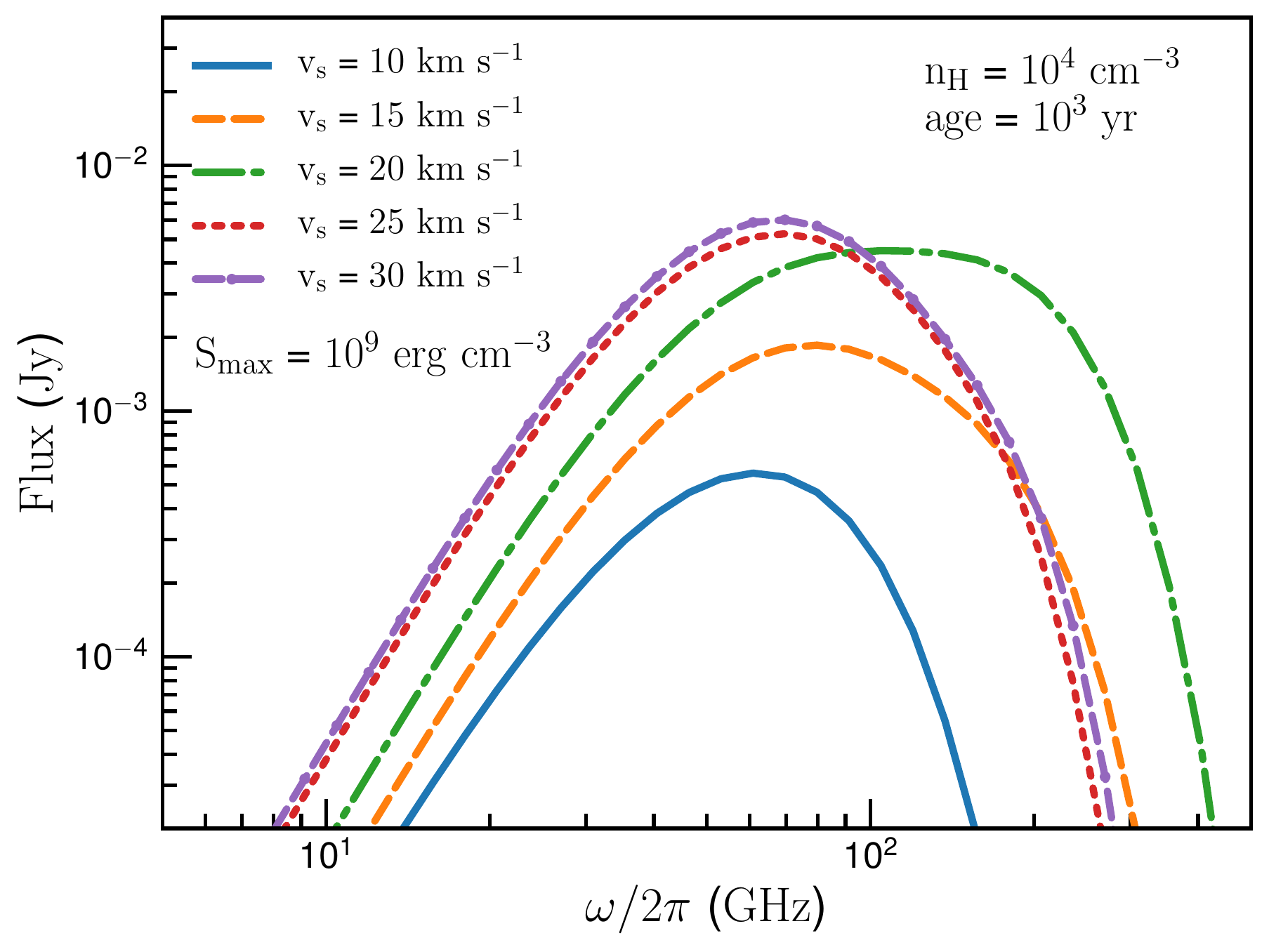}
    \includegraphics[width=0.45\textwidth]{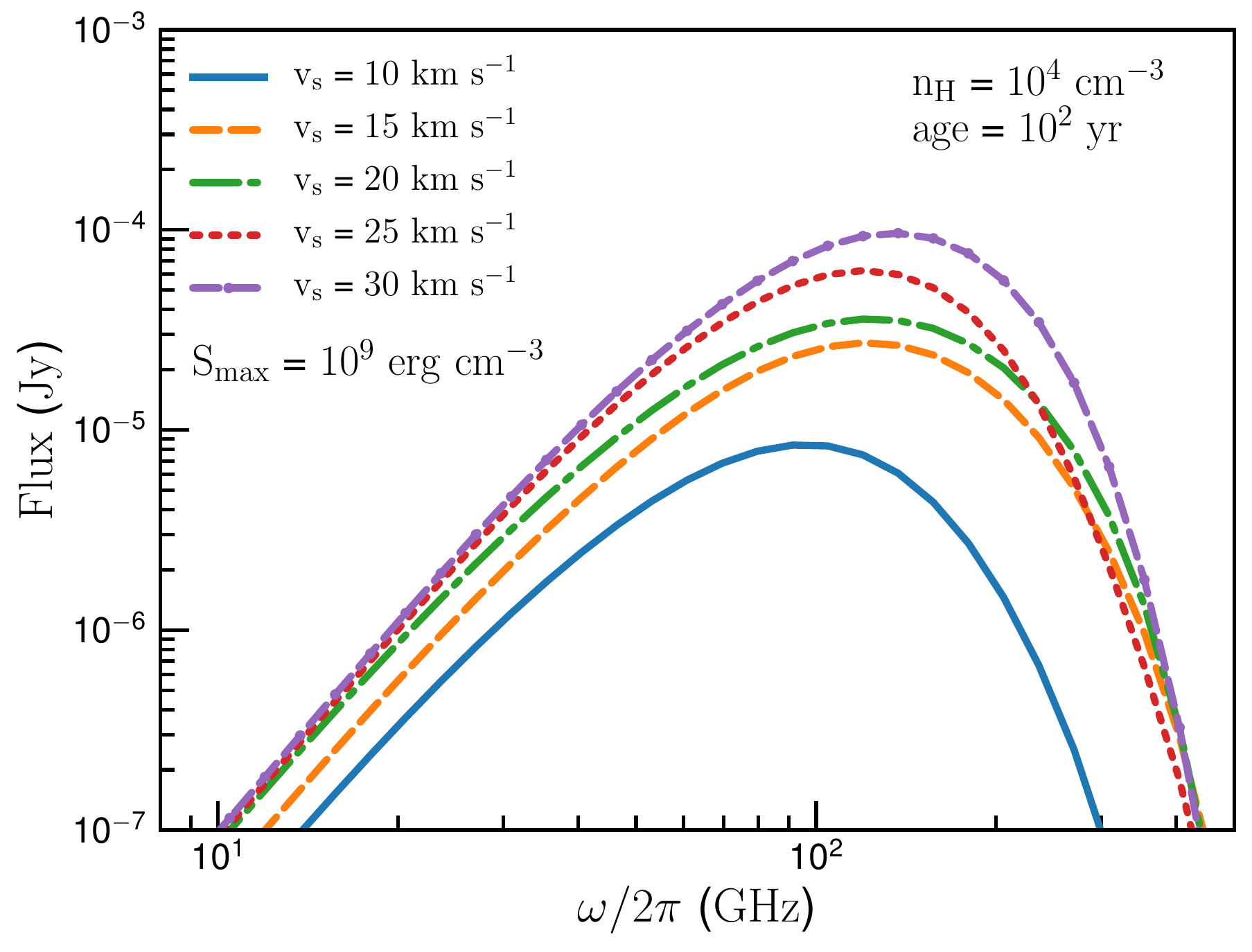}
    }\\
    \subfloat{
    \includegraphics[width=0.45\textwidth]{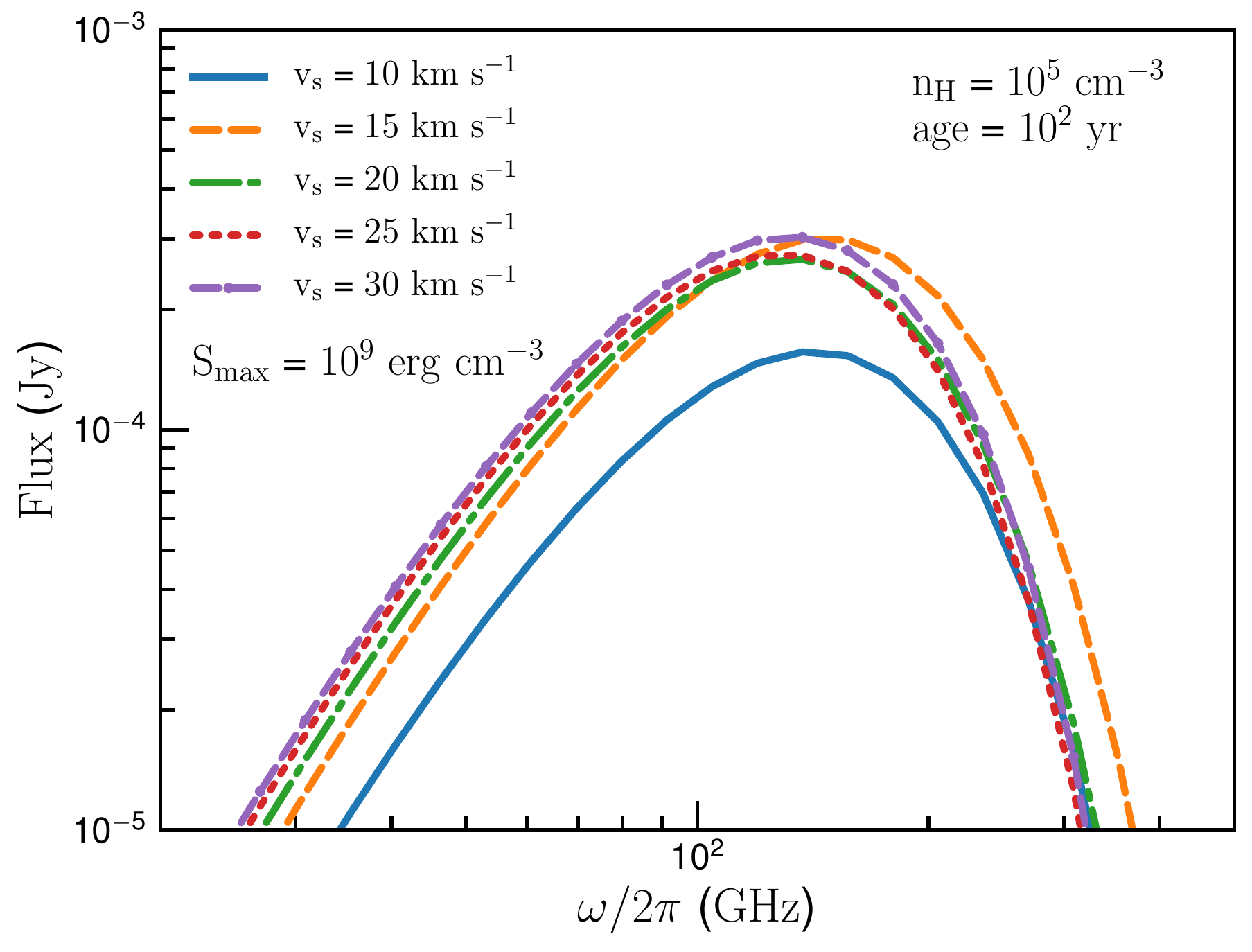}
    \includegraphics[width=0.45\textwidth]{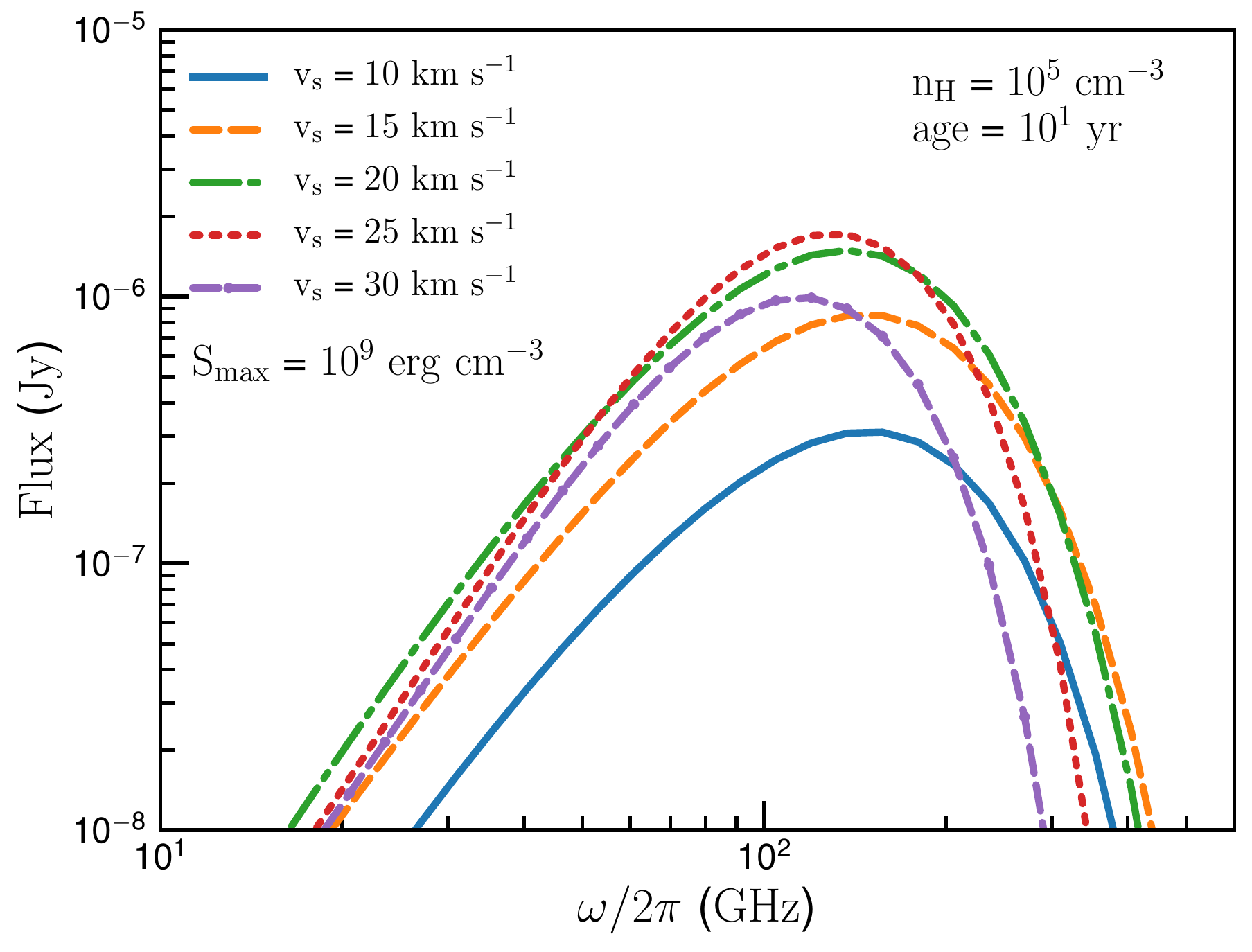}
    }\\
    \subfloat{
    \includegraphics[width=0.45\textwidth]{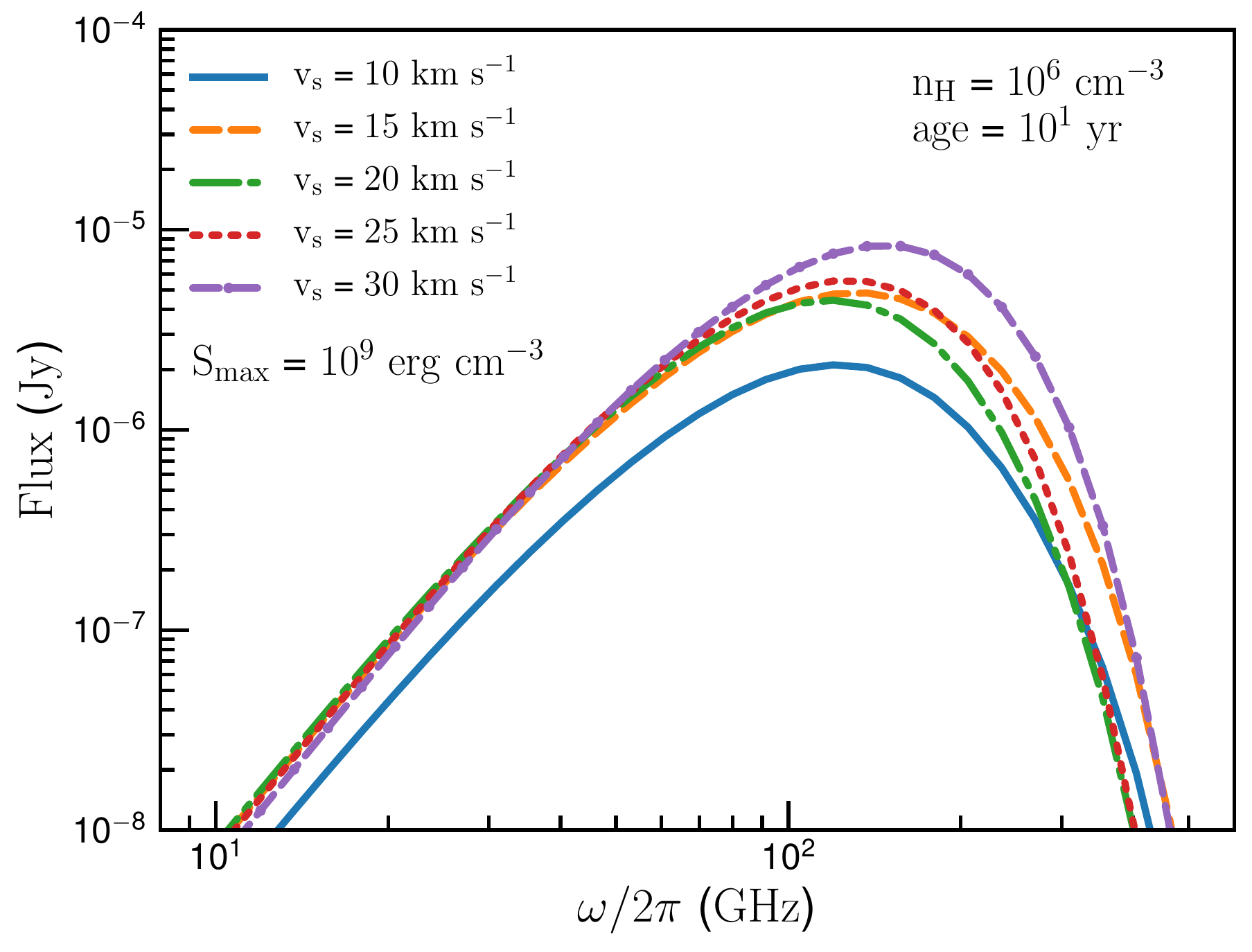}
    \includegraphics[width=0.45\textwidth]{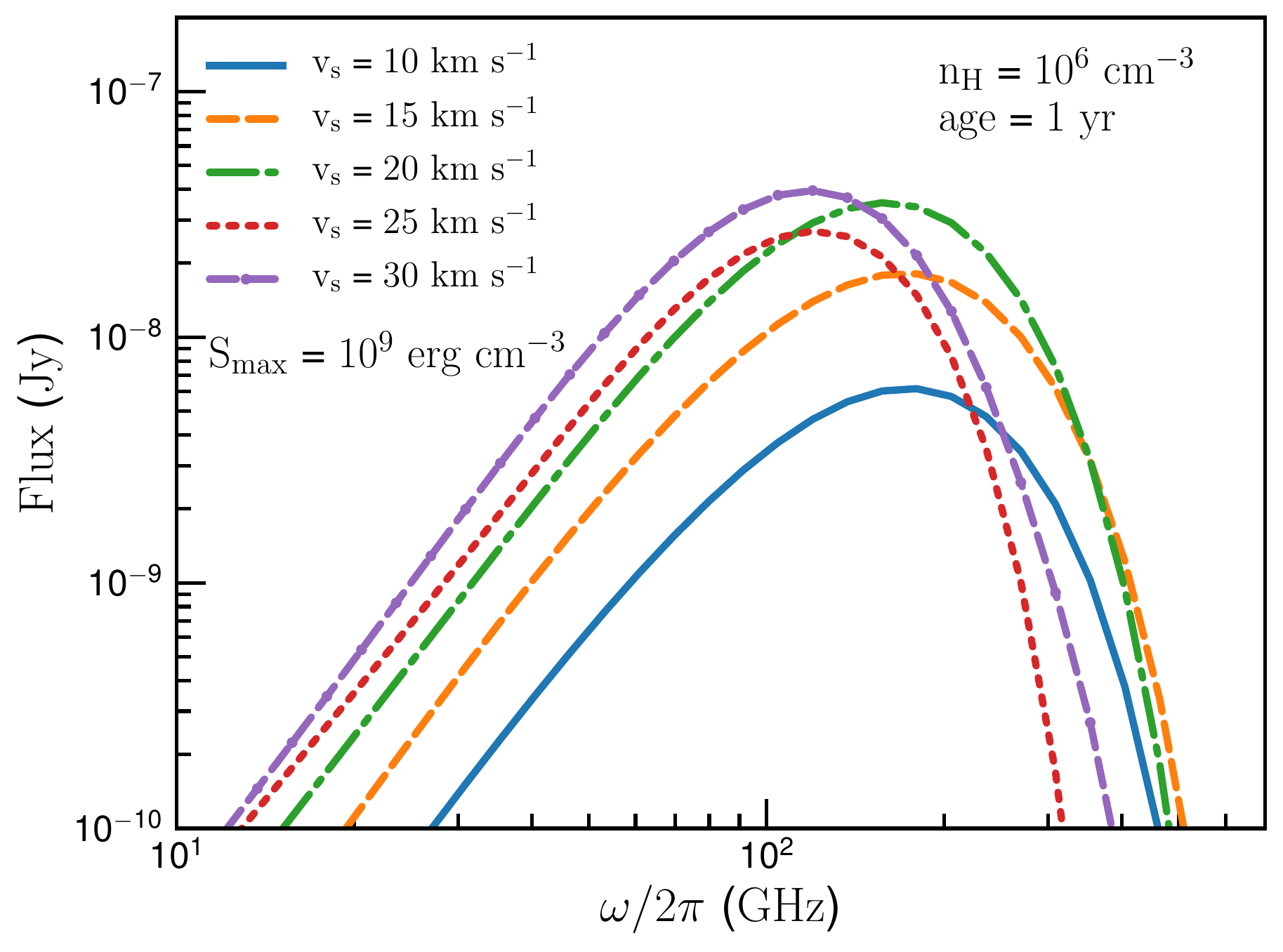}
    }
    \caption{Spectral flux of spinning dust emission for $n_{\H}=10^{4}\cm^{-3}$ (top panel), $n_{\H}=10^{5}\cm^{-3}$ (middle panel), and $n_{\H}=10^{6}\cm^{-3}$ (bottom panel). The right hand side represents for younger age. The spectral flux significantly decreases with increasing the pre-shock density and decreasing the shocked age. $S_{\max}=10^{9}\,$erg$\,$cm$^{-3}$ is assumed, and $D=100\pc$ is taken.}
    \label{fig:flux_S9}
\end{figure*}

\begin{figure*}
    \centering
    \subfloat{
    \includegraphics[width=0.45\textwidth]{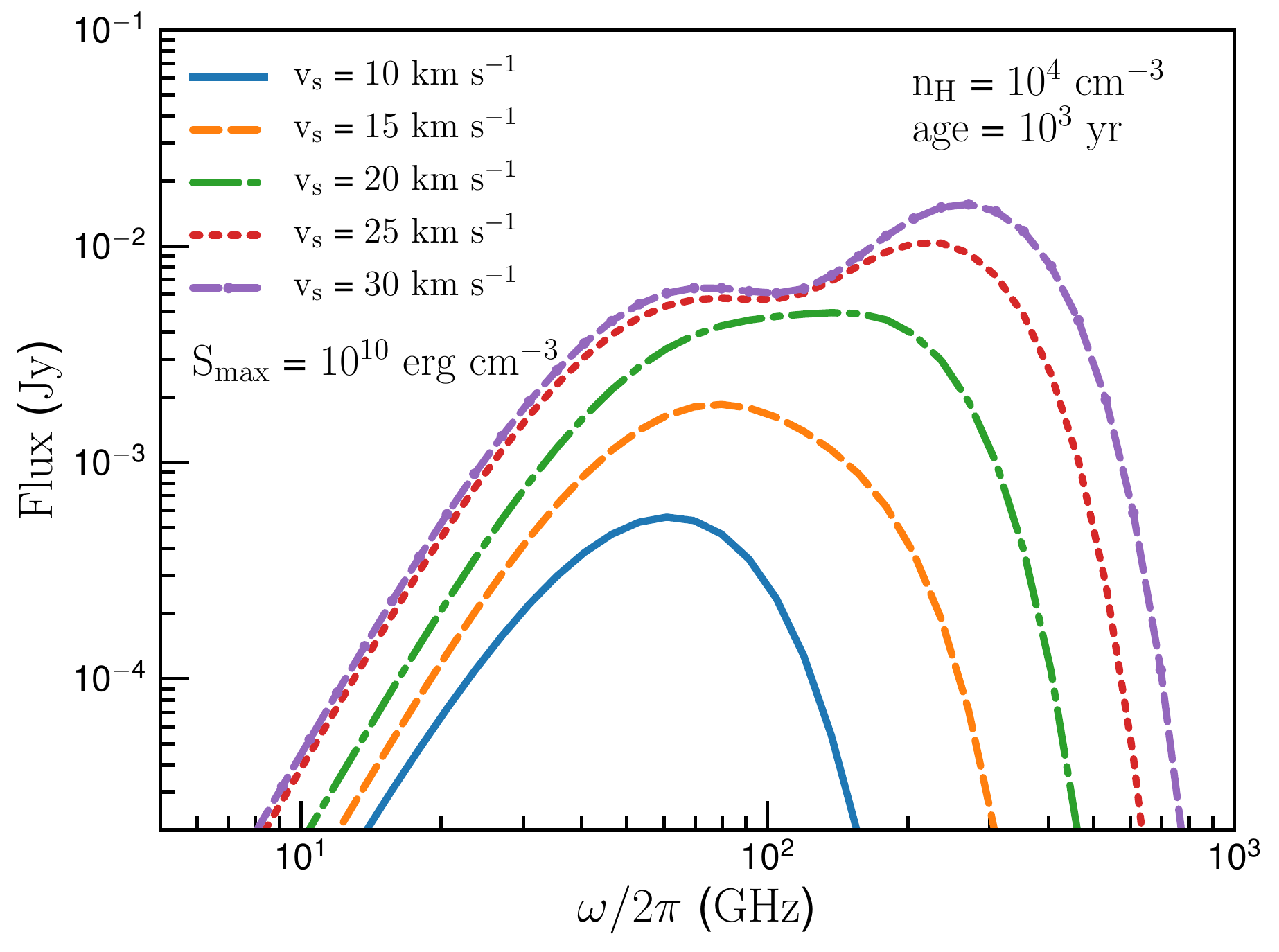}
    \includegraphics[width=0.45\textwidth]{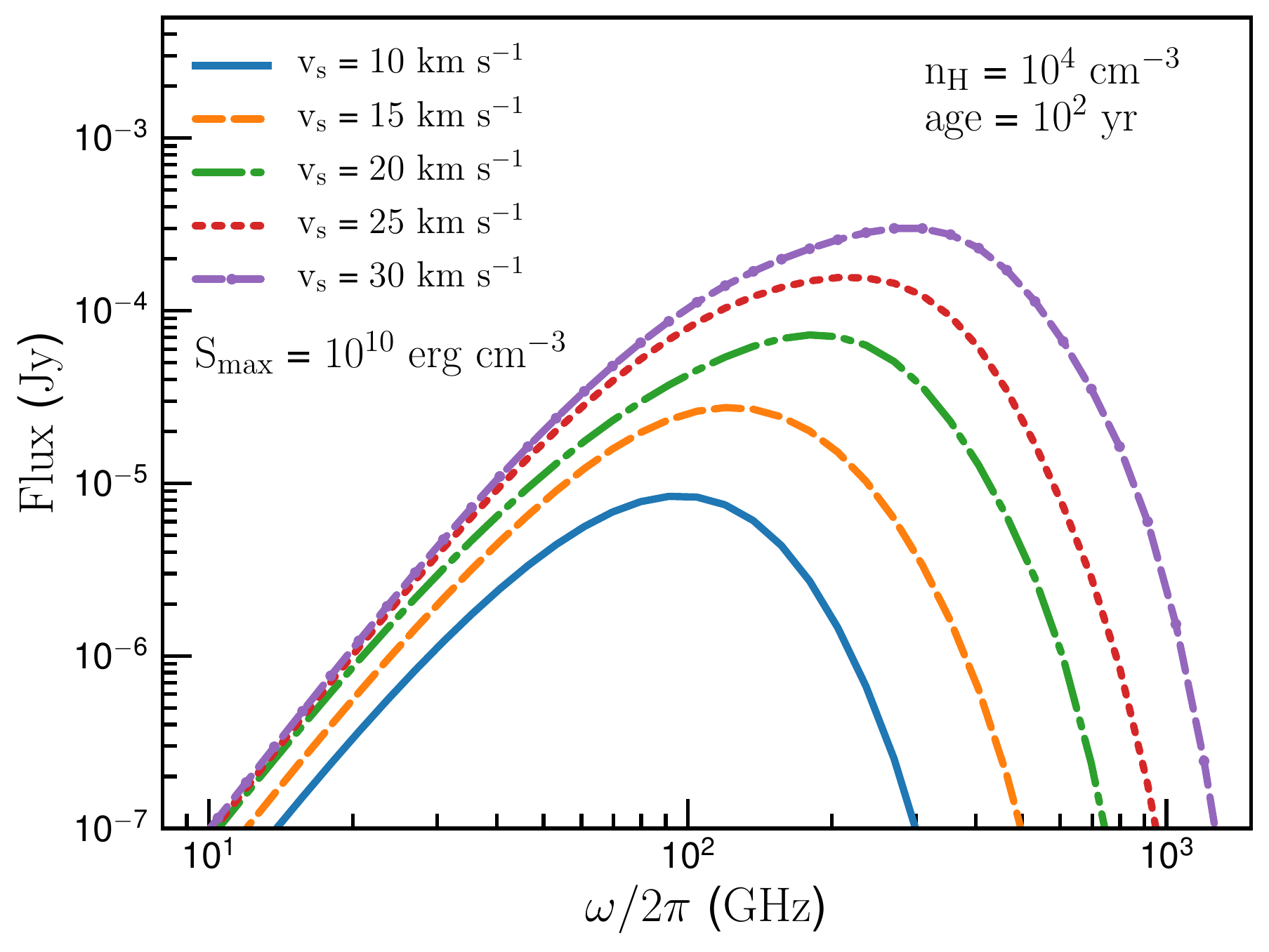}
    }\\
    \subfloat{
    \includegraphics[width=0.45\textwidth]{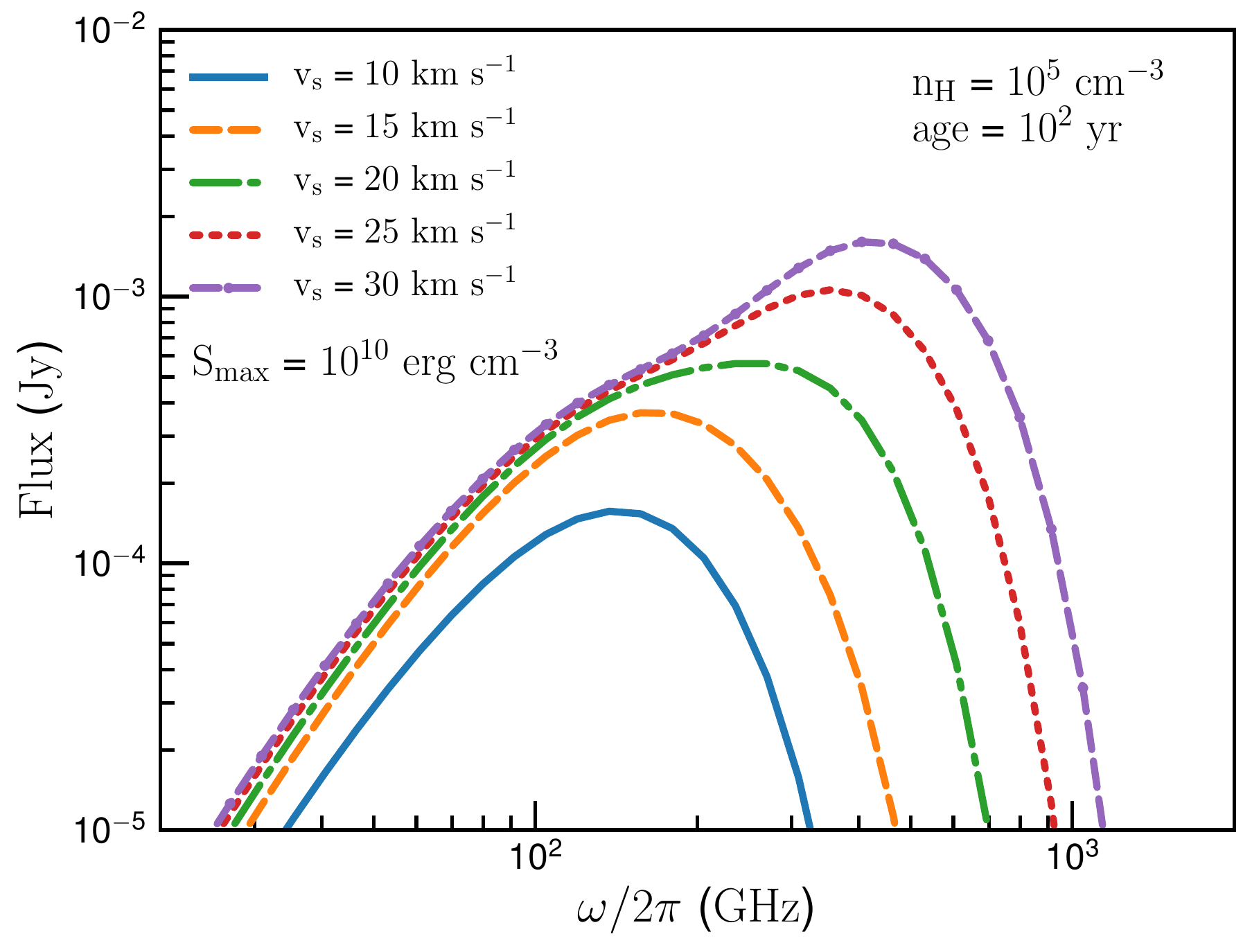}
    \includegraphics[width=0.45\textwidth]{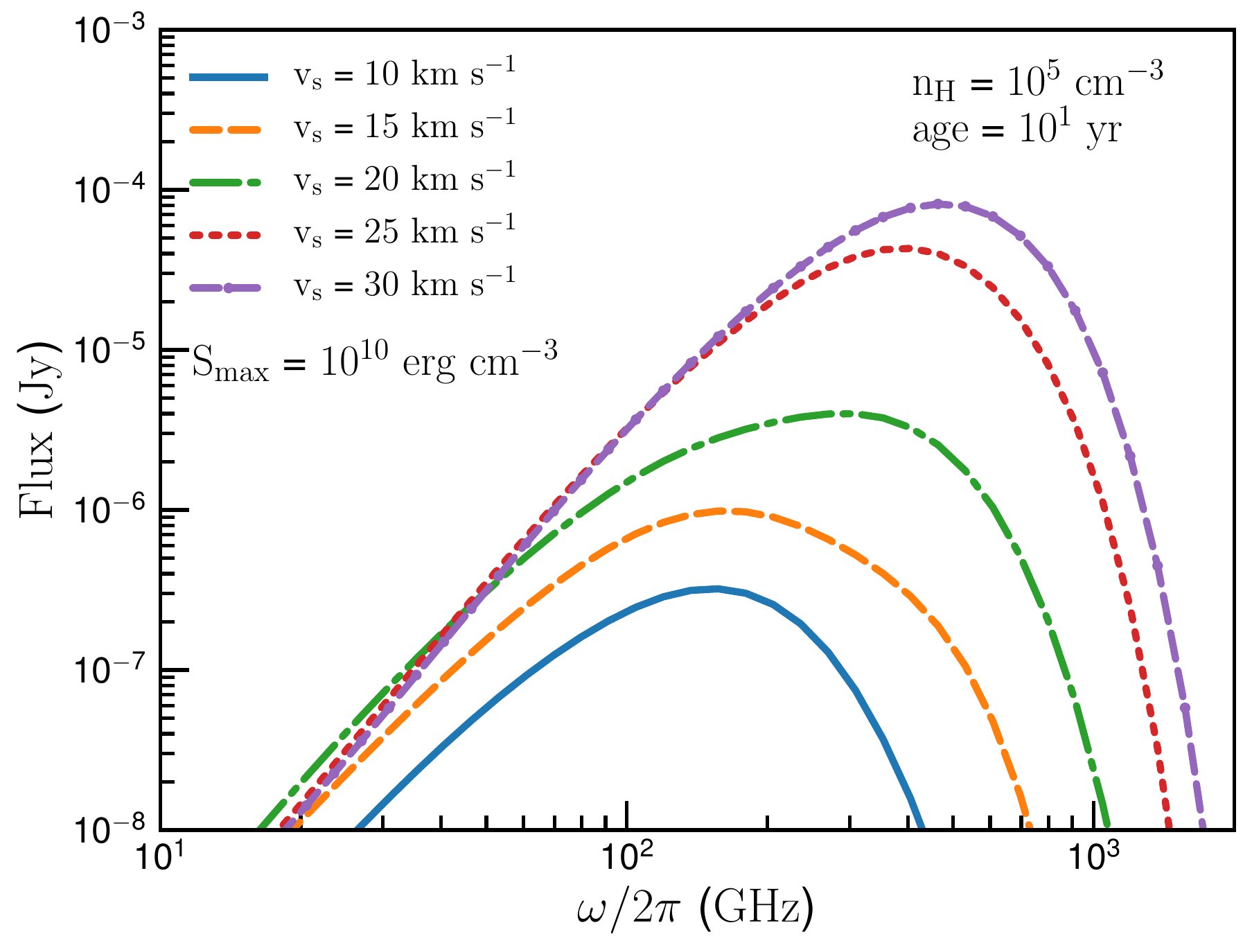}
    }\\
    \subfloat{
    \includegraphics[width=0.45\textwidth]{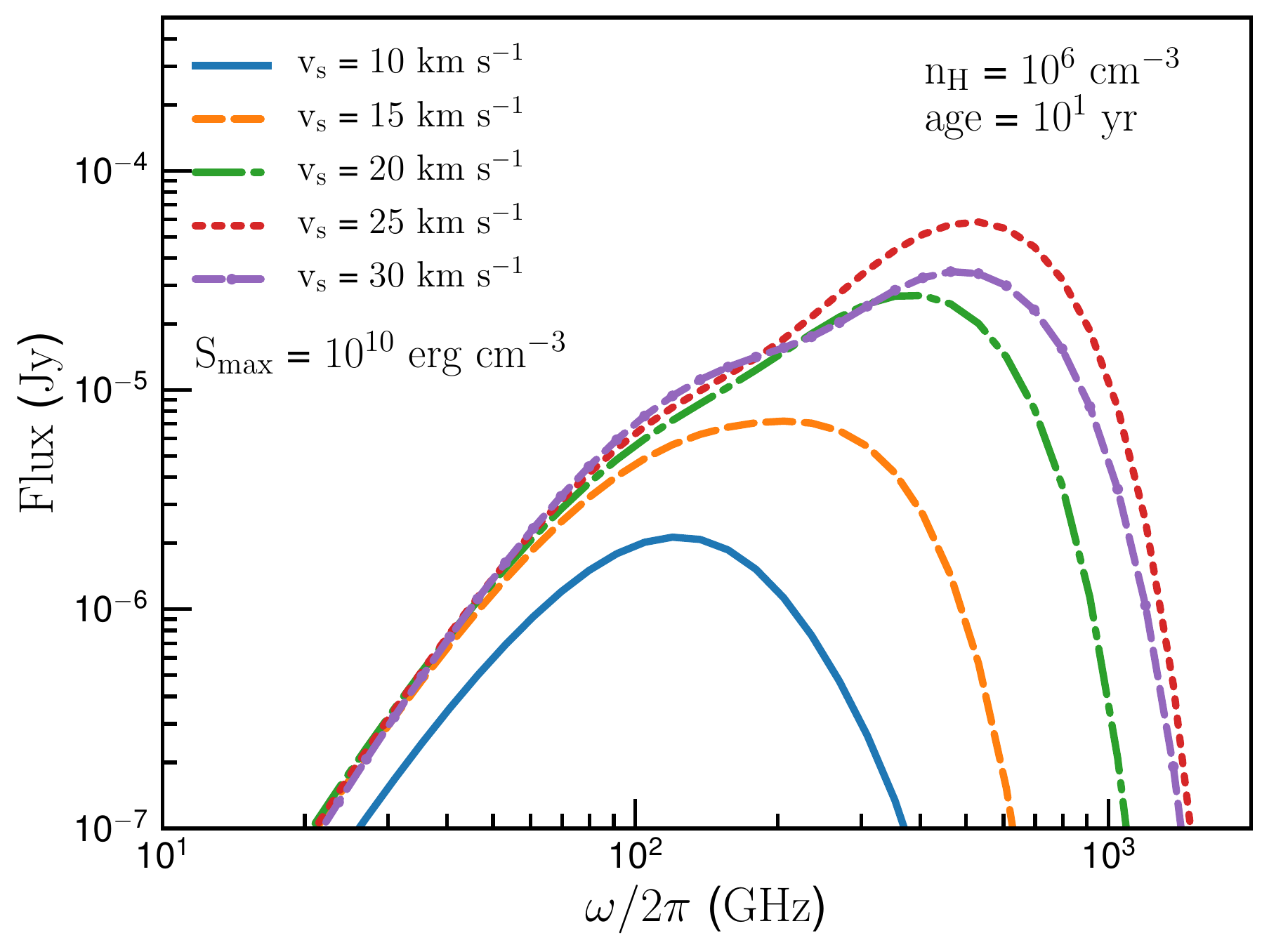}
    \includegraphics[width=0.45\textwidth]{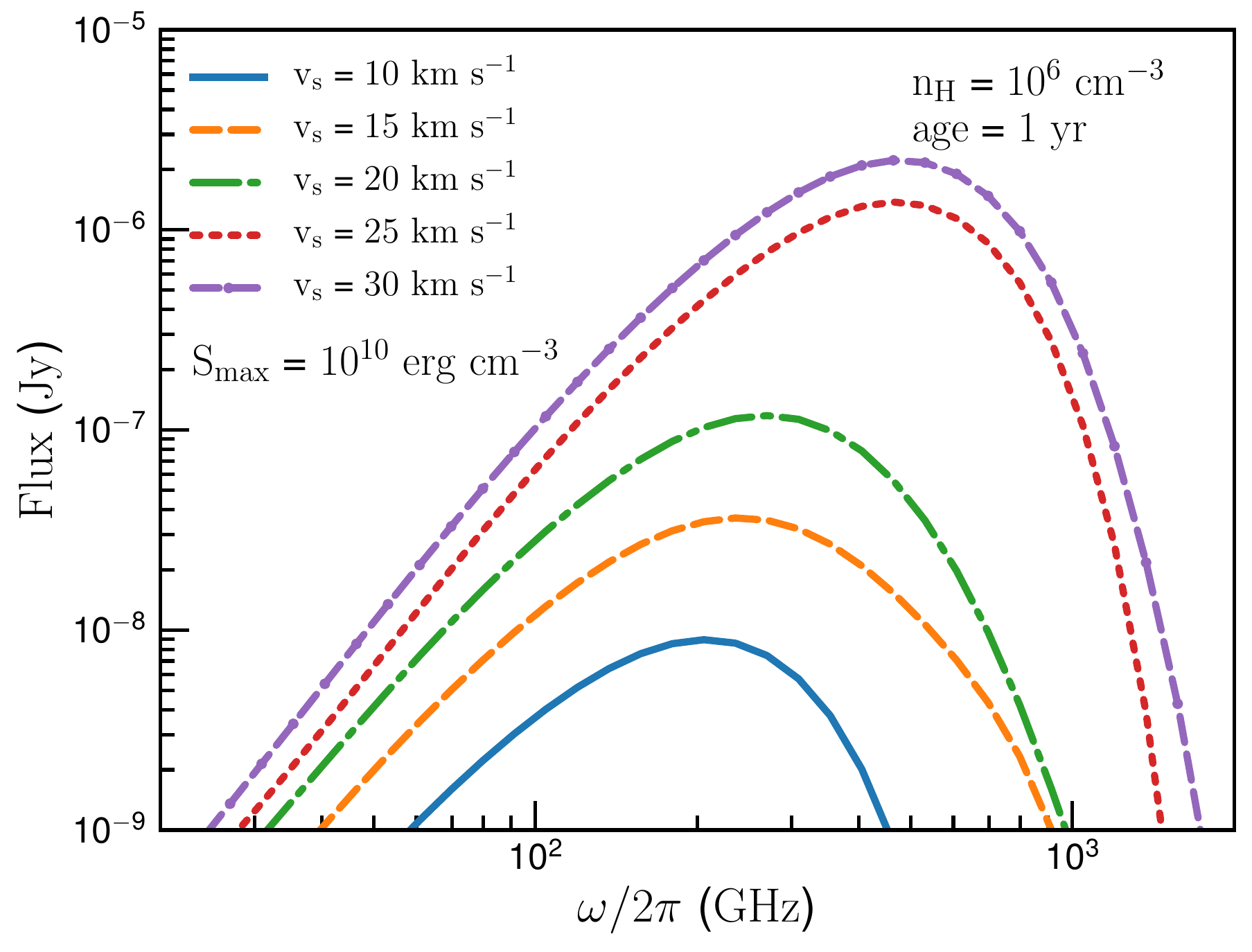}
    }
    \caption{Same as Figure \ref{fig:flux_S9} but for stronger materials with $S_{\max}=10^{10}\,$erg$\,$cm$^{-3}$. The spectral flux also decreases with increasing the pre-shock density and decreasing the shock age. However, the amplitude of the spectral flux of the strong materials is higher than of the weak materials.}
    \label{fig:flux_S10}
\end{figure*}

\subsection{Emission spectral flux}
Assuming a spherical geometry for the shocked region, the spectral flux of spinning dust emission from the shocked region is:
\bea
F_{\nu}= \frac{4\pi}{D^{2}}\int_{z_{\min}}^{z_{\max}} z^{2} n_{\H}(z)\left(\frac{j_{\nu}}{n_{\H}}\right)dz,
\ena
where $D$ is the distance from observer to the shocked region of radius equal to the shock length (see Section \ref{sec:model}).

Figure \ref{fig:flux_S9} shows the spectral flux calculated with different shock velocities for three different CJ-shock models (see Table \ref{tab:ISM}), assuming the tensile strength $S_{\rm max}=10^{9}\erg\cm^{-3}$. From the top to the bottom, the emission flux is significantly decreased with increasing of density due to both the decrease of shocked volume (see Figures \ref{fig:profile_temp}) and the increase in the disruption size as the gas temperature in higher density is hotter than in lower density (see Section \ref{sec:model}). From the left to the right, the emission flux is also significantly decreased due to both the decrease of shocked volume and the increase of the rotational disruption as the gas temperature in the young CJ-shock is higher than in the older ones (see Section \ref{sec:model}). Furthermore, stronger shocks (i.e., higher shock velocity) induce stronger disruption effect, thus the peak frequency decreases due to the removal of smallest nanoparticles. The value of this peak is around $\sim$ 100 GHz.

Figure \ref{fig:flux_S10} shows the similar results but for stronger materials. For this case, since rotational disruption is less efficient, both the peak of the spectral flux and peak frequency are higher than the results shown in Figure \ref{fig:flux_S9} for weaker materials.

\section{Discussion}\label{sec:discuss}
\subsection{Rotational disruption of nanoparticles in J-shocks vs. C-shocks}
In shocked regions, two popular mechanisms that are previously known to destroy dust grains include sputtering and grain-grain collisions (\citealt{1996ApJ...469..740J}). \cite{2019ApJ...877...36H} proposed a new mechanism of destruction for very small grains (i.e., nanoparticles), so-called rotational disruption, which appears to be the fastest mechanism working in steady C-shock regions (see Table \ref{tab:destr}). The key points of this new mechanism are the followings: (i) the dominance of gas collisional excitation due to high gas density in shocks makes nanoparticles rotate thermally, which corresponds to high rotation rates due to high gas temperatures, (ii) the supersonic drift between neutrals and charged nanoparticles is then able to spin-up charged nanoparticles to suprathermal rotation. As a result, the centrifugal stress induced by extremely fast rotation can exceed the maximum tensile strength of grain materials, disrupting the nanoparticle into smaller fragments. We also found that weak grains are efficiently disrupted, while the strong ones are hardly destroyed.

In this paper, we extended our previous study for the non-stationary shocks, which are driven by outflows and young SNRs. This type of shocks is called CJ-shock because it approximately composes of the C-type and the J-type shocks. We found the same mechanism in the C-shock component of the CJ-shock as reported in \cite{2019ApJ...877...36H}. Nevertheless, nanoparticles, in the J-shock component of the CJ-shock, rotate thermal/subthermally because rotational excitation cannot overcome the rotational damping, leading to its rotational temperature equal/lower than the gas temperature. However, as J-shocks can heat gas up to very high temperatures, the grain rotational rate is thus still high enough to disrupt the smallest nanoparticles. We demonstrate that this process is also the most efficient mechanism to disrupt nanoparticles in the J-shock component (see Table \ref{tab:destr}). Note that spherical nanoparticles are assumed and rotational excitation is only considered by stochastic mechanical torques in this study. The efficiency of the disruption mechanism for the realistic, irregular shapes would be increased due to stronger mechanical torques (\citealt{2007ApJ...669L..77L}; \citealt{2018ApJ...852..129H}).

\begin{table}
\begin{center}
\caption{Grain destruction in CJ-shocks}\label{tab:destr}
\begin{tabular}{ll} \hline\hline
{\it Mechanism} & {Timescales (yr)}\cr
\hline\\
& {\bf C-shock part}\cr
Rotational disruption & $0.5 a_{-7}^{4}n_{4}^{-1}v_{\rm drift,1}^{-3}S_{\rm max,10}$\cr
Thermal sputtering & $3.1\times 10^{3}a_{-7}n_{4}^{-1}T_{3}^{-1/2}({10^{-4}}/{Y_{\rm sp}})$\cr
Non-thermal sputtering & $2.4\times 10^{3}a_{-7}n_{4}^{-1}v_{\rm drift,1}^{-1}({10^{-4}}/{Y_{\rm sp}})$\cr
Grain-grain collision & $76a_{-5}n_{4}^{-1}v_{\rm drift,1}^{-1}$\cr
\cr
\hline
\cr
& {\bf J-shock part}\cr
Rotational disruption & $24 a_{-7}^{4}n_{4}^{-1}T_{3}^{-3/2}S_{\rm max,10}$\cr
Thermal sputtering & $3.1\times 10^{3}a_{-7}n_{4}^{-1}T_{3}^{-1/2}({10^{-4}}/{Y_{\rm sp}})$\cr
Grain-grain collision  \footnote{In spite of moving with same velocity, grain-grain collision can occur because of turbulence. The turbulence velocity v$_{\rm gg}$ is about few $\rm km\s^{-1}$} & $76a_{-5}n_{4}^{-1}v_{\rm gg,1}^{-1}$\cr
\cr
\hline
\cr
\multicolumn{2}{l}{{\it Notes}:~$v_{\rm drift,1}=v_{\rm drift}/10\rm km\s^{-1}$}\cr 
\multicolumn{2}{l}{$S_{\rm max,10}=S_{\max}/10^{10} \erg \cm^{-3}$}\cr
\multicolumn{2}{l}{$T_{3}=T_{\gas}/10^{3}\K$} \cr
\multicolumn{2}{l}{$v_{\rm gg,1} =v_{\rm gg}/10\rm km\s^{-1}$} \cr\cr
\hline\hline
\end{tabular}
\end{center}
\end{table}

\subsection{Implications for mid-IR emission from shock regions}
As summarized in Table \ref{tab:destr}, rotational disruption is the most efficient mechanism to destroy smallest nanoparticles such as small PAHs ($a\lesssim 1$ nm) in shocks at low velocity (i.e., v$_{\s}<50\km\s^{-1}$), whereas sputtering is subdominant (see extended discussion in \citealt{2019ApJ...877...36H}). This rotational disruption effect can explain the lack of mid-IR PAH emission in most of SNRs (see \citealt{Kaneda:2011jd} for a review). On the other hand, nanoparticles of size $a\gtrsim 1$ nm can survive throughout the shock, which might reproduce the ubiquitous mid-IR emission features at $9\mum$ and $21\mum$ observed toward SNRs \citep{Rho:2018ee}. For instance, the authors can reproduce these mid-IR features by emission from hot dust grains of SiC and silica with size $a=10$nm, respectively.

\subsection{Constraining the shock velocity in dense regions with spinning dust}
\begin{figure}
\includegraphics[width=0.45\textwidth]{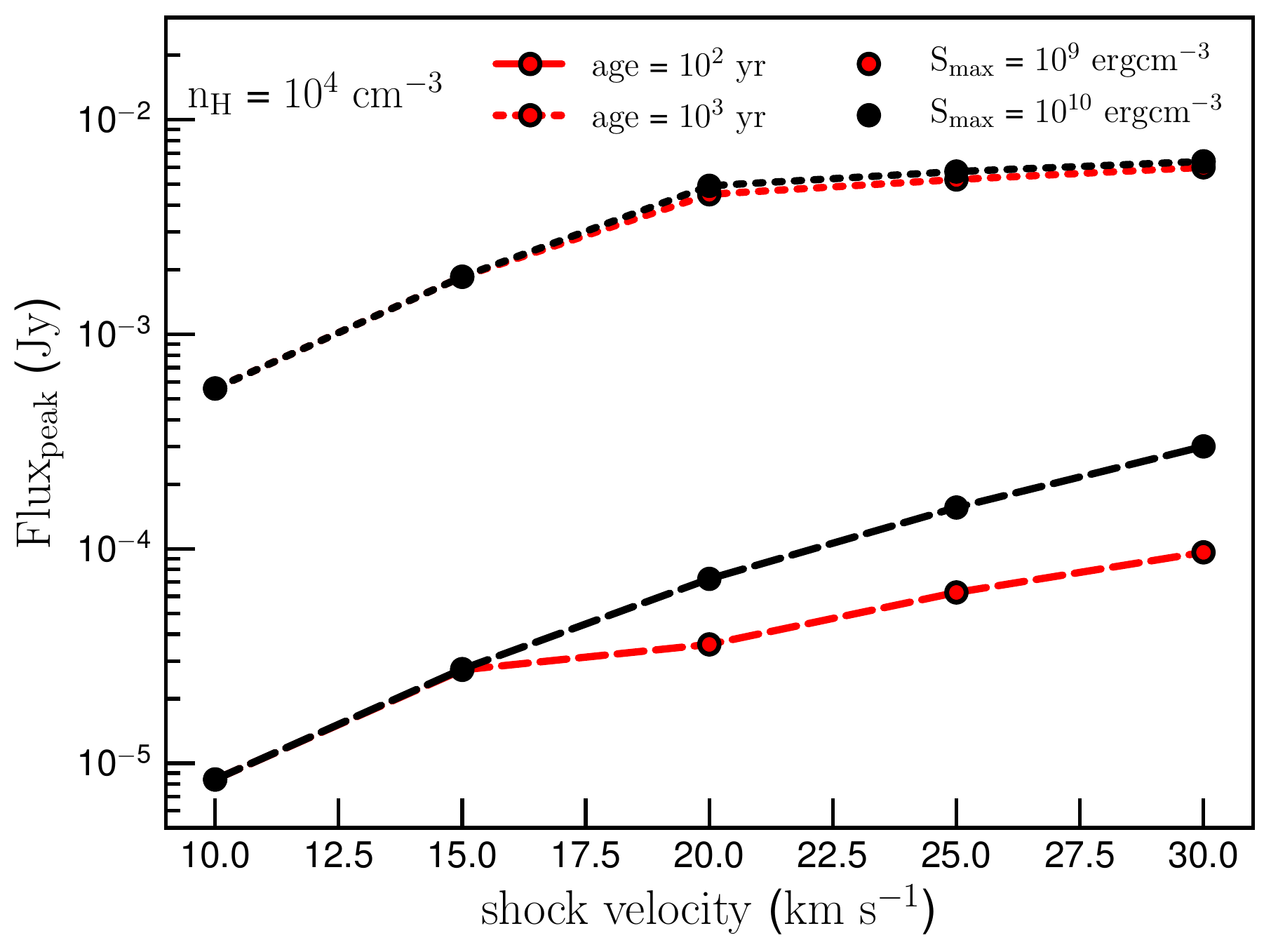}
\includegraphics[width=0.45\textwidth]{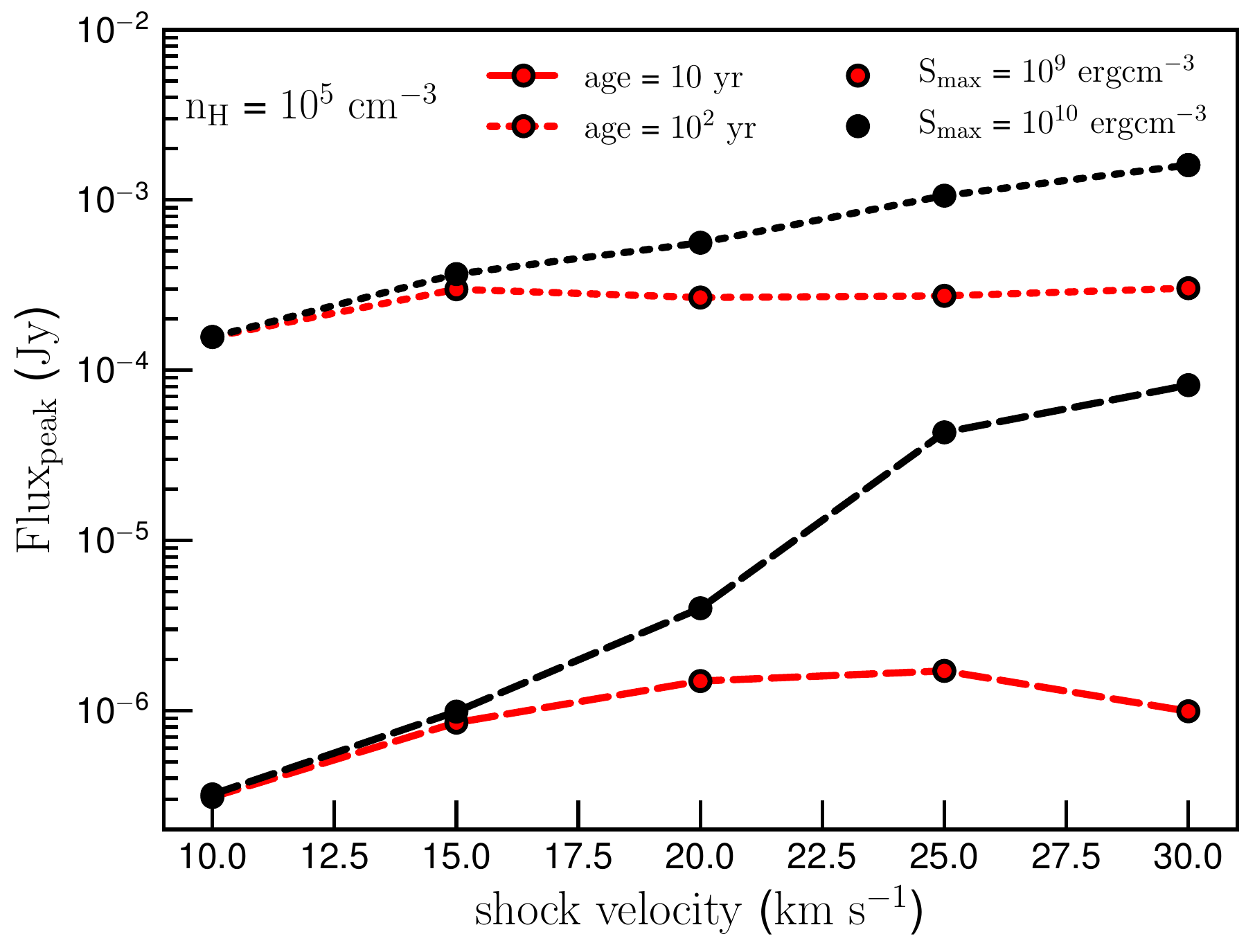}
\includegraphics[width=0.45\textwidth]{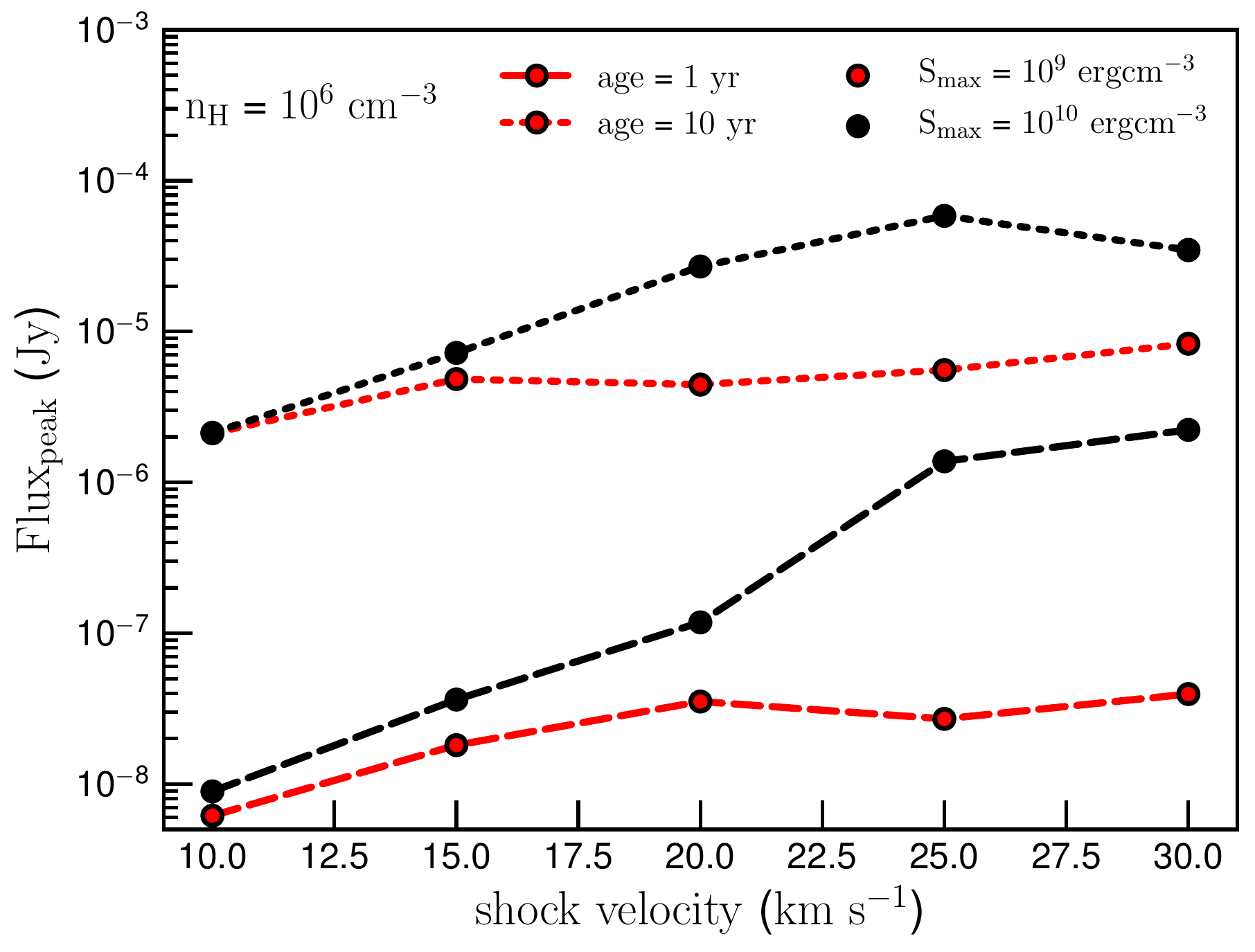}
\caption{Peak emission flux as a function of the shock velocity for $n_{\H}=10^{4}\cm^{-3}$ (top), $n_{\H}=10^{5}\cm^{-3}$ (middle), and $n_{\H}=10^{6}\cm^{-3}$ (bottom). Two values of shock ages and tensile strengths are considered. The peak flux increases rapidly with increasing $v_{\rm s}$ for strong materials (black lines).}
\label{fig:fluxmax_vs}
\end{figure}

\begin{figure}
\includegraphics[width=0.45\textwidth]{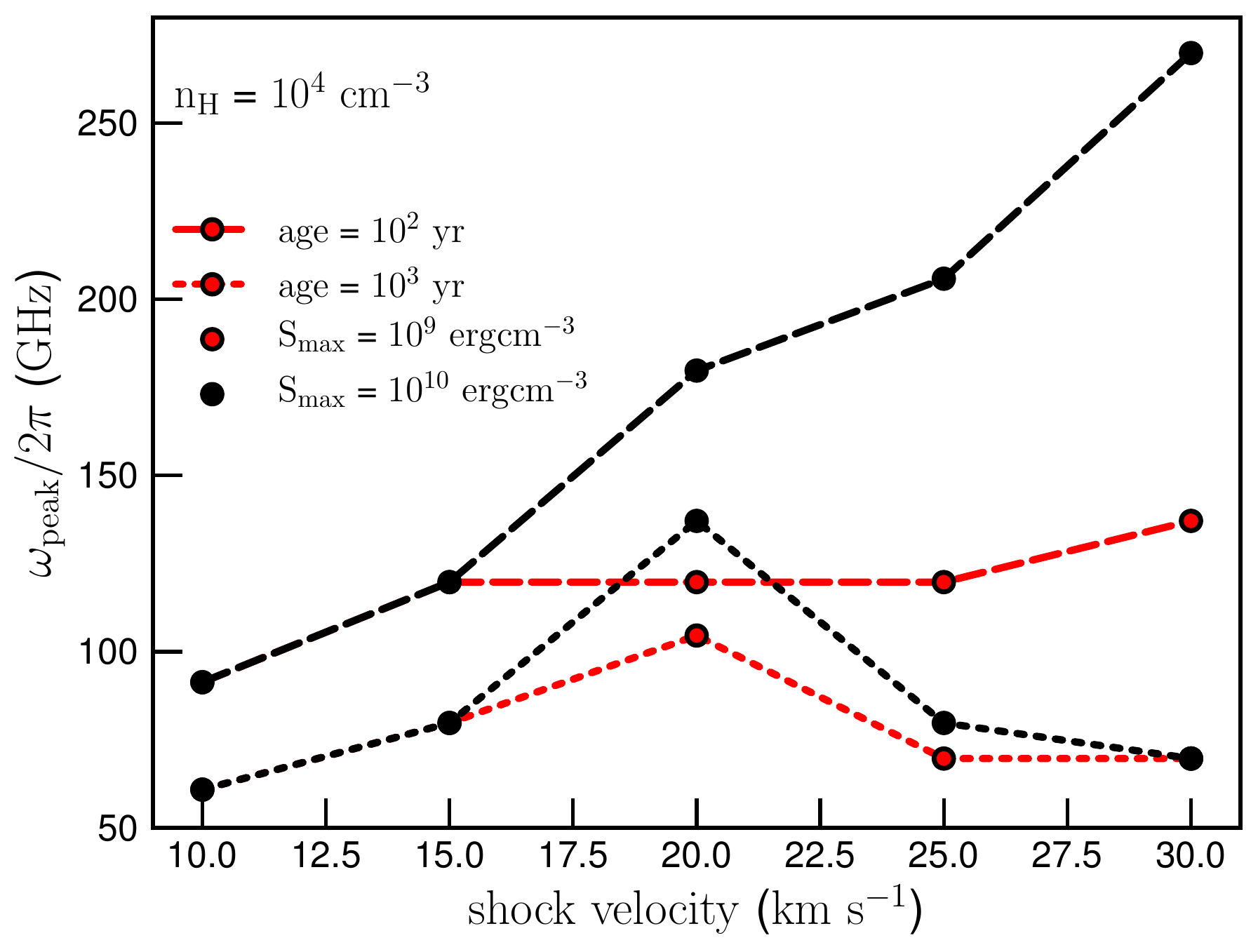}
\includegraphics[width=0.45\textwidth]{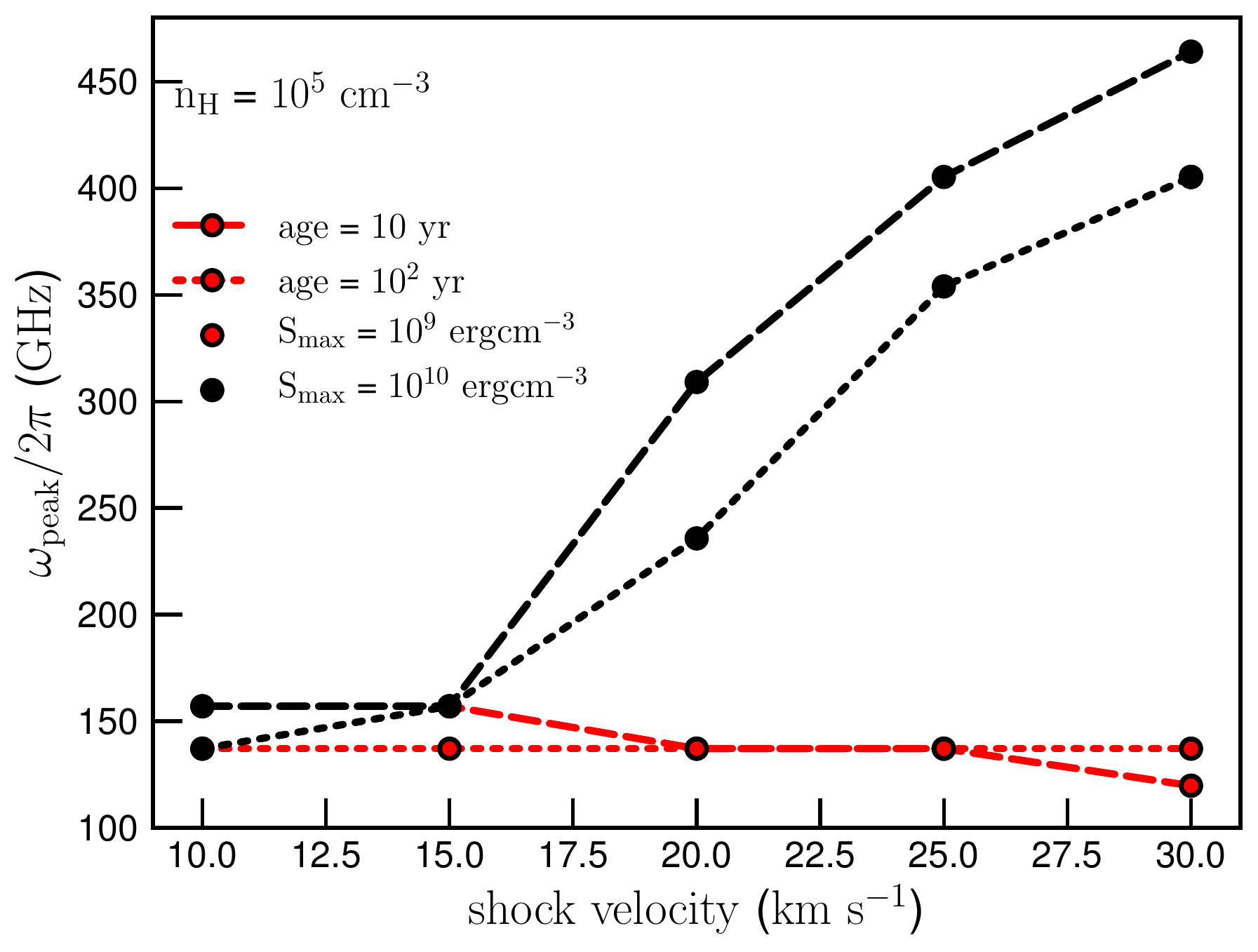}
\includegraphics[width=0.45\textwidth]{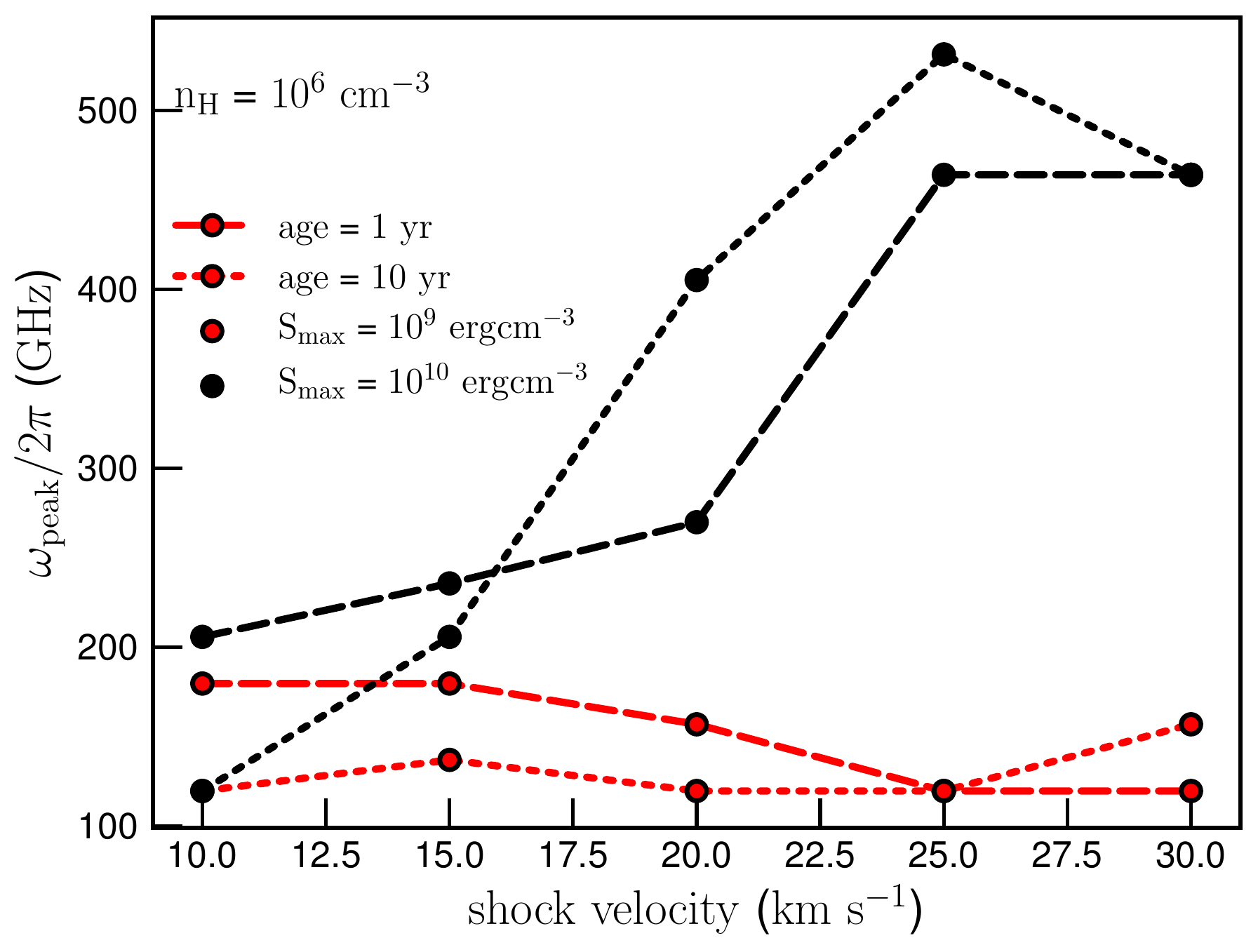}
\caption{Peak emission frequency corresponding to Figure \ref{fig:fluxmax_vs}. Peak frequency tends to increase with $v_{s}$.}
\label{fig:omegamax_vs}
\end{figure}

Similar to a previous study \citep{2019ApJ...877...36H}, we propose a new technique to trace shocks, which uses continuum microwave emission from spinning dust beside the popular technique, i.e., observing molecular emission lines in outflows and young SNRs.

Figure \ref{fig:fluxmax_vs} shows the maximum flux (peak flux) of spinning dust emission as a function of the shock velocity for the different shock models and tensile strengths. For strong grain materials, the peak flux increases rapidly with the shock velocity due to increased drift velocity and gas temperature (black lines). For weaker grain materials, the peak flux first increases rapidly with $v_{s}$ and then slowly varies for high velocities of $v_{s}>15-20\km\s^{-1}$ where rotational disruption occurs to remove smallest nanopaticles (red lines). The peak flux in the older shock is higher than in the younger shock due to dominance of J-shock component in younger one. 

Figure \ref{fig:omegamax_vs} shows the peak frequency corresponding to peak flux as a function of the shock velocity for the different shock models. For strong grain materials, the peak frequency increases with increasing shock velocity up to $v_{s}\sim 25\km\s^{-1}$ (black lines). For weaker grain materials, the peak frequency first increases rapidly with $v_{s}$ and then decreases slowly for higher velocities of $v_{s}>15-20\km\s^{-1}$ due to rotational disruption (red lines). The peak frequency of strong dust grains consequently longer than of the weak ones.

We note that the peak flux decreases while the peak frequency increases with increasing the gas density (see Figures \ref{fig:fluxmax_vs} and \ref{fig:omegamax_vs}) which arises from the fact that the shock length is narrower for higher gas density (see Figures \ref{fig:profile_temp}). The reason for this feature is particularly obvious at low velocities of $v_{s}< 15\km\s^{-1}$ where the spectral flux just only depends on the shock length, because the rotational disruption is ineffective at such low shock velocities.

\subsection{Tracing nanoparticles in shocks with spinning dust}

In shock regions, despite the fact PAHs and nanoparticles are thought to be abundant (\citealt{1996ApJ...469..740J}; \citealt{2011A&A...527A.123G}), there is no direct observations to support their formation mechanism. Possibly, we show that spinning dust comes out as a very strong emission in shock regions. In comparison with thermal dust emission, the spectral flux of spinning dust emission is dominant over at frequencies below $\sim 100$ GHz, as well as this emissivity is several orders of magnitude higher than thermal one when the rotational disruption is disregarded. Therefore, we can trace nanoparticles in outflows from young stellar objects using spinning dust. This technique can be unique because of the lack of optical/UV-photons to trigger mid-IR emission in dense medium, and would be tested by future radio observations with ALMA Band and ngVLA. 

\subsection{Implication for grain chemistry: rotational desorption}
Icy grain mantles play a very important role in interstellar chemistry. Ice mantles catalyze chemical reactions at its surface, which allows for the formation of molecules. However, how newly formed molecules are returned to the gas from the ice mantle is not well understood. Several mechanisms have been proposed, including evaporation/sublimation, desorption (i.e., thermal, non-thermal, cosmic ray, chemical desorption), and sputtering (see, e.g., \citealt{2016A&A...585A..24M} and references therein). Here, we discuss a new mechanism that may be important for releasing molecules in shocks termed rotational desorption.

In shocks propagating through dense clouds, nanoparticles can be spun-up to suprathermal rotation, at rates $\omega \gtrsim 10^{11}\s^{-1}$ (see Figure \ref{fig:omega_cri}). For nanoparticles made of strong materials ($S_{\max}\gtrsim 10^{10}\erg\cm^{-3}$) which can withstand the rotational disruption, the centrifugal force acting on a molecule can exceed its binding force to the grain surface, resulting in the ejection of the molecule. Note that the ejection of individual molecules assumed here is appropriate for monolayer ices, which is plausible for nanoparticles. For large grains, the entire mantle can be disrupted into fragments followed by rapid evaporation of molecules from the fragments if the ice mantle is thick (\citealt{Hoang:2019td}) or the centrifugal force can assist the thermal sublimation via so-called ro-thermal desorption effect (\citealt{Hoang:2019wra}).

The centrifugal force acting on a molecule of mass $m$ when the nanoparticle rotates with an angular velocity $\omega_{rot}$ is:
\bea
    F_{cen}=m\omega_{rot}^{2}a,
\ena
where $\omega_{rot}$ is defined in Equation (\ref{eq:omega_Trot}). 

The binding force of the molecule to the grain surface is estimated as:
\bea
    F_{bind}\sim \frac{\partial U}{\partial r}\sim \frac{U}{r},
\ena
where $U$ is the binding energy between the molecule and grain surface, and $r\sim 3\AA$ is the interaction distance between the molecule and the grain surface. 

The molecule is ejected when and only when $F_{cen}\gtrsim F_{bind}$, which yields:
\bea
    a \lesssim 7.27 \left(\frac{m}{m_{CO}}\right)^{1/4} \left(\frac{T_{rot}}{10^{4} K}\right)^{1/4}\left(\frac{U}{300\K}\right)^{-1/4}\ \mbox{(\AA).}
\ena

The rotational temperature $T_{rot}$ varies with grain radius and its location in the shock. It is also different in the C-shock or J-shock component (see Fig. \ref{fig:Trot_a}). Regardless of all these details, however, the average value of $T_{rot}$ is order of $10^{4}\K$. The value of the binding energy $U$ of several molecules with an icy grain surface is listed in Table \ref{tab:rot_desorp}. Adopting all these quantities, we can estimate the critical grain size of which molecules can be desorbed due to centrifugal force. The results are shown in Table \ref{tab:rot_desorp}. For instance, the centrifugal force on a nanoparticle of size $a\leq 3.31\AA$ can be large enough to eject H$_2$, or that of size $a\leq 5.21\AA$ for N$_{2}$. 

\begin{table}
\begin{center}
\caption{Critical grain size of rotational desorption for some molecules (\citealt{2016A&A...585A..24M}).}\label{tab:rot_desorp}
\begin{tabular}{lll} \hline\hline
{\it Molecules} & {$\rm U_{ice}\ (K)$} & {$\rm a_{critical}\ (\AA)$}\cr
\hline\\
$\rm H_{2}$    & 500  & 3.31\cr
$\rm OH$       & 4600 & 3.24\cr
$\rm H_{2}O$   & 4800 & 3.25\cr
$\rm CO$       & 1300 & 5.04\cr
$\rm CO_{2}$   & 2300 & 4.89\cr
$\rm CH_{3}O$  & 3700 & 3.98\cr
$\rm CH_{3}OH$ & 3700 & 4.01\cr
$\rm N_{2}$    & 1140 & 5.21\cr
\cr
\hline
\cr
\end{tabular}
\end{center}
\end{table}

\subsection{Effects of charge fluctuations on grain rotation in CJ-shocks}\label{sec:chargefluc}
In the present paper, we have assumed that dust grains in the shock can be statistically described by two populations of neutral and charged grains as in \cite{2019ApJ...877...36H}. The neutral grains only experience rotational excitation by thermal collisions with the neutral gas, whereas the latter charged grains coupled to the magnetic field can experience excitation by supersonic drift of neutral atoms. The damping and excitation coefficients are then averaged by multiplying with the charge distribution $f(Z)$ that varies with the grain location in the shock (see Equations \ref{eq:F} and \ref{eq:G}). Note that a detailed treatment of grain rotational dynamics should take into the effect of charge fluctuations of grains in the shock, as previously studied by \cite{2007A&A...476..263G} for the translational dynamics. Therefore, we now consider to what extent our assumption can hold and the impact of charge fluctuations on our obtained results.

Let $\tau_{0}$ be the charging time, which is equivalent to the timescale on which the grain stays in the neutral charge state before being attached by an electron or an ion. When the shock speed is much larger than the sound speed (i.e, the March number $M\gg 1$), one has $T_e/m_e \approx T_i/m_i$ where $T_{e}$ and $T_{i}$ are electron and ion temperatures, which implies that electrons and ions move at the same speed (\citealt{2015A&A...579A..13V}). For a neutral grain, the Coulomb effect is absent, and the neutral-timescale can be roughly estimated as follows (see \citealt{1987ApJ...320..803D}):
\bea \label{eq:tau_0}
    \tau_0 &\simeq& \frac{1}{2\times s_e n(e) \pi a^{2} \left(8k_{B}T_e/\pi m_e\right)^{1/2}}\\ \nonumber
    &\simeq& 2.3 \times a^{-2}_{-7} \left( \frac{T_e}{5\times 10^3 \K}\right)^{-1/2} \left[\frac{x(e)}{10^{-6}}\right]^{-1} \left(\frac{n_H}{10^4\cm^{-3}}\right)^{-1} \rm{yr}
\ena
where the sticking coefficient of electrons $s_e=0.5$ is assumed (see \citealt{2001ApJS..134..263W}; \citealt{2007A&A...476..263G}), $x(e)=n(e)/n_H$ is the abundance of electrons. Note that $T_e>T_n$ (see \citealt{1996MNRAS.280..447F}; \citealt{2004A&A...427..147L}).

The gas drag timescale is defined as the time required for the grain to loose its momentum, which is essentially the same as the time required to collide with the gas of the same grain mass. For completely inelastic gas-neutral grain collisions, the gas drag time is equal to
\bea \label{eq:tau_drag}
    \tau_{\rm drag} &\simeq& \frac{m_{grain}}{1.3\times m_{\H}n_{\H} \pi a^{2} v_{drift}} =\frac{\rho (4\pi a^{3}/3)}{1.3\times m_{\H}n_{\H} \pi a^{2} v_{drift}} \\ \nonumber
    &\simeq& 0.22 \times a_{-7} \left(\frac{n_H}{10^4 \cm^{-3}}\right)^{-1} \left(\frac{v_{drift}}{20 \km \s^{-1}}\right)^{-1} \rm{yr}.
\ena

From Equation \ref{eq:tau_0} and Equation \ref{eq:tau_drag}, one can derive the grain size at which $\tau_{0}=\tau_{\rm drag}$ as follows
\bea
     a_0 = 2.2 \times \left(\frac{T_e}{5\times 10^3\K}\right)^{-1/6} \left[\frac{x(e)}{10^{-6}}\right]^{-1/3} \\ \nonumber \times \left(\frac{v_{drift}}{20\km \s^{-1}}\right)^{1/3} \rm{nm}.
\ena

Therefore, for very small grains of size $a<a_0$ with $\tau_0>\tau_{\rm drag}$, the charge states of dust grains hardly change before they are stopped by the gas drag force. For this regime, the average of the rotational damping and excitation coefficients ($F$ and $G$) over the charge distribution $f(Z)$ are valid. For larger grains of $a>a_0$ with $\tau_0<\tau_{\rm drag}$, on contrary, the charge state of grains change rapidly. In this regime, the grains with initially neutral charge state quickly become charged and then coupled to ions. Therefore, our assumption of two neutral and charged populations become inapplicable. Practically, our calculations show that $f(Z=0)$ is rather small for $a>2 nm\sim a_{0}$, such that the $F$ and $G$ coefficients computed with Equations (\ref{eq:F}) and (\ref{eq:G}) and consequently $T_{rot}/T_{n}$ (Figure \ref{fig:Trot_a}) become nearly flat for $a>a_{0}$. Moreover, the disruption sizes that we obtained with this approach is mostly below $\sim 2$ nm (see, e.g., Figure \ref{fig:a_cri}). As a result, we expect that switching from our two separate neutral and charged grain populations to the one grain population (i.e., neutral and charged grains are coupled) for $a>a_{0}$ would not considerably change our final results on $a_{\rm disr}$ and then spinning dust emissivity.



\section{Summary}\label{sec:sum}
We study rotational dynamics of nanoparticles in non-stationary shocks (namely CJ-shock) driven by outflows and supernova remnants in dense molecular clouds. Our principal results are summarized as follows:

\begin{itemize}
\item[1] For the first time, we study the rotation dynamics of dust grains in CJ-shocks driven by outflows from young stars and by young supernova remnants in dense clouds, which consist of the C-shock and J-shock components. In the C-shock part, nanoparticles can be rapidly spun-up to suprathermal rotation due to supersonic drift of neutral relative to charged nanoparticles. In contrast, in the J-shock part, nanoparticles subthermally rotate due to the lack of supersonic drift between neutral and charged grains.

\item[2] We show that, in both C-shock and J-shock components, smallest nanoparticles ($a<1$ nm) of weak materials (tensile strength $S_{\rm max}\lesssim 10^{9}\erg\cm^{-3}$) can be disrupted by centrifugal stress due to extremely fast rotation, which is caused by suprathermal rotation in the C-shock part and thermal rotation with enhanced gas temperatures in the J-shock part, respectively. However, strong materials (e.g., nanodiamonds) can withstand the rotational disruption in CJ-shocks of velocity $v_{s}\le 30\km\s^{-1}$.

\item[3] We compare the timescale of rotational disruption with other destruction mechanisms in CJ-shocks and find that rotational disruption is the most efficient mechanism. Hence, the minimum size of nanoparticles should be constrained by the rotational disruption rather than by thermal sputtering. This rotational mechanism might play an important role in dust destruction in shock regions.

\item[4] We model the microwave emission from spinning nanoparticles in CJ-shocks where their minimum size is determined by the rotational disruption. In the absence of rotational disruption, the peak frequency and emissivity are found to increase with increasing the shock velocity. In the presence of rotational disruption, the peak frequencies decrease with the shock velocity because of removal of smallest nanoparticles.

\item[5] We find a new way to eject molecules from surface of strong nanoparticles that can withstand rotational disruption, so-called rotational desorption, which is based on centrifugal force within extremely fast rotating nanoparticles in CJ-shocks.

\item[6] We propose that spinning dust emission could be a new technique to probe nanoparticles and shock velocities in outflows and supernova remnants. 
\end{itemize}

\acknowledgments
 We are grateful to the anonymous referees for useful comments that improved our paper. We would like to thank P. Lesaffre, A. Gusdorf, and V. Guillet for all helpful comments and discussions. This work was supported by the Basic Science Research Program through the National Research Foundation of Korea (NRF) grants, funded by the Ministry of Education (2017R1D1A1B03035359) and the Korea government (MSIT) (2019R1A2C1087045).



\bibliography{ms.bbl}

\end{document}